\documentclass[acmtog, screen, nonacm]{acmart}

\usepackage{booktabs} 
\usepackage{graphicx}
\graphicspath{{figs/}{./}}

\usepackage{subcaption} 

\usepackage[format=plain,
labelfont={bf},
textfont=it]{caption}

\usepackage{color, colortbl}
\definecolor{LightCyan}{rgb}{0.95,1,1}

\usepackage{makecell}
\usepackage{multirow}
\usepackage[makeroom]{cancel}
\usepackage{empheq}

\usepackage{algorithm2e} 
\SetAlFnt{\small}
\SetAlCapFnt{\small}
\SetAlCapNameFnt{\small}
\SetAlCapHSkip{0pt}

\copyrightyear{2024}
\acmYear{2024}

\usepackage{booktabs}
\usepackage{bbold}

\newcommand{\secref}[1]{\S~\ref{sec:#1}}
\newcommand{\tabref}[1]{Tab.~\ref{tab:#1}}
\newcommand{\figref}[1]{Fig.~\ref{fig:#1}}
\renewcommand{\eqref}[1]{Eq.~(\ref{eq:#1})}
\newcommand{\algref}[1]{Algo.~(\ref{alg:#1})}
\newcommand{\alglref}[1]{Line~(\ref{lin:#1})}
\newcommand{\appref}[1]{Appendix~\ref{app:#1}}
\newcommand{\etal}{\emph{et al.~}}
\newcommand{\eg}{\emph{e.g.}}
\newcommand{\ie}{\emph{i.e.}}
\newcommand{\etc}{\emph{etc.}}
\newcommand{\cf}{\emph{cf.}}

\newcommand{\mbJ}{\ensuremath{\mathbf{J}}}
\newcommand{\mbR}{\ensuremath{\mathbf{R}}}

\newcommand{\mbS}{\ensuremath{\mathbf{S}}}

\newcommand{\mbV}{\ensuremath{\mathbf{V}}}
\newcommand{\mbU}{\ensuremath{\mathbf{U}}}

\newcommand{\mbQ}{\ensuremath{\mathbf{Q}}}

\newcommand{\mbI}{\ensuremath{\mathbf{I}}}

\newcommand{\mbM}{\ensuremath{\mathbf{M}}}
\newcommand{\mbN}{\ensuremath{\mathbf{N}}}

\newcommand{\mbc}{\ensuremath{\mathbf{c}}}
\newcommand{\mbe}{\ensuremath{\mathbf{e}}}
\newcommand{\mbf}{\ensuremath{\mathbf{f}}}
\newcommand{\mbx}{\ensuremath{\mathbf{x}}}
\newcommand{\mbg}{\ensuremath{\mathbf{g}}}

\newcommand{\mbq}{\ensuremath{\mathbf{q}}}
\newcommand{\mbn}{\ensuremath{\mathbf{n}}}

\newcommand{\mbd}{\ensuremath{\mathbf{d}}}
\newcommand{\mbs}{\ensuremath{\mathbf{s}}}

\newcommand{\vecOp}{\ensuremath{\mathrm{vec}}}

\newcommand{\realNum}{\ensuremath{\rm I\!R}}

\DeclareMathOperator*{\argmin}{arg\,min}

\begin{document}
	
	\title{GIPC: Fast and stable Gauss-Newton optimization of IPC barrier energy }
	
	\author{Kemeng Huang}
	\orcid{0000-0001-9147-2289}
	\email{kmhuang@connect.hku.hk}
	\email{kmhuang819@gmail.com}
	\affiliation{%
		\institution{The University of Hong Kong, TransGP}
		\country{Hong Kong}
	}
	
	\author{Floyd~M.~Chitalu}
	\authornote{Currently unaffiliated. Was at The University of Hong Kong while contributing to this work.}
	\orcid{0000-0001-9489-8592}
	\email{floyd.m.chitalu@gmail.com}
	\affiliation{%
		\institution{The University of Hong Kong}
		\country{Hong Kong}
	}
	
	\author{Huancheng Lin}
	\orcid{0000-0003-4446-1442}
	\email{lamws@connect.hku.hk}
	\affiliation{%
		\institution{TransGP, The University of Hong Kong}
		\country{Hong Kong}
	}
	
	\author{Taku Komura}
	\orcid{0000-0002-2729-5860}
	\email{taku@cs.hku.hk}
	\affiliation{%
		\institution{The University of Hong Kong, TransGP}
		\country{Hong Kong}}
	
	\renewcommand\shortauthors{Huang \etal}
	
	\begin{abstract}
		{
			Barrier functions are crucial for maintaining an intersection and inversion free simulation trajectory but existing methods which directly use distance can restrict implementation design and performance. We present an approach to rewriting the barrier function for arriving at an efficient and robust approximation of its Hessian. The key idea is to formulate a simplicial geometric measure of contact using mesh boundary elements, from which analytic eigensystems are derived and enhanced with filtering and stiffening terms that ensure robustness with respect to the convergence of a Project-Newton solver. A further advantage of our rewriting of the barrier function is that it naturally caters to the notorious case of nearly-parallel edge-edge contacts, for which we also present a novel analytic eigensystem. Our approach is thus well suited for standard second order unconstrained optimization strategies for resolving contacts, minimizing nonlinear nonconvex functions where the Hessian may be indefinite. The efficiency of our eigensystems alone yields a {3}$\times$ speedup over the standard IPC barrier formulation. We further apply our analytic proxy eigensystems to produce an entirely GPU-based implementation of IPC with significant further acceleration. 
			
		}
	\end{abstract}

	%
	%


	%
	%


	\begin{teaserfigure}
		\centering
		\begin{subfigure}[b]{0.245\textwidth}
			\centering
			\includegraphics[width=\columnwidth, trim=600 150 400 100, clip]{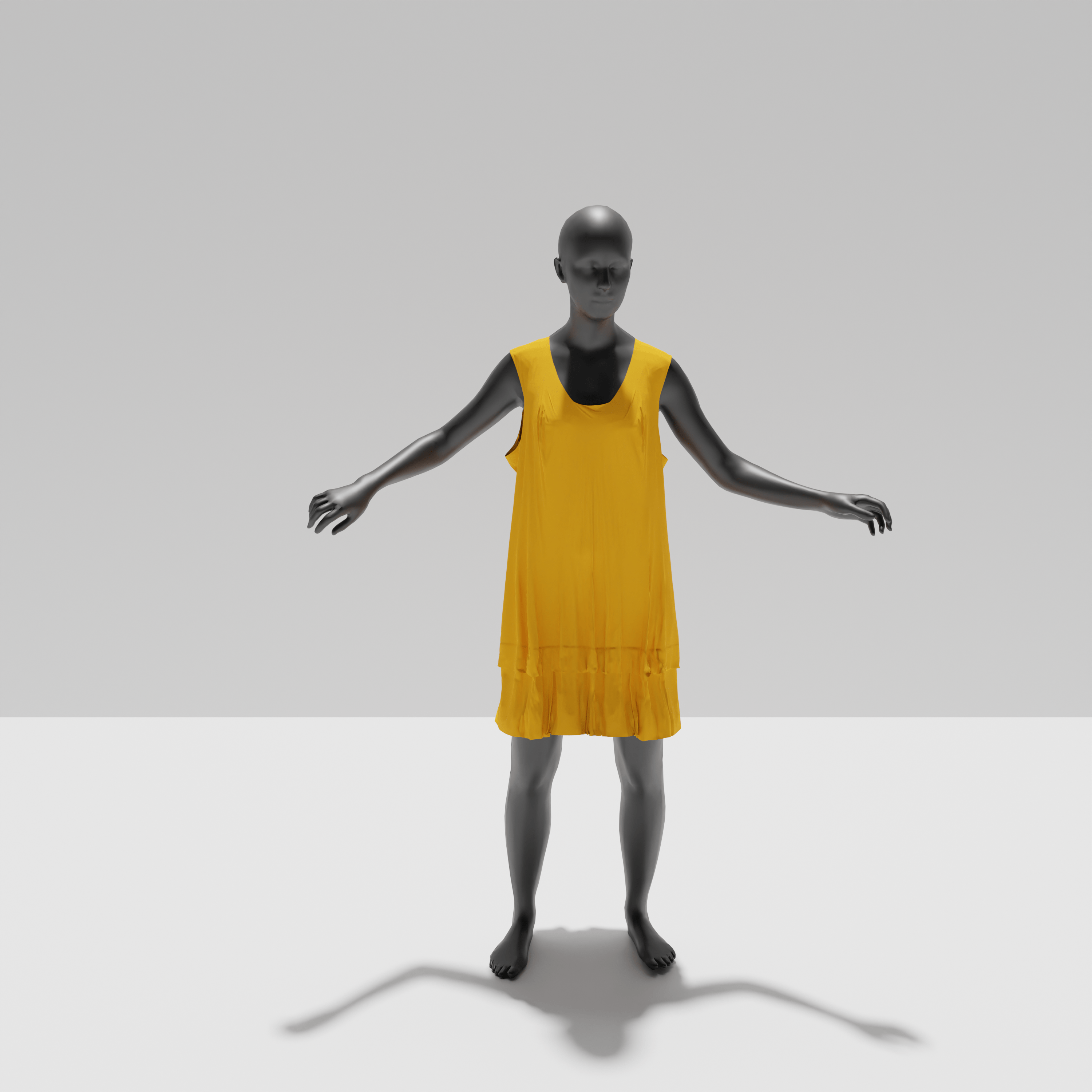}
		\end{subfigure}
		\begin{subfigure}[b]{0.245\textwidth}
			\centering
			\includegraphics[width=\columnwidth, trim=200 150 800 100, clip]{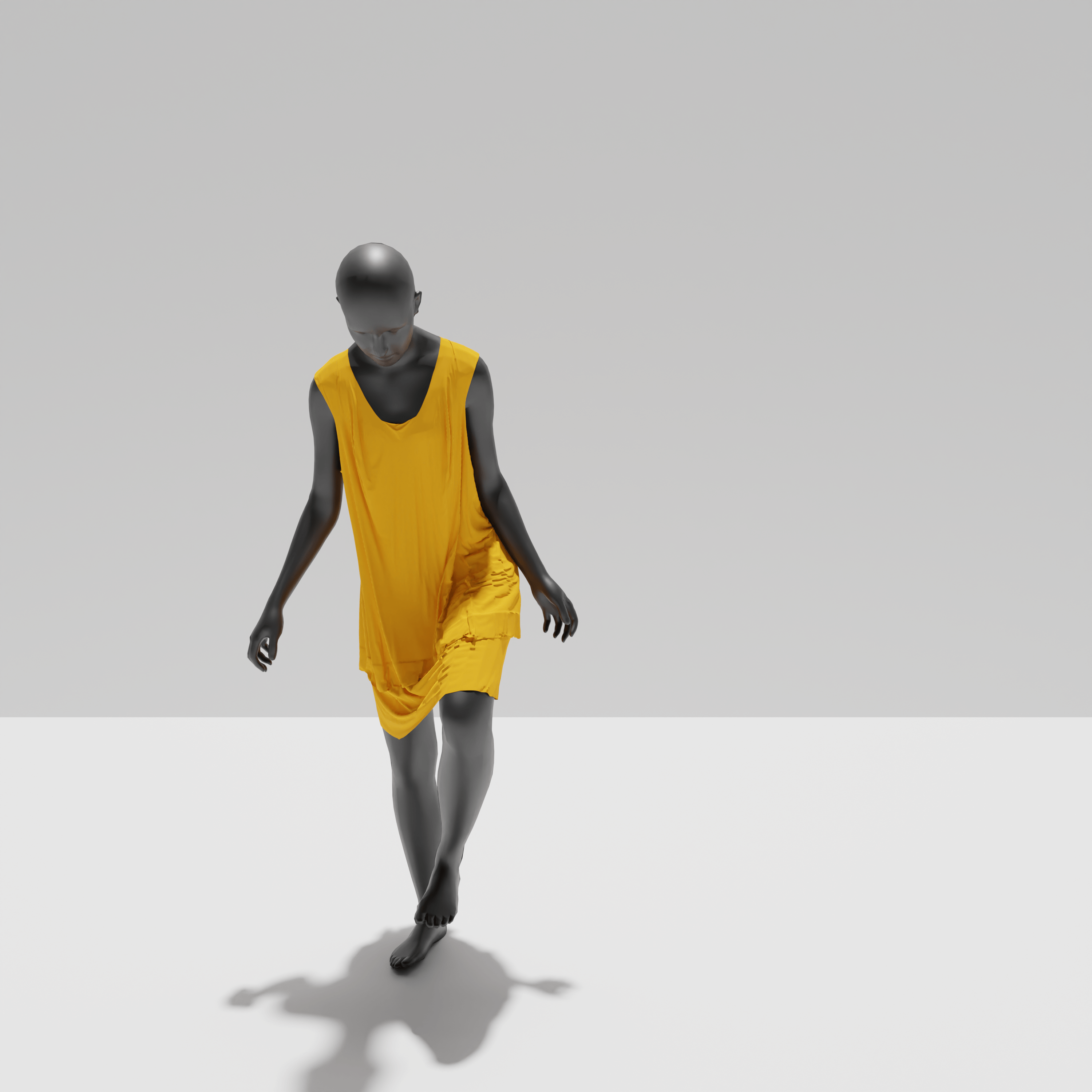}
		\end{subfigure}
		\begin{subfigure}[b]{0.245\textwidth}
			\centering
			\includegraphics[width=\columnwidth, trim=350 150 650 100, clip]{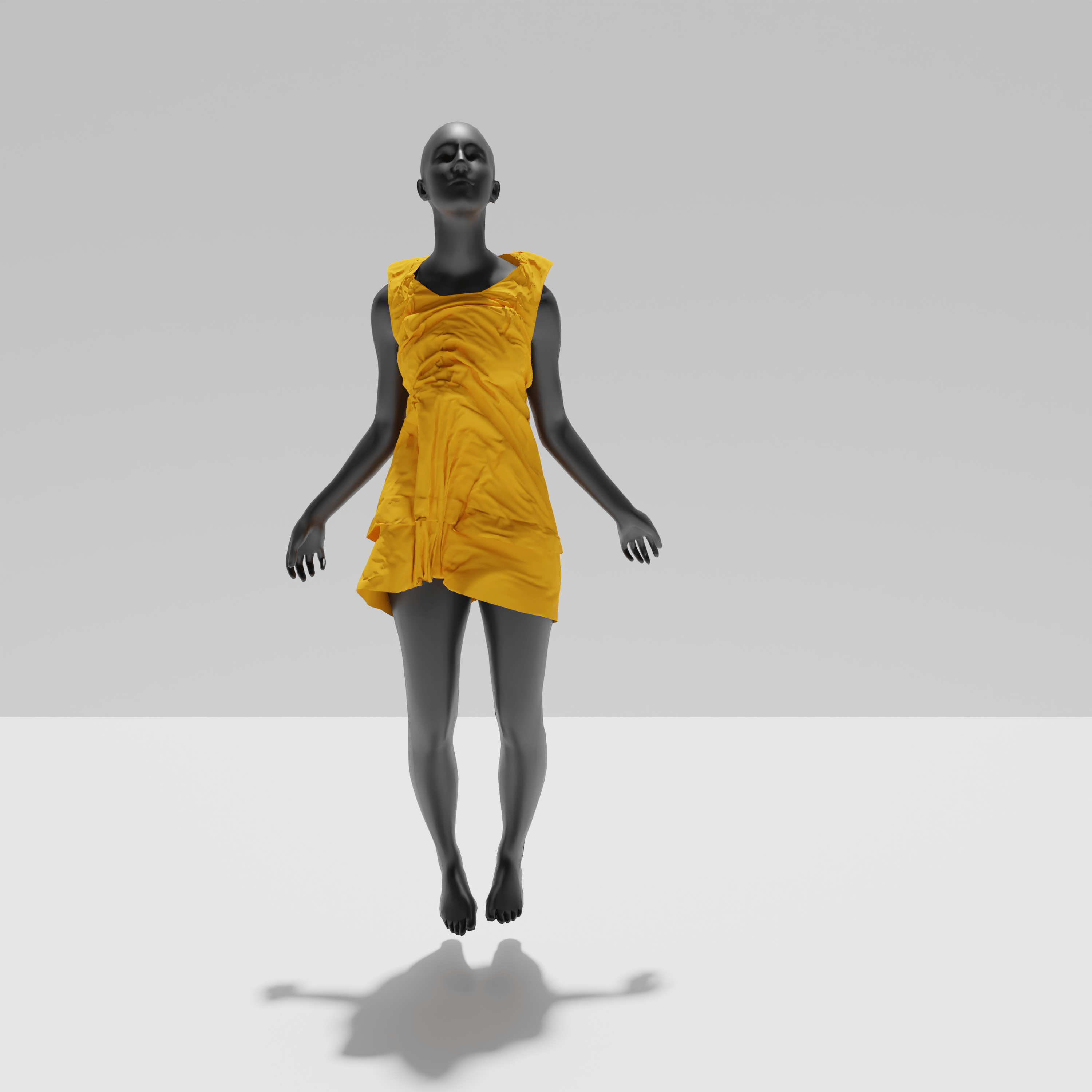}
		\end{subfigure}
		\begin{subfigure}[b]{0.245\textwidth}
			\centering
			\includegraphics[width=\columnwidth, trim=450 150 550 100, clip]{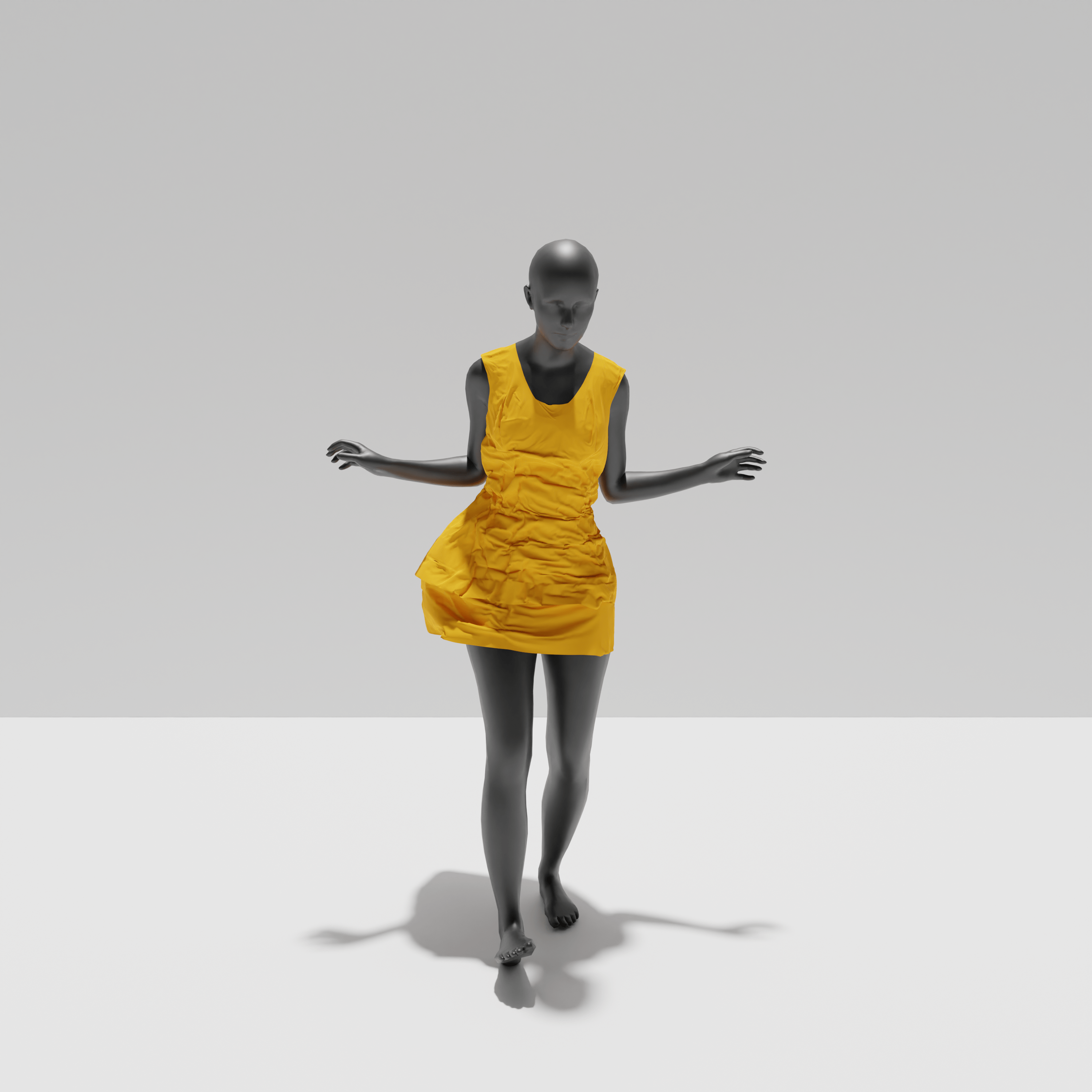}
		\end{subfigure}
		\caption{\label{fig:cloth-sim}We offer a robust method enabling simultaneous construction and projection of approximated IPC barrier Hessians to positive semi-definite state for faster implicit time integration. A multilayered cloth animation is shown, where contacts are resolved using our method, which ensures fast convergence rates without numerical eigendecompositions of local barrier Hessians.
		}
	\end{teaserfigure}
	
	\maketitle

	\section{Introduction}
	Incremental Potential Contact (IPC) \cite{10.1145/3386569.3392425} is the state-of-the-art for simulating intersection and inversion free updates using implicit integration with large time steps. The main idea of IPC is to formulate non-penetration constraints of mesh boundary elements using distances for constructing a simulator with non-penetration guarantees. These \textit{unsigned} distances are inputs to barrier potentials, representing the contact constraints added to the global {potential} function that is minimized during the updates.

	However, physical simulation with robust contact dynamics remains difficult to optimize numerically due to the occurrence of nonconvex local energy landscapes. In particular, second order Newton methods can diverge or stall when the global energy Hessian is indefinite by containing negative eigenvalues \cite{NoceWrig06}. Of particular interest is the Projected Newton (PN) method that projects local energy Hessians to the positive semi-definite (PSD) cone prior to assembly of the global Hessian (see \eg~\citet{10.1145/1073368.1073394} and \citet{10.1145/3241041}). PN is especially relevant since IPC barrier functions produce locally indefinite Hessians with unknown conditions under which this indefiniteness occurs \cite{SiggraphContact22}.

	Numerical eigendecomposition of the IPC barrier Hessian is the primary solution available to address indefiniteness when using the PN method, which on the other hand hinders the acceleration of the IPC runtime. Numerical eigendecomposition requires iterative computation, which is rather slow and also difficult to parallelize on the GPU. As a result, existing massively parallel implementations of IPC require passing the data between the CPU and GPU, which adds extra overhead to the system.
	
	Analytic eigenanalysis, which is often applied in the simulation of deformable objects, can be an alternative for systematically projecting Hessian to PSD. Indeed, in a recent analysis of penalty-based collision energies \cite{10.1145/3606934}, closed-form eigensystems were proposed for both signed and unsigned barrier energies, aiming to circumvent numerical eigendecomposition. Nevertheless, this approach encounters severe convergence problems due to the lack of further consideration of nearly parallel edge-edge contacts (see \cite{10.1145/3386569.3392425} for details) and the vanishing displacement problem that we encounter when simulating relatively stiff materials, where the contact primitives get trapped in configurations with minuscule distances, limiting their effectiveness. At such configurations,  the second-order derivative of the barrier energy increases rapidly compared to the corresponding first-order derivative, resulting in stalled displacements during Newton iterations. 
	
	In this paper, we propose a novel IPC framework that is based on a revised barrier function that forms a Gauss-Newton solver, which overcomes the issues faced by existing approaches. We revisit the log barrier function of IPC, augmenting its definition to arrive at an approximate Hessian, where the eigensystem is derived and evaluated analytically for amortizing the cost of assembly and PSD projection. The resulting expressions apply to all configurations of contact (\ie~\textit{point-edge}, \textit{point-triangle}~\etc ), where we also present a novel analytic eigensystem for the notorious nearly-parallel \textit{edge-edge} case. Following our analysis to tackle issues that relate to vanishing displacements, we propose a method for further enhancing the eigensystem of our approximated Hessian to exhibit robust behavior when dealing with distances close to zero. This method involves applying eigenvalue filtering before computing search directions and elevating the barrier profile by steepening its gradient.
	
	Our method provides up to 3$\times$ speedup over the distance based formulation of full-space IPC, which is significant given that a few minutes' worth of simulation typically requires tens of hours to compute and that we obtain visually matching results. We separately demonstrate the flexibility of our analytic eigensystems by engineering a massively parallel implementation of IPC running \textit{entirely} on the GPU (since we bypass need for numerical eigendecompositions) to achieve significant speedups over the publicly-available CPU implementation of \citet{10.1145/3386569.3392425}. Within this system, we also apply a matrix-free Preconditioned Conjugate Gradient solver (PCG) with a carefully designed preconditioner that is amenable to GPU data-parallelism for further accelerating nonlinear optimization with Newton's method. We also demonstrate that by using PCG with a relatively low tolerance (\eg~$1\mathrm{e}{-4}$), our system can achieve results that are comparable to a direct solver within Newton's method.

	A summary of our contributions is as follows:
	\begin{itemize}
		\item A formulation of the IPC barrier function to allow efficient Hessian approximations for optimization, {which gives local search directions that are aligned with the contact normal. We combined this with our solution to the problem of vanishing displacements for improving convergence of during IPC optimization (\secref{barrier}).}
		\item A novel construction of the contact constraint Jacobian from which we measure penetration state. This Jacobian can be extended to handle near-parallel edge-edge cases, enabling approximated eigenanalysis in such scenarios.
		\item A new mollified barrier function expressed in terms of our constraint Jacobian (\secref{par-ee}), where we provide analytic eigensystems of its approximate Hessian for fast and robust minimization (\secref{moll-bar-Hess-ana}).
		\item A detailed analysis showing that by using a relatively large  tolerance, PCG can still efficiently achieve accurate results in nonlinear optimization while using Newton's method.
		\item The first full-GPU implementation of IPC, without sacrificing accuracy, that is combined with extensive and rigorous comparisons. 
	\end{itemize}
	
	\section{Related Work}
	\label{sec:relatedwork}
	In this section, we review collision detection and response methods covering IPC, eigenanalysis techniques and GPU accelerated simulation of deformable objects. 
	
	\paragraph{Collision Detection}
	Culling techniques based on spatial hashing~\cite{pabst2010fast,tang2018cloth,tang2018pscc} and Bounding Volume Hierarchies (BVH)~\cite{ERICSON2005235} are the basis for accelerating the collision processing with deformable objects (\ie~the broad/mid-phase). A popular representation is the linear-BVH (LBVH) \cite{LauterbachGSLM09} which reduces standard construction to a sorting problem by using spatial Morton codes and with several extensions \cite{Karras12,Apetrei14,WangTMT18,https://doi.org/10.1111/cgf.13948}. We refer readers to \citet{https://doi.org/10.1111/cgf.142662} for a review of recent developments in acceleration structures. 
	
	Continuous Collision Detection (CCD) also plays a vital role in the IPC algorithm and represents part of the the narrow-phase: CCD~\cite{ccd1997,WangFSJAP21} calculates the maximum feasible steps, thereby preventing inter-penetration during line search with simulations involving implicit time integration. We adopt additive CCD (ACCD)~\cite{Li2021CIPC}, a numerically stable scheme to iteratively accumulate a lower-bound to convergence-time of impact before collision, which improves efficiency in our parallel implementation on the GPU.
	
	\paragraph{Collision Response}
	When simulating the contact with deformable objects, the next state must be computed such that all the  penetrations are resolved. Methods based on impulse computation~\cite{bridson2002robust,sifakis2008globally}, impact zone methods~\cite{RTSC,provot1997collision} and constraint solvers~\cite{otaduy2009implicit,li2015deformable,kaufman2008staggered} are proposed. Earlier works are based on iterative solvers that resolve collisions by local adjustments~\cite{bridson2002robust,RTSC,provot1997collision}. Such local adjustment can result in adding energy into the system, thus potentially causing instability as a simulation proceeds.  
	
	Constraint-based methods describe the combinatorial nature of all possible contact states \cite{otaduy2009implicit,li2015deformable,verschoor2019efficient,10.1145/3338695}. For example, \citet{otaduy2009implicit} propose an implicit Linear Complementarity Problem (LCP) based approach that globally resolves the constraints. Li et al.~\shortcite{li2015deformable} present an efficient gradient projection method for computing contact responses by decoupling constraints, while Verschoor and Jalba~\shortcite{verschoor2019efficient} tackle the contact problem by directly solving a Mixed LCP (MLCP) and omitting the construction of an LCP matrix. These methods may also make use of signed distance functions (SDF) or their approximations for constructing the constraints. The linearized constraints can be invalidated when the output displacement of the nodes are large ~\cite{erleben2018methodology}. To cope with these issues of SDFs, \citet{10.1145/3386569.3392425} instead use unsigned distances which then parameterize barrier potentials that are added to a global objective function, resulting in an unconstrained optimization problem of contact. We follow a similar path but address the computational challenge of deriving and simulating with a fast and robust Gauss-Newton approximation scheme that remains efficient with a large number of collision stencils.   
	
	Alternatively, volume-based approaches are proposed~\cite{allard2010volume,muller2015air,sifakis2008globally,jiang2017simplicial} where constraints based on the open-space between the collision pairs are imposed. \citet{muller2015air} pre-compute a volumetric mesh in the open-space and impose positive volume constraints. \citet{jiang2017simplicial} propose a similar shape editing framework to avoid self-penetration. These approaches are only valid when the configuration does not change significantly over time; otherwise an expensive re-meshing process is needed. \citet{sifakis2008globally} (see also \cite{KANE19991}) construct tetrahedra between the colliding pairs every frame and preserve their volume to avoid penetration, where artificial ghost contact forces may appear when sheared to give false positives due to rotations. Our method resembles \citet{sifakis2008globally} but we eliminate influence of extraneous rotation by measuring volume via a equivalent strain-like metric that is invariant to such rotations.
	
	\paragraph{Incremental Potential Contact}
	IPC is proposed by \citet{10.1145/3386569.3392425} as a variational method (\ie~formulates a nonlinear system that is solved as an optimization problem) based on \citet{kane2000variational}'s incremental potential formulation to update a physical system while ensuring an intersection- and inversion-free simulation trajectory over the duration of the timestep. It has been extended to globally injective 3D shape deformation~\cite{Fang2021IDP}, intersection-free rigid body dynamics~\cite{RigidIPC,DBLP:journals/corr/abs-2201-10022} and simulation of co-dimensional objects~\cite{Li2021CIPC}. Unfortunately, IPC has a high computational cost, which may at-times require hundreds-of-seconds to compute one time step~\cite{10.1145/3386569.3392425}. \citet{MedialIPC} propose Medial-IPC to accelerate the standard model of IPC by using a reduced model to represent the object. Significant speedup is achieved by simulating with less collision pairs in the reduced search space but details of the geometric deformation are lost. { \citet{10.1145/3528223.3530069} combine IPC with Projective Dynamics (PD) to implement a full-space penetration-free simulation on GPU, replacing IPC distance-barrier energy with a formulation based on projected target-positions for their barrier constraint. Their framework is efficient but details of deformation are still lost compared with full-space IPC using finite-elements because PD does not guarantee complete physical accuracy and it is also sensitive to time-step size while our approach is practically unaffected (see \secref{results}, \figref{funnel}).}

	\paragraph{Eigen Analysis for Energy  Minimization}
	Minimizing contact constrained energies for physical problems is at the core of robust simulations involving elastic body dynamics. Classic 2nd order Newton methods \cite{NoceWrig06} can diverge or stall when the energy Hessian is indefinite by containing negative eigenvalues - thus projection to positive semi-definite state is required. For isotropic and anisotropic finite element (FE) energy Hessians, ~\citet{10.1145/3241041} and \citet{10.1145/3306346.3323014} (see also \cite{10.1145/3180491,10.1111/cgf.14111,https://doi.org/10.48550/arxiv.2008.10698,OurArapPaper}) present analytic expressions for the eigensystems of a wide range of distortion energies. Inline with the analyses of \citet{10.1145/3306346.3323014} and \citet{10.1145/3606934}, we extend and rewrite the barrier function to derive analytic eigensystems of its approximate Hessian for a fast and robust Gauss Newton optimization.

	\paragraph{GPU optimization}
	The GPU optimization of physical simulation is sought for accelerating high-quality and computationally demanding animations~\cite{MullerCG03,tang2016cama,tang2018cloth,li2020p,lauterbach2010gproximity,wang2021gpu}. For deformable objects, researchers have considered accelerating collision response based on zone impacts ~\cite{RTSC} on the GPU (see \eg~ \citet{tang2018cloth} and \citet{li2020p}), while others (\eg~\citet{lauterbach2010gproximity,https://doi.org/10.1111/cgf.13948}) accelerate proximity queries. Parallel techniques like \citet{wang2021gpu} also simulate sub-millimeter level cloth deformation with regular grids fitted to the garment. 
	
	In general, GPU parallelism has been under-utilized for optimizing IPC and its variants. The Affine Body Dynamics (ABD) method ~\cite{DBLP:journals/corr/abs-2201-10022} is presented as a data-parallel GPU method but a hybrid scheme is described, where \eg~ linear system solves are computed on the CPU. Medial IPC~\cite{MedialIPC} also partially adopts the GPU acceleration, \eg~ for vertex updates through time integration, but a significant portion of computation is still executed on the CPU - such as eigendecomposition, projection and construction of barrier Hessians, as well as solving of large linear systems which we instead compute with a novel data-parallel conjugate-gradient algorithm that scales optimally (\cf~\secref{implementation}). 
	
	\section{Background and preliminaries}  \label{sec:background}
	
	\subsection{IPC optimization and barrier functions}\label{sec:dist-and-bar-funcs}
	
	IPC optimization \cite{10.1145/3386569.3392425} is about finding a minimiser
	\begin{equation}\label{eq:energy-min}
	\mbx^{t+\Delta{t}}\approxeq\argmin_{\mbx \in \realNum^{{3n}}} \mathcal{I}(\mbx),
	\end{equation}
	representing the new state of a contact-constrained physical system over a period $\Delta{t}$ from time $t$. This new state is the minimiser of a global potential  $\mathcal{I}(\mbx)$, where this potential\footnote{The global potential function $\mathcal{I}(\mbx)$ is in general non-convex, which implies that numerical methods like PN cannot guarantee a global minimum, meaning that a local minimum is often reached instead as an \textit{approximation} of the true solution.} is defined by the total momentum, elastic, barrier, friction and other (\eg~external loading) terms for $n$ vertex locations in {3}-dimensional space stored in vector $\mbx$. To do this, an iterated approximation and stepping scheme is applied with local (quadratic) approximation of the potential
	\begin{equation}\label{eq:energy-approx}
	\mathcal{I}_i(\mbx) = \mathcal{I}(\mbx_i) + \left(\mbx - \mbx_i\right)^T\nabla \mathcal{I}(\mbx_i) + \frac{1}{2}\left(\mbx - \mbx_i\right)^T\nabla^2 \mathcal{I}(\mbx_i) \left(\mbx - \mbx_i\right).
	\end{equation}
	For each iteration $i$, the terms in \eqref{energy-approx} are evaluated: A linear solve $\nabla^2 \mathcal{I}(\mbx_i) \cdot \mbd = -\nabla \mathcal{I}(\mbx_i)$ determines the stationary point $\mbx_i^* = \argmin_{\mbx} \mathcal{I}_i(\mbx)$, where $\nabla^2 \mathcal{I}(\mbx_i)$ is symmetric and must be positive definite. The solution $\mbd = \mbx_i^* - \mbx_i$ is the direction of (probable) energy descent with which the new iterate $\mbx_{i+1} = \mbx_i + \alpha\mbd$ is computed, where $\alpha$ is a length parameter from a bounded line-search. The iterated approximation and stepping scheme is terminated when some quantity associated with the stationary point approaches zero - like the solution norm $\|\mbd\| \le \varepsilon_d$.

	In the remainder, we will focus on the total contribution to the global potential due to barrier energy, which is expressed as a sum over contacts $k$ in a collection $\mathcal{C}$ (of \textit{point-triangle}, \textit{point-edge} pairs \etc~ depending on $\mbx$)
	\begin{equation}\label{eq:total-barrier}
	\mathcal{B}(\mbx) = \sum_{k\in \mathcal{C}} b(d_k(\mbx)).
	\end{equation}
	(The remaining terms of the global potential $\mathcal{I}(\mbx_i)$ are treated in the same manner as \citet{10.1145/3386569.3392425}, where we also summarise our treatment of friction in the technical supplement). The local barriers $b(d_k(\mbx))$ are functions of distance $d_k(\mbx)$ that is evaluated with vertex locations in the stencil of a contact. This distance may be understood as between a point $v_{p}$ and a triangle $T = (v_{T1}, v_{T2}, v_{T3})$ 
	\begin{align}\label{eq:pt-dist-orig}
		\mathcal{D}^{PT} =& \min_{\beta_1,\beta_2}|| v_{p} - (v_{T1}+ \beta_1(v_{T2} - v_{T1}) + \beta_2(v_{T3} - v_{T1}))|| \nonumber\\
		&\textrm{s.t.} \quad \beta_1 \ge 0, \beta_2 \ge 0, \beta_1 + \beta_2 \le 1,
	\end{align}
	and between edges $v_{11} - v_{12}$ and $v_{21} - v_{22}$ 
	\begin{align}\label{eq:ee-dist-orig}
		\mathcal{D}^{EE} =& \min_{\gamma_1,\gamma_2}|| v_{11} + \gamma_1(v_{12} - v_{11}) - (v_{21} + \gamma_2(v_{22} - v_{21}))|| \nonumber\\
		&\textrm{s.t.} \quad 0 \le \gamma_1, \gamma_2 \le 1.
	\end{align}
	Then, a given combination of active constraints in \eqref{pt-dist-orig} and \eqref{ee-dist-orig} will determine the specific distance measure that is used between two potentially intersecting triangles (\eg~\textit{point-edge} distance).  
	This unsigned distance $d$ together with a computational accuracy target $\hat{d}$ then fully parameterizes the smoothly-clamped local barrier potential (\cf~Eq. (6) in \cite{10.1145/3386569.3392425})
	\begin{equation}\label{eq:ipc-barrier-orig}
	b(d, \hat{d})  = \begin{cases}
	- (\hat{d} - d)^2 \ln \left(\frac{d}{\hat{d}}\right), \quad 0 < d < \hat{d}\\
	0 \qquad d \ge \hat{d}.
	\end{cases}
	\end{equation}

	\subsection{Minimizing the barrier function} \label{sec:barrier-derivatives}
	Minimizing \eqref{ipc-barrier-orig} (summed over all contact pairs as in \eqref{total-barrier}) together with accompanying terms that define the global potential $\mathcal{I}(\mbx)$ gives the solution for contact-constrained dynamics with unconstrained optimization as in \eqref{energy-min}, where the order of differentiable energies is $C^2$ continuous. A solution method for this optimization will employ the approach outlined in \secref{dist-and-bar-funcs}. 
	
	We seek a solution method resembling PN and thereby require  
	forces  $\nabla_\mbx b(d, \hat{d})  \Rightarrow {\partial b(d, \hat{d})}/{\partial \mbx}$ and their Jacobian ${\nabla_\mbx^2} b(d, \hat{d}) \Rightarrow {\partial^2 b(d, \hat{d})}/{\partial \mbx^2}$, where this Jacobian must be PSD. Evaluating this local barrier force Jacobian is relatively straight-forward but \textit{always} indefinite to require numerical eigendecomposition. \citet{10.1145/3386569.3392425} as well as \citet{10.1145/3606934} explore Gauss-Newton approximations to eliminate additional projections but at the cost of reduced solver convergence rates. We present a novel barrier formulation permitting analytic eigensystems of a comparably approximate Hessian but without reduced convergence rates. 
	
	\section{Measuring contact distance}\label{sec:constraint-jac-gp-fn}
	
	We re-write the barrier energy in terms of a constraint Jacobian, which is a matrix representing the distortion of a simplex (tetrahedron, triangle or line) defined using vertex locations in the stencil of a contact. We describe in this section our approach to measuring distance using a novel construction of this constraint Jacobian and the contact normal vector from the vertices.

	\subsection{The gap function} \label{sec:barrier-energy} 
	We adopt the following function
	\begin{align}\label{eq:invariant}
		g(\mbx,\hat{d}) &= \|\mbJ(\mbx,\hat{d})\mbn(\mbx)\|_2^2 \nonumber\\ &\equiv \mbn(\mbx)^T\mbJ(\mbx,\hat{d})^T\mbJ(\mbx,\hat{d})\mbn(\mbx),
	\end{align}
	as our \textit{gap function} \cite{SiggraphContact22} for measuring distance between two contact primitives like a point and a triangle.  
	This function measures distance as a weighted vector-norm with the unit-length normal vector $\mbn(\mbx) \in \realNum^{3}$ of local contact, where $\mbx \in \realNum^{3{s}}$ is the vector of stacked positions of ${s}$ vertices {($s=2, 3$ or $4$)} defining the simplex. $\mbJ(\mbx,\hat{d})$ is the {constraint Jacobian} matrix encoding the directions of local contact forces and torques, which will be defined w.r.t the computational accuracy target $\hat{d}$: Example constructions are also provided in our technical supplement for reference but we detail our specific implementation in \secref{cp-dg}. 
	
	\eqref{invariant} is particularly appealing because the eigensystem of an energy expressed solely in its terms can be stated in closed-form (see \eg~\citet{10.1145/3306346.3323014}
	). It is also well suited to the task of measuring the barrier energy in contact configurations that prescribe volume (\textit{point-triangle}, \textit{edge-edge}), area (\textit{point-edge}) or length (\textit{point-point}), which we make use of.

	\subsection{Evaluating the gap function}\label{sec:cp-dg}
	Evaluating the gap function in \eqref{invariant} requires that we compute the Jacobian $\mbJ(\mbx,\hat{d})$ and normal $\mbn(\mbx)$ from the vertices of the stencil that have locations $\mbx$ . We will describe a method to compute $\mbJ(\mbx,\hat{d})$, which also gives in an implicit expression for $\mbn(\mbx)$. This implicit expression for $\mbn(\mbx)$ comes from the observation that the nearest two points between \textit{any} pair of contact primitives (\eg~a point and an edge/triangle from-which we will construct $\mbJ(\mbx,\hat{d})$) prescribe the normal $\mbn(\mbx)$ as the nearest direction of contact (\cf~\eqref{pt-dist-orig} and \eqref{ee-dist-orig}).
	Moreover, we seek a construction of $\mbJ(\mbx,\hat{d})$ satisfying 
	\begin{equation}
	\label{eq:J-constraint}
	\mbJ(\mbx,\hat{d})\mbn(\mbx) = \sigma\mbn(\mbx), 
	\end{equation}
	as the zero tangent force condition, where the normal $\mbn(\mbx)$ is a singular vector of this $\mbJ(\mbx,\hat{d})$ and $\sigma, 0\le \sigma \le 1$ is the corresponding singular value.

	\paragraph{The singular value perspective}
	Our solution is based on reducing $\mbJ(\mbx,\hat{d})$ to a diagonal matrix and simplifying $\mbn(\mbx)$ to a standard basis vector, thereby satisfying \eqref{J-constraint} to arrive at an efficient evaluation of $g(\mbx,\hat{d})$ in \eqref{invariant}. We derive a construction from singular value decomposition (SVD) $\mbJ(\mbx, \hat{d}) = \mbU \Sigma \mbV^T$. From this perspective, we can view $g(\mbx, \hat{d})$ as a {rotationally invariant} measure of the norm of $\mbn(\mbx)$ when projected into the (squared) principle stretch space $\Sigma^2$, which is crucial for arriving at our solution. Moreover, we have $$\mbJ(\mbx, \hat{d})^T\mbJ(\mbx, \hat{d}) = \left(\mbS^T\underbrace{\mbR^T\mbR}_{\mbI}\mbS\right) = \mbS^2  = \mbV\Sigma^2\mbV^T,$$ with $\mbR\mbS \coloneqq \mbJ(\mbx, \hat{d})$ as the polar decomposition from which one obtains the rotation $\mbR = \mbU\mbV^T$ for highlighting that $\mbn(\mbx)$ is merely transformed into the column space of $\Sigma^2$ before a dot product in this space. We use this property to define the diagonal \textit{re}-presentation of $\mbJ(\mbx, \hat{d})  \Rightarrow \Sigma(\mbx,\hat{d})  \in \realNum^{m\times{m}}$ and $\mbn(\mbx) \Rightarrow [\text{const}]^T \in \realNum^m$ satisfying \eqref{J-constraint}, where $\Sigma(\mbx,\hat{d})$ will now be evaluated directly from positions $\mbx \in \realNum^{3{s}}$ without any calculation of SVD. The variable $m = 1, 2, 3$ denotes the simplex dimension.
	\begin{figure}[htbp]
		\centering
		\includegraphics[width=0.8\columnwidth]{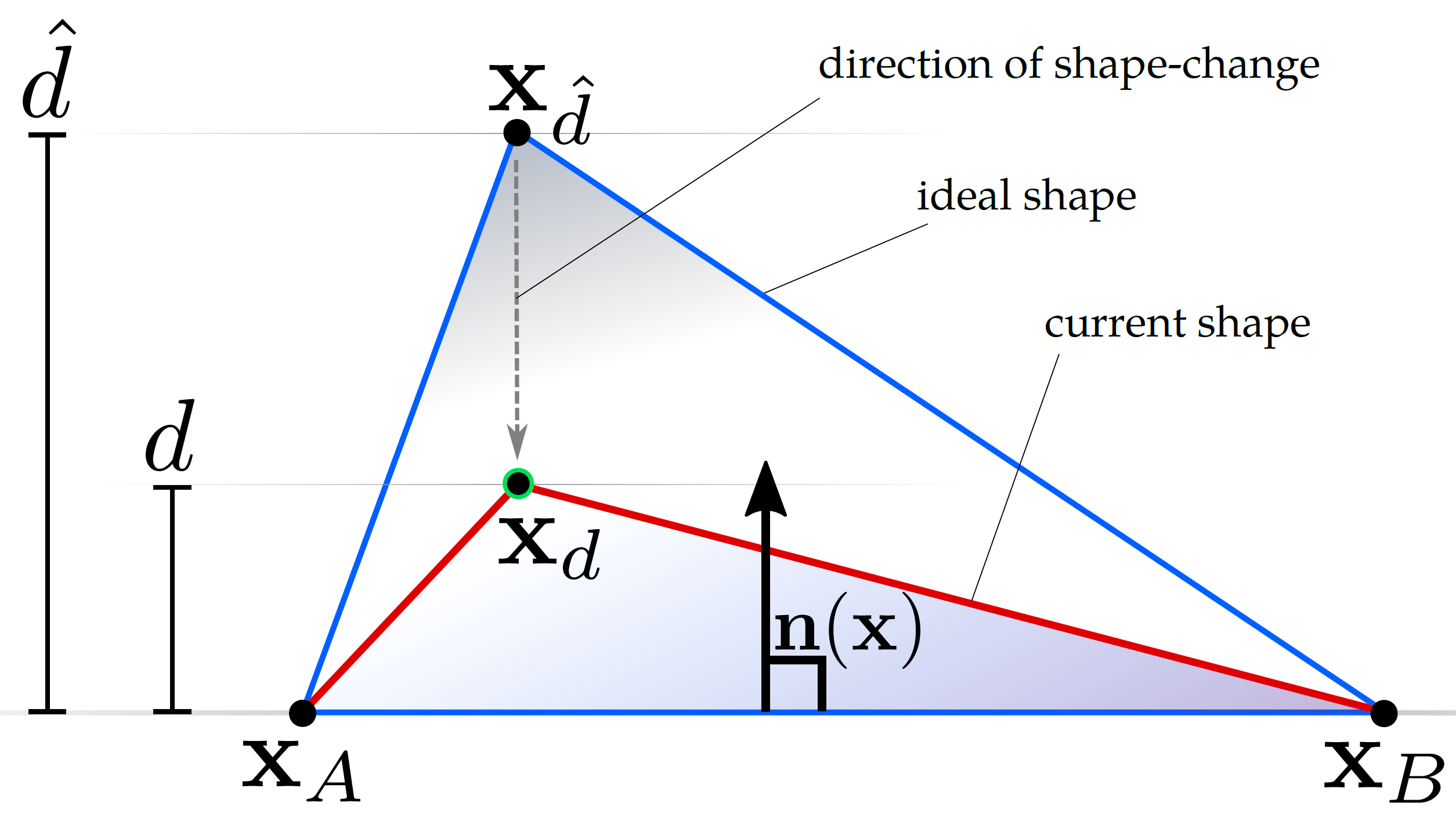}
		\caption{\label{fig:force-simplification}The ideal shape subtended by vertices $\mbx_{A}$, $\mbx_{B}$ and $\mbx_{\hat{d}}$ is deformed into the current shape (with $\mbx_d$ instead of $\mbx_{\hat{d}}$) due to proximal contact where $d < \hat{d}$. Deformation is understood to be in the  direction $-\mbn(\mbx)$, which is opposite to the normal vector that is computed from vertex positions $\mbx$ in the stencil of the local contact. The positions of the vertices in the stencil here are $\mbx_{A}$, $\mbx_{B}$ and $\mbx_{{d}}$, which come from the mesh boundary elements at the current step. One can always construct the ideal/undeformed shape of the simplex by duplicating a subset of the vertices $\in \mbx$ and shifting them by an offset $ \mbs = (\hat{d}-d)\mbn(\mbx)$ from the position of their original copy. Only $\mbx_d$ is duplicated and shifted by the offset in the example illustrated here. The linear map $\mbJ(\mbx)$ is then a transformation from the ideal to the current shape, which is a pure compression (without shear) due to the uniaxial change-in-shape with $\mbs$. 
	}\end{figure}
	
	\paragraph{The diagonal Jacobian} 
	To simplify $\mbJ(\mbx, \hat{d})$ for our intended purpose, we view the simplex as changing shape by deforming along the direction $\mbn(\mbx)$ from its ideal configuration (see~\figref{force-simplification}). This means that all but the smallest singular value of $\mbJ(\mbx, \hat{d})$ are one, which will be 
	\begin{equation}\label{eq:sigma-sum}
	f(\mbx,\hat{d}) = \det(\mbJ(\mbx, \hat{d})) = \det(\Sigma) = \prod_{i=1}^{{m} }\sigma_{i} = \frac{d(\mbx)}     {\hat d}, 
	\end{equation}
	based on the understanding that $\mbJ(\mbx, \hat{d})$ is a linear map from the ideal to the current shape of the simplex. We also have $\mathrm{diag}(\Sigma) = (\sigma_{1}, \ldots, \sigma_{{m}})$ as the singular values, and $d(\mbx), 0< d(\mbx) < \hat{d}$ as the closest distance between the pair of contact primitives considered. 
	
	\eqref{sigma-sum} follows directly from our construction of the deformed simplex shape which implies \textit{compression} along $-\mbn(\mbx)$, where the determinant is the ratio of the space (\ie~volume, area or length) subtended by the vertex positions of the current shape to those of the ideal shape. This determinant will be equivalent to the distance ratio precisely because we have uniaxial compression to give $\mbJ(\mbx, \hat{d}) = \mbV\Sigma\mbV^T = \Sigma$ since $\mbV=\mbI$. The diagonal factor will thus reduce to  
	\begin{align}\label{eq:reduced-F}
		{\Sigma}(\mbx, \hat{d}) = 
		\begin{cases}
			\begin{bmatrix}
				1&0&0\\
				0&1&0\\
				0&0&f
			\end{bmatrix} & \textsf{Point-Triangle and Edge-Edge}\\
			\begin{bmatrix}
				1&0\\
				0&f
			\end{bmatrix} & \textsf{Point-Edge} \\
			\begin{bmatrix}
				f
			\end{bmatrix}  & \textsf{Point-Point}, 
		\end{cases}	
	\end{align}
	where $f$ is a substitute for $f(\mbx,\hat{d})$, and $\sigma_{i=1} \ge \ldots \ge \sigma_{{m}}$ since the (clamped) barrier function is non-zero when $d(\mbx) \le \hat{d}$. We can use this to rewrite \eqref{invariant} by
	\begin{equation}\label{eq:invariant-simplified}
	\bar{g}(\mbx, \hat{d}) =\mbe_{m}^T{\Sigma}(\mbx, \hat{d})^T{\Sigma}(\mbx, \hat{d}){\mbe_{m}} =             
	\mbe_{m}^T{\Sigma}^2(\mbx, \hat{d}){\mbe_{m}},
	\end{equation}
	where the natural basis vector
	\begin{align*}
		{\mbe}_{m} = \begin{cases}
			\begin{bmatrix}
				0\\
				0\\
				1
			\end{bmatrix} & \textsf{Point-Triangle and Edge-Edge}\\
			\begin{bmatrix}
				0\\
				1
			\end{bmatrix} & \textsf{Point-Edge} \\
			\begin{bmatrix}
				1
			\end{bmatrix}  & \textsf{Point-Point} 
		\end{cases},
	\end{align*}
	then selects for change-of-shape along the contact normal $\mbn(\mbx) \coloneqq \mbe_{m}$ with $\mbJ(\mbx, \hat{d}) \coloneqq {\Sigma}(\mbx, \hat{d})$ to give $g(\mbx, \hat{d}) = \bar{g}(\mbx, \hat{d}) = f^2(\mbx, \hat{d})$, which we use in our implementation\footnote{{Intuitively, the normal vector has different dimensions (other than just three) because we are measuring distance in the principle-stretch space of $\mbJ$, which is equivalent to the local space at a contact with the same dimensions as the simplex, $m$.}}.

	\section{Barrier function} \label{sec:barrier}
	In this section, we describe our barrier function, which is constructed with the gap function described in \secref{constraint-jac-gp-fn} for obtaining the analytic eigensystems of the approximate Hessian. 
	
	\subsection{Notation}
	For conciseness, we will adopt a convention that functions of one or more variables will be referred to by symbolic name except when first encountered \eg~ $\mbJ(\mbx, \hat{d}) \rightarrow \mbJ$. For eigenanalysis, we will write eigenpairs {of the barrier Hessian} as $(\lambda_{i}, \mbQ_{i}),$ using  \emph{eigenmatrix} notation (as in \eg~ \citet{10.1145/3241041}) instead of the usual notation $(\lambda_{i}, \mbq_{i})$. This form permits easier derivation of the analytic eigensystems we seek. The usual form is gotten via the relation $\mbq_{i} = \vecOp(\mbQ_{i})$, where the \textit{vectorization} operator $\vecOp\left(\cdot\right)$ denotes a stacking of matrix columns in the simplest case. We also refer readers to the tutorial of  \citet{10.1145/3388769.3407490} for a gist of the tensor notation and analysis methods that we adopt in this paper. A more general treatment of tensor \textit{matricization} and vectorization can be found in the work of \eg~\citet{KoBa09}.
	
	\subsection{Our barrier, its derivatives and Hessian approximation}\label{sec:bar-fn-and-derivs}
	We advocate for a barrier potential of the form 
	\begin{equation}\label{eq:ivc-barrier}
	b(\mbx, \hat{d}) = -\left(\hat d^2-\hat{d^2}g(\mbx, \hat{d})\right)^2\ln\left(g(\mbx, \hat{d})\right),
	\end{equation} 
	which is expressed in terms of the gap function $g$.
	Using the constraint Jacobian $\mbJ$ as the primary variable that we will be analyzing, the gradient of this energy  
	\begin{align}\label{eq:barrier-contact-force}
		\frac{\partial b}{\partial \mbx} 
		=  \frac{\partial \mbJ}{\partial \mbx} \colon \frac{\partial b}{\partial \mbJ} &=
		\frac{\partial \mbJ}{\partial \mbx} \colon  \frac{\partial b}{\partial g}\frac{\partial g}{\partial \mbJ} 
	\end{align} 
	is evaluated by applying the change-of-basis tensor
	${\partial \mbJ}/{\partial \mbx}$ via double contraction,  
	with 
	\begin{equation}
	\label{eq:dg-dJ}
	\frac{\partial g}{\partial \mbJ} = 2\mbJ\mbN,
	\end{equation}
	where $\mbN = \mbn\mbn^T$. 
	The second-order derivative (force Jacobian) 
	follows from \eqref{barrier-contact-force} by
	\begin{align}\label{eq:hess-tens}
		\frac{\partial^2 b}{\partial \mbx^2}  &=  {\frac{\partial \mbJ}{\partial \mbx}}: \frac{\partial^2 b}{\partial \mbJ^2}:\frac{\partial \mbJ}{\partial \mbx} + \frac{\partial^2 \mbJ}{\partial \mbx^2}\colon \frac{\partial b}{\partial \mbJ},
	\end{align}
	which will be the focus of our analysis. 
	
	\subsection{Analytic eigensystems of the approximate Hessian}\label{sec:analytic hess}
	Our goal is to ensure that the barrier Hessian in the first term of \eqref{hess-tens}
	\begin{align} \label{eq:4th-ord-tens}
		\frac{\partial^2 b}{\partial \mbJ^2} &= 
		\frac{\partial b}{\partial g} \frac{\partial^2 {g}}{\partial \mbJ^2} 
		+ \frac{\partial^2 b}{\partial {g}^2} \left(\frac{\partial {g}}{\partial\mbJ} \otimes \frac{\partial {g}}{\partial\mbJ} \right),
	\end{align}
	is PSD but without actually performing \textit{any} numerical eigendecomposition, where $\otimes$ denotes the {tensor} product operator.
	The equivalent vectorized form is given by
	\begin{align} 
		\vecOp\left(\frac{\partial^2 b}{\partial \mbJ^2}\right)
		&=
		\frac{\partial b}{\partial {g}}\vecOp\left(\frac{\partial^2 {g}}{\partial \mbJ^2} \right) + \frac{\partial^2 b}{\partial {g}^2} \vecOp\left(\frac{\partial {g}}{\partial\mbJ}\right) \vecOp\left(\frac{\partial {g}}{\partial\mbJ} \right)^{T}, \label{eq:4th-ord-tens-vec}
	\end{align}
	which is what we use to perform eigenanalysis. 
	Applying the analysis of \citet{10.1145/3306346.3323014} to \eqref{4th-ord-tens-vec}, we find that three eigenpairs define our barrier Hessian form {(see also Appendix A.2 in \cite{10.1145/3606934} for a related approach)}. The eigenvalues 
	\begin{align}\label{eq:lambda1}
		\lambda_1 &= 4g\frac{\partial^2 b}{\partial {g}^2}+2\frac{\partial b}{\partial {g}},\\
		\lambda_{2,3}&=2\frac{\partial b}{\partial {g}},\label{eq:lambda23}
	\end{align} 
	are the same for all configurations of contact (\eg~\textit{edge-edge}, \textit{point-point} \etc) and only the primary \ie~ $\lambda_1$ is ever positive (see technical supplement). 
	We can therefore construct the approximate Hessian of \eqref{ivc-barrier} by
	\begin{equation}\label{eq:psd-gen-hess}
	\frac{\partial^2 b}{\partial \mbJ^2} \approxeq \lambda_1 {\mbQ_{1}}\otimes{\mbQ_{1}},
	\end{equation}
	which is PSD where the eigenmatrix is given by 
	\begin{equation}\label{eq:q1}
	\mbQ_1=\frac{1}{\sqrt{g}}\mbJ\mbN.
	\end{equation}
	
	\paragraph{Neglected terms} Our Hessian approximation is based on the fact that we do not consider the second term of the force Jacobian given in \eqref{hess-tens}.  
	This omission follows a simplifying assumption on $\mbJ$, which is that the ideal shape of a simplex is constant w.r.t time to give ${\partial^2 \mbJ}/{\partial \mbx^2} = 0$ (even though this ideal shape is constructed from $\mbx(t)$~as shown in \figref{force-simplification}).
	The effect is a cancelling out/removal of the second term in \eqref{hess-tens} to arrive at a form resembling the Gauss-Newton approximation of the barrier Hessian form of \citet{10.1145/3386569.3392425}, which is given by
	\begin{equation}\label{eq:bar-hess-orig}
	\frac{\partial^2b}{\partial d^2} \nabla_{\mbx} d  (\nabla_{\mbx} d)^T + \cancel{\frac{\partial b}{\partial d} \nabla_{\mbx}^2d},
	\end{equation} 
	where $b$ is \eqref{ipc-barrier-orig}. We share the same subspace of the search direction but not the magnitude of this search direction since our expression in~\eqref{lambda1} incorporates information from the omitted second term of \eqref{bar-hess-orig}. Readers are referred to \appref{gn-appr-comp} for a more detailed treatment.
	\begin{figure}[t]
		\centering
		\includegraphics[width=0.9\columnwidth]{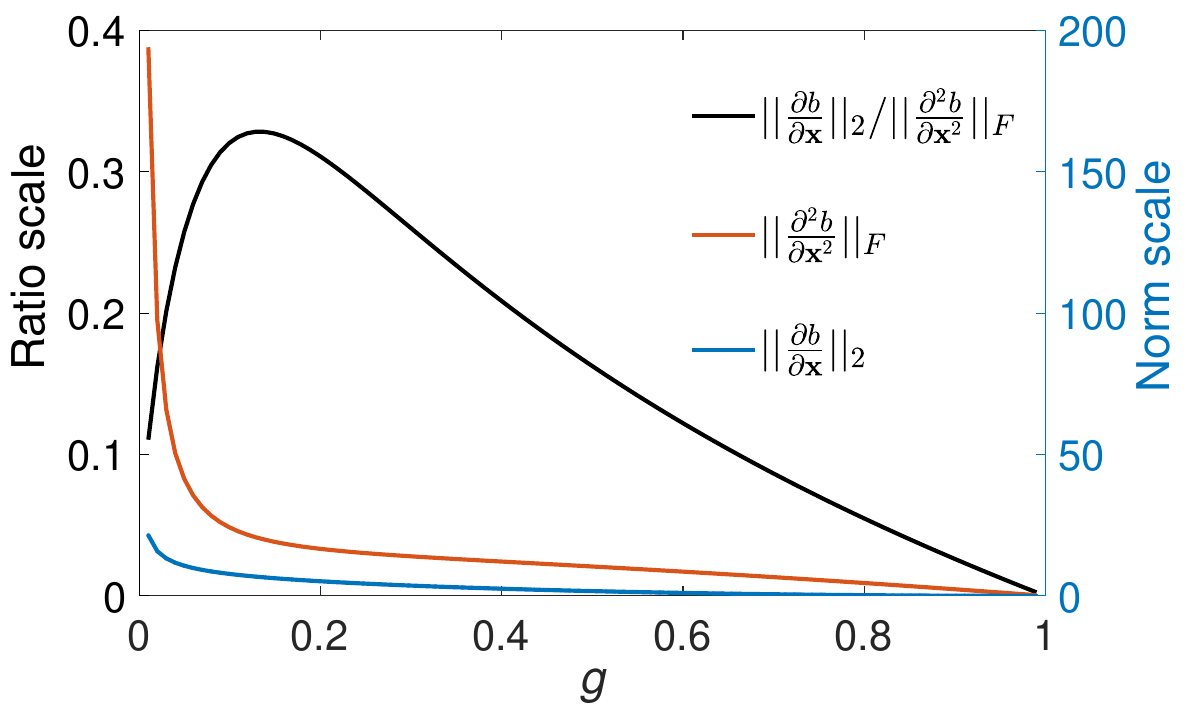}
		\caption{A plot of the norms of the barrier gradient and Hessian together with their ratio for illustrating the source of vanishing displacements $\mbd$ as $g \rightarrow 0$. 
			The scalar expressions used to measure the respective norms are described in \appref{magnitude_evaluation}. A threshold of $\hat{d} = 1$ is assumed here to aid visualization. 
		}
		\label{fig:gradient-mag-diff}
	\end{figure}
	
	\subsection{Filtered Hessian}
	The method presented in \secref{bar-fn-and-derivs} and \secref{analytic hess} can be used to produce a generally working implementation but is particularly inefficient when the barrier energy is evaluated with a very small distance $d \ll \hat{d}$. Such configurations are rare if simulating with soft elastic materials but occur frequently with high stiffness to affect solver performance. The specific effect is an increase in the number of solver iterations or even convergence failure in the worst case. The source of this problem is the dependence of the barrier function on the $\log$ function, {which implies that the magnitude of its second-order derivative increases significantly faster than the magnitude of its first-order derivative, to give $$\|{\partial b}/{\partial \mathbf{x}}\|_2/\|{\partial^2 b}/{\partial \mathbf{x}^2}\|_F \rightarrow 0,$$} as shown in \figref{gradient-mag-diff}. This fact causes the search direction magnitude $\|\mbd\|_2$ to tend toward zero as $g \rightarrow 0$ since { $\mbd=({{\partial^2 b}/{\partial \mbx^2}})^{-1} \cdot {{\partial b}/{\partial \mbx}}$}, thereby leading to an extremely small step size during optimization with bounded line search when $d \ll \hat{d}$. \citet{10.1145/3386569.3392425} multiply each barrier with a dynamically updated global coefficient $\kappa$ to address this issue. While this suffices when simulating with the full Hessian (\eqref{hess-tens}), application is limited if using a GN approximation because the omitted terms of the Hessian incorporate additional components into $\mbd$ that also work to minimize the energy when $d \ll \hat{d}$ holds\footnote{It is noteworthy too that these additional components in $\mbd$ are also associated with a subtle issue of introducing undesirable deviations in motion trajectories over time as shown in \figref{search-dir}.}. Furthermore, the magnitude of a search direction calculated with the full hessian can still diminish when evaluating distances near or at a barycentre, which typically occurs when the closest contact-point is near the centroid of a triangle or edge (see supplementary video). We describe in this section an approach to prevent vanishing displacements as $g \rightarrow 0$ by applying a filter directly to \eqref{lambda1} and thereby minimize the barrier energy {without} reduced convergence rates. 
	
	Our solution is to apply a correction on $\lambda_1$, resembling force filtering \cite{10.1145/1028523.1028541} to give
	\begin{equation}\label{eq:psd-gen-hess2}
	\frac{\partial^2 b}{\partial \mbJ^2} \approxeq 
	\begin{cases}  
	\lambda_1 {\mbQ_{1}}\otimes{\mbQ_{1}}, &g\geq \epsilon_g  \\  
	\lambda^{\text{thr}}_1 {\mbQ_{1}}\otimes{\mbQ_{1}}, &    g<\epsilon_g
	\end{cases},  
	\end{equation}
	as the filtered Hessian that is defined using a proximal limit $\epsilon_g=(d_{\text{thr}}/\hat{d})^2 $ via the distance  threshold that we set to $d_{\text{thr}} = 0.1\hat{d}$, where $\lambda_1^{\text{thr}}$ is \eqref{lambda1} evaluated at $d=d_{\text{thr}}$. 
	This correction ensures that the magnitude $\|\mbd\|_2 > 0$ is sufficiently larger than zero for a stable GN optimization. Although effective, this solution has yet one drawback that we must address. 
	
	\begin{figure}[t]
		\centering
		\includegraphics[width=0.7\columnwidth]{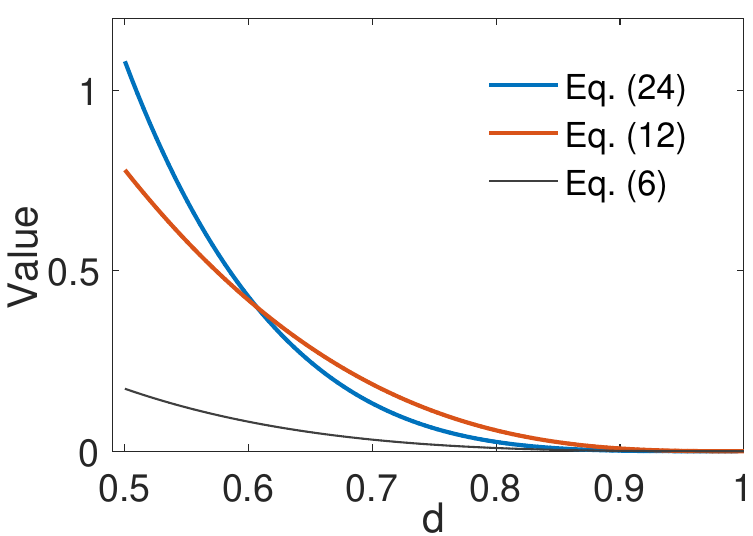}
		\caption{\label{fig:energy_jpg}{Barrier function comparison, where a threshold of $\hat{d} = 1$ is assumed to aid visualization. Note that the form $b(d^2, \hat{d}^2)$ is actually used in the source code of \citet{10.1145/3386569.3392425} to avoid squared roots, which results in an energy with a profile that is identical to \eqref{ivc-barrier}}.}
	\end{figure} 
	
	\paragraph{Quadratic stiffening} While \eqref{psd-gen-hess2} prevents vanishing displacement, the barrier gradient (\cf~\eqref{barrier-contact-force}) will have insufficient magnitude to get enough repulsion when the adaptive conditioning parameter $\kappa$ is too small. We propose modifying the barrier function in \eqref{ivc-barrier} to
	\begin{equation}\label{eq:ivc-barrier2}
	b(\mbx, \hat{d}) = \left(\hat d^2-\hat{d}^2g(\mbx, \hat{d})\right)^2\ln^2\left(g(\mbx, \hat{d})\right),
	\end{equation} 
	where we now use a quadratic log term, which guarantees sufficient force magnitude for minimizing the energy without direct dependence on $\kappa$. 
	{Stiffening the energy allows for a more locally-adaptive increase in repulsion force, which we found to be sufficient for eliminating vanishing displacements when combined with our filtered Hessian (\eqref{psd-gen-hess2}), even when simulating challenging elastic materials with extremely high stiffness as shown in \secref{results}. An illustrative comparison is also provided in \figref{energy_jpg}.}
	All forms in \secref{bar-fn-and-derivs} remain unchanged and therefore the expression of the positive eigenvalue in~\eqref{lambda1} and its eigenvector remain the same for evaluating \eqref{psd-gen-hess2}. \eqref{ivc-barrier2} is the barrier that we use throughout our implementation.

	\section{mollification} \label{sec:par-ee}
	In this section we describe our approach to handling the notorious case of nearly-parallel edges, which is crucial for minimizing the number of Newton solver iterations and preventing intersections in rare but significant instances. 
	{Handling near-parallel edges is especially critical when simulating with accurate Newton tolerance, wherein Newton iteration is highly sensitive to the continuity/smoothness of the energy. Managing near-parallel edges becomes particularly important in such scenarios.}
	
	Close-to-parallel \textit{edge-edge} contacts are difficult failure modes because the distance is nonsmooth, where numerical rounding error will also exacerbate barrier energy gradient- and Hessian values. 
	There are two causal problems: The first is that vertices forming any \textit{edge-edge} pair will reduce to a \textit{point-edge} case when the edges are parallel (\cf~Fig.9 in \cite{10.1145/3386569.3392425} and \S 7.1 in \cite{10.1145/3386569.3392425-ts}): 
	The consequence is that constraint culling (\ie~dismissal of contact pairs that are farther than $\hat d$ apart) may then oscillate between the two possible \textit{point-edge} cases at the parallel configuration during line-search.
	The effect is long convergence times and even solver divergence because components of the state vector $\mbx$ that is updated during line-search are determined by contact-force contributions from the three vertices of the \textit{point-edge} case evaluated \textit{prior} to solving the system matrix (see line 8 of Algorithm 1 in \cite{10.1145/3386569.3392425}). 
	The second problem is that the limited precision of floating point numbers prevents a sufficiently accurate calculation of normal vectors (between the edges) to make optimization intractable.
	
	\citet{10.1145/3386569.3392425} have addressed these issues by multiplying all \textit{edge-edge} barrier energies by a piecewise polynomial mollifier
	for smoothening their discontinuous distance gradients at the parallel configuration. 
	This has the effect of coupling all four vertices for assisting line-search. 
	We follow in their steps but tackle the problem from an unconventional-and-yet-intuitive perspective, which is consistent with our approach thus far and permits analytic description of the ensuing approximate Hessian.
	
	\subsection{The mollified barrier and its derivatives}
	Following from \secref{bar-fn-and-derivs}, we propose 
	\begin{equation}\label{eq:update-millifier}
	{e}_k(\mbx, \hat{d}) = 
	\begin{cases}  
	-\frac{1}{\epsilon_x^2}\gamma(\mbx, \hat{d})^2+\frac{2}{\epsilon_x}\gamma(\mbx, \hat{d}), &c<\epsilon_x  \\  
	1, &    c\geq\epsilon_x
	\end{cases},  
	\end{equation} 
	as our mollifier for a contact $k$ comprised of nearly-parallel edges. 
	We compute the measure of parallelness $c=\|(\mbx_{12}-\mbx_{11})\times(\mbx_{22}-\mbx_{21})\|^2$ from the vertices of the edges (labelled as in \eqref{ee-dist-orig}) at the current step; and $\epsilon_x=10^{-3}\|\bar{\mbx}_{12}-\bar{\mbx}_{11} \|^2\|\bar{\mbx}_{22} -\bar{\mbx}_{21} \|^2$ is the tolerance defined using reference positions $\bar{\mbx}$ from before the first timestep.
	We introduce the auxiliary function (\cf~\eqref{invariant-simplified})
	\begin{equation}\label{eq:invariant-simplified-gamma}
	\gamma(\mbx, \hat{d}) 
	= \mbn_{\gamma}^T{\Sigma}^2(\mbx, \hat{d}){\mbn_{\gamma}},
	\end{equation}
	for decoupling mollifier terms from our barrier energy to define
	\begin{equation}
	{e}_k(\mbx, \hat{d})b(\mbx, \hat{d}),
	\label{eq:ee-barrier-energy}
	\end{equation} 
	as the mollified barrier function. This energy is in-turn evaluated with
	\begin{equation}\label{eq:reduced-F-pe}
	\mbJ =\Sigma= \begin{bmatrix}
	1&0&0\\
	0&\sqrt{c}&0\\
	0&0&f
	\end{bmatrix},
	\end{equation}
	which encodes the state of contact w.r.t the normal (\cf~\secref{cp-dg}) and mollifier.
	Moreover, we have $\mbn_\gamma = \begin{bmatrix} 0 & 1 & 0 \end{bmatrix}^T$ as the vector selecting for the mollification measure $\sqrt{c}$ in $\mbJ$. We also have $\mbn_g = \mbe_m
	$ as the normal vector that is used to evaluate the barrier term $b$ in \eqref{ee-barrier-energy}, selecting for the distance ratio (\cf~\secref{cp-dg}).
	Written out explicitly, we have 
	\begin{equation}\label{eq:mollified-barrier2}
	\Tilde{b}(\mbx, \hat{d}) = {e}_k(\mbx, \hat{d}) \left(\left(\hat d^2-\hat{d}^2g(\mbx, \hat{d})\right)^2\ln^2\left(g(\mbx, \hat{d})\right)\right),
	\end{equation} 
	with $\gamma(\mbx, \hat{d}) = \|\mbJ(\mbx,\hat{d})\mbn_\gamma(\mbx)\|_2^2 = c$, which is the mollified barrier function that we use in our implementation.

	\paragraph{Derivatives} Derivatives of the mollified barrier are of the form provided in \eqref{barrier-contact-force} and \eqref{hess-tens}, where 
	\begin{equation}\label{eq:ee-barrier-gradient}
	\frac{\partial\Tilde{b}}{\partial \mbJ} = {\frac{\partial \Tilde{b}}{\partial \gamma}\frac{\partial \gamma}{\partial \mbJ}} + 
	{\frac{\partial \Tilde{b}}{\partial g}\frac{\partial g}{\partial \mbJ}}, 
	\end{equation} 
	and the mollified barrier Hessian is then 
	\begin{align}\label{eq:ee-barrier-hessian}
		\frac{\partial^2 \Tilde{b}}{\partial \mbJ^2} & =\left[\frac{\partial \Tilde{b}}{\partial \gamma} \frac{\partial^2 \gamma}{\partial \mbJ^2}+\frac{\partial^2 \Tilde{b}}{\partial \gamma^2} \left(\frac{\partial \gamma}{\partial\mbJ} \otimes \frac{\partial \gamma}{\partial\mbJ} \right) \right] + 
		\left[\frac{\partial \Tilde{b}}{\partial g} \frac{\partial^2 g}{\partial \mbJ^2}+\frac{\partial^2 \Tilde{b}}{\partial g^2} \left(\frac{\partial g}{\partial\mbJ} \otimes \frac{\partial g}{\partial\mbJ} \right) \right]
		+ \nonumber\\
		&\, \,
		\frac{\partial^2 \Tilde{b}}{\partial \gamma\partial g}\left(\frac{\partial g}{\partial\mbJ} \otimes \frac{\partial \gamma}{\partial\mbJ} \right) +\frac{\partial^2 \Tilde{b}}{\partial g\partial \gamma}\left(\frac{\partial \gamma}{\partial\mbJ} \otimes \frac{\partial g}{\partial\mbJ} \right)
	\end{align}
	which we use for eigenanalysis (in vectorized form like \eqref{4th-ord-tens-vec}). 
	
	\subsection{Eigenanalysis} \label{sec:moll-bar-Hess-ana}
	Here we describe the analytic eigensystem of our mollified barrier Hessian in \eqref{ee-barrier-hessian}, which we analyse on a term-by-term basis (there are four terms).
	
	\begin{figure}[htbp]
		\centering
		\begin{subfigure}[b]{0.495\columnwidth}
			\centering
			\includegraphics[width=\columnwidth ]{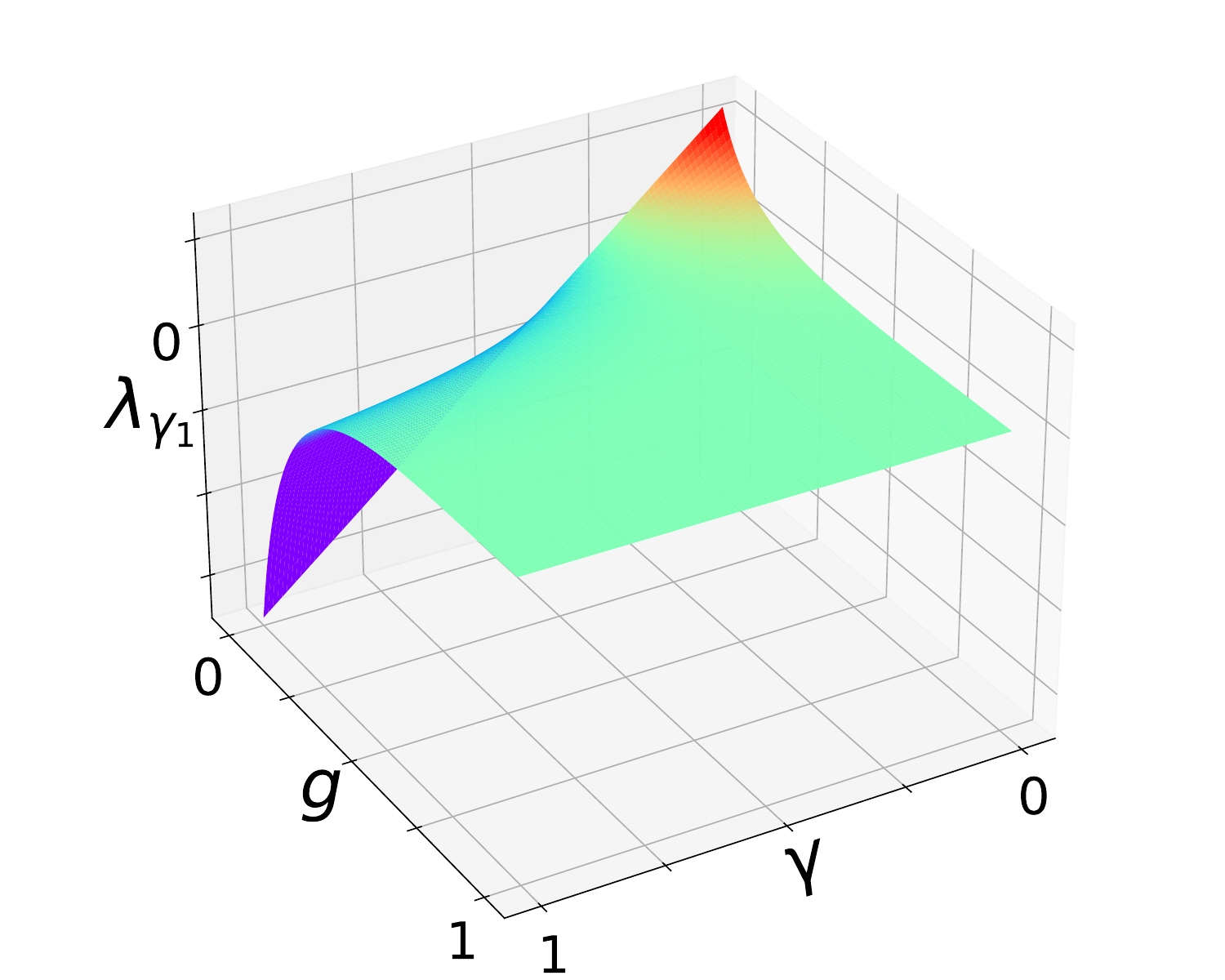}
			\caption{$\lambda_{\gamma1}$}
			\label{fig:lambda-alpha1}
		\end{subfigure}
		\begin{subfigure}[b]{0.495\columnwidth}
			\centering
			\includegraphics[width=\columnwidth ]{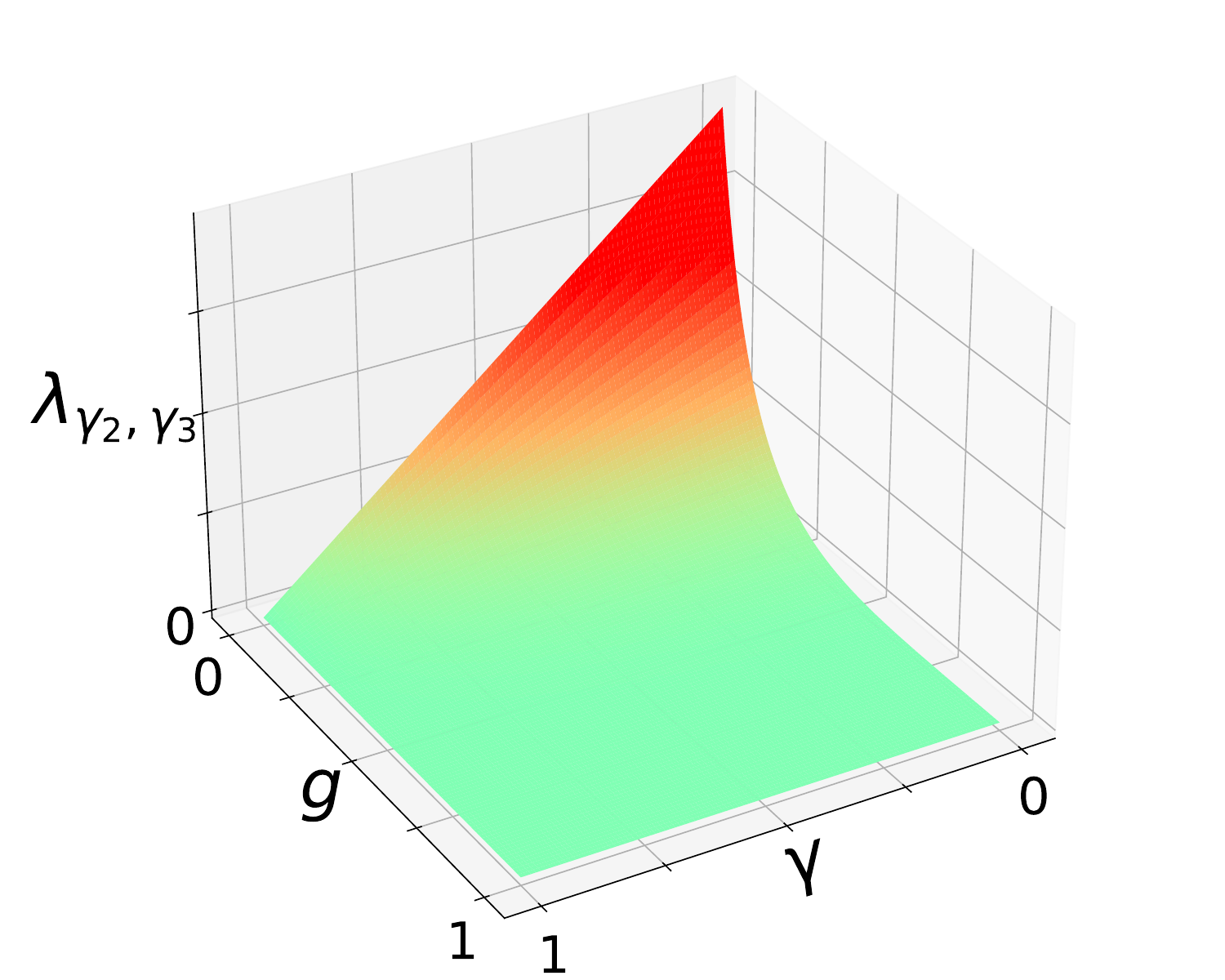}
			\caption{$\lambda_{\gamma3,\gamma2}$}
			\label{fig:lambda-alpha23}
		\end{subfigure}
		
		\begin{subfigure}[b]{0.495\columnwidth}
			\centering
			\includegraphics[width=\columnwidth]{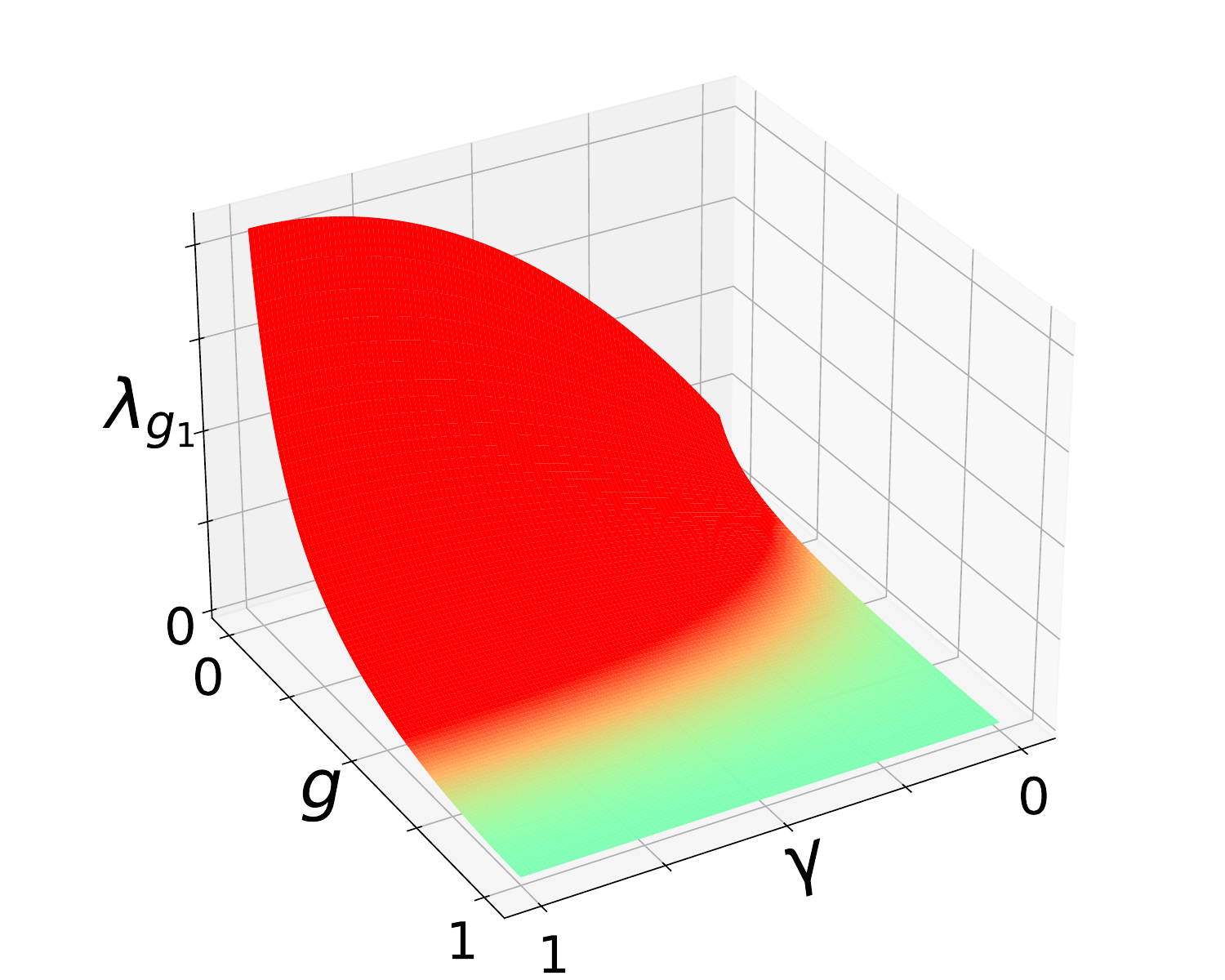}
			\caption{$\lambda_{g1}$}
			\label{fig:lambda-beta1}
		\end{subfigure}
		\begin{subfigure}[b]{0.495\columnwidth}
			\centering
			\includegraphics[width=\columnwidth ]{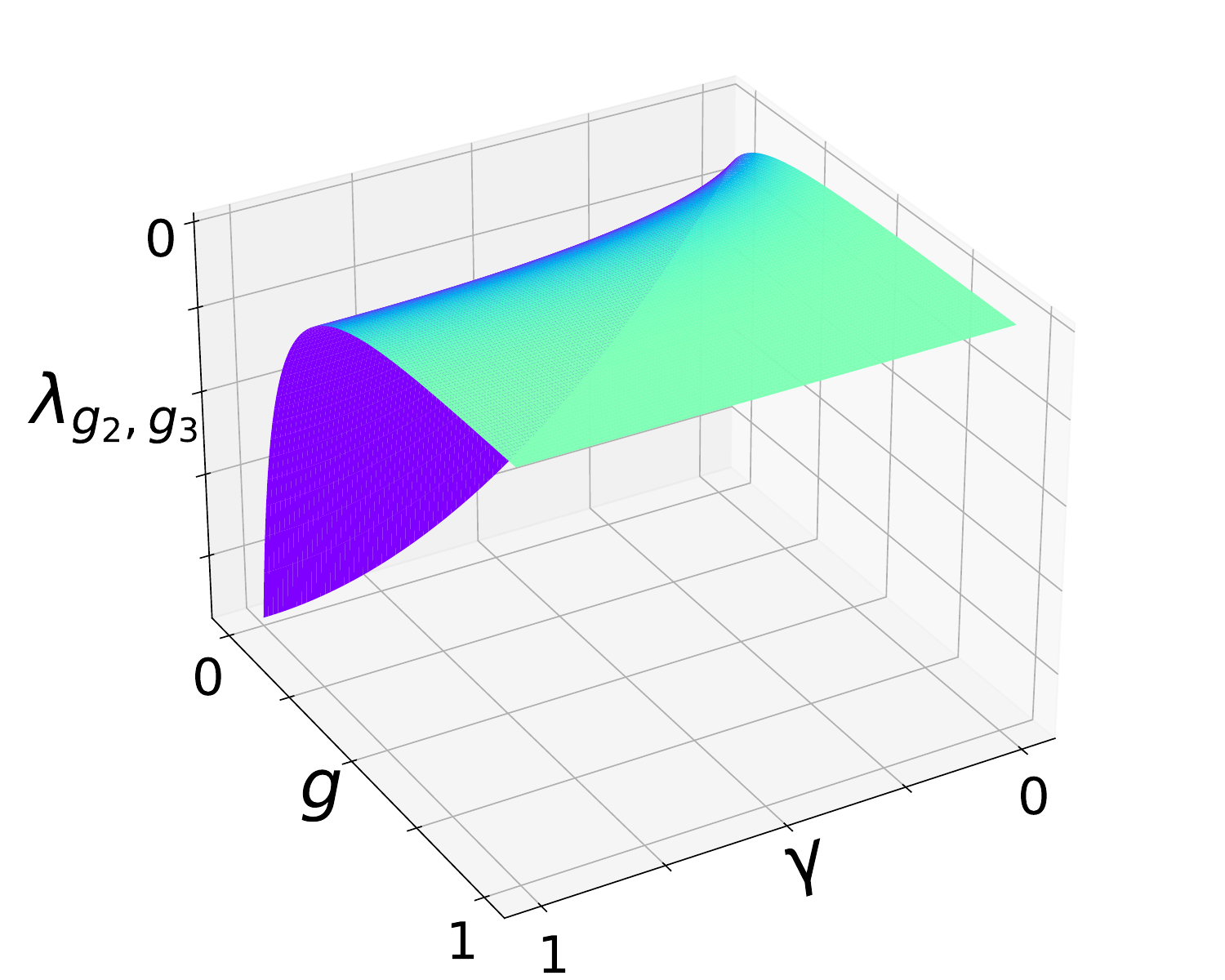}
			\caption{$\lambda_{g2,g3}$}
			\label{fig:lambda-beta23}
		\end{subfigure}
		\begin{subfigure}[b]{0.495\columnwidth}
			\centering
			\includegraphics[width=\columnwidth]{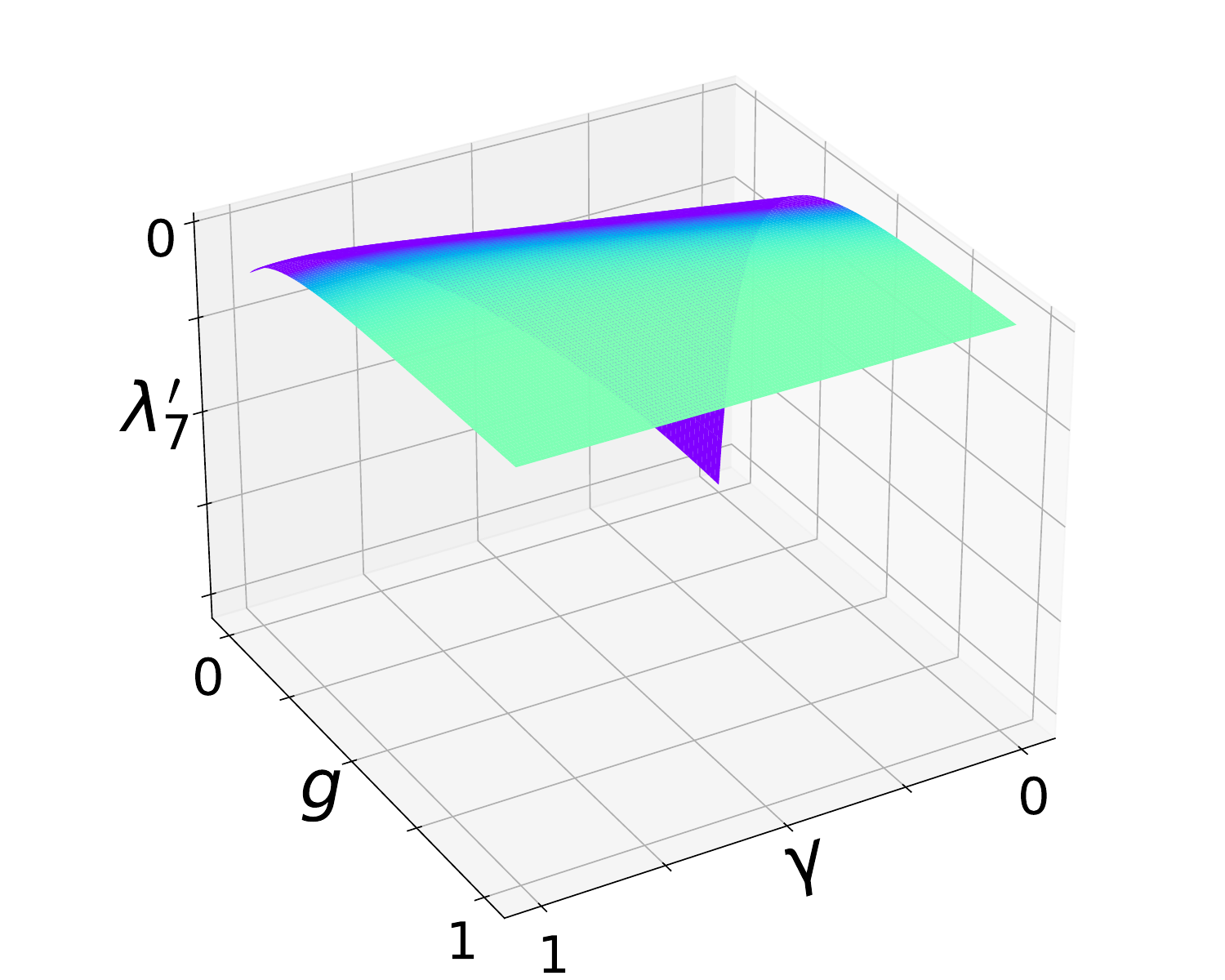}
			\caption{$\lambda_7^\prime$}
			\label{fig:lambda-prime7}
		\end{subfigure}
		\begin{subfigure}[b]{0.495\columnwidth}
			\centering
			\includegraphics[width=\columnwidth]{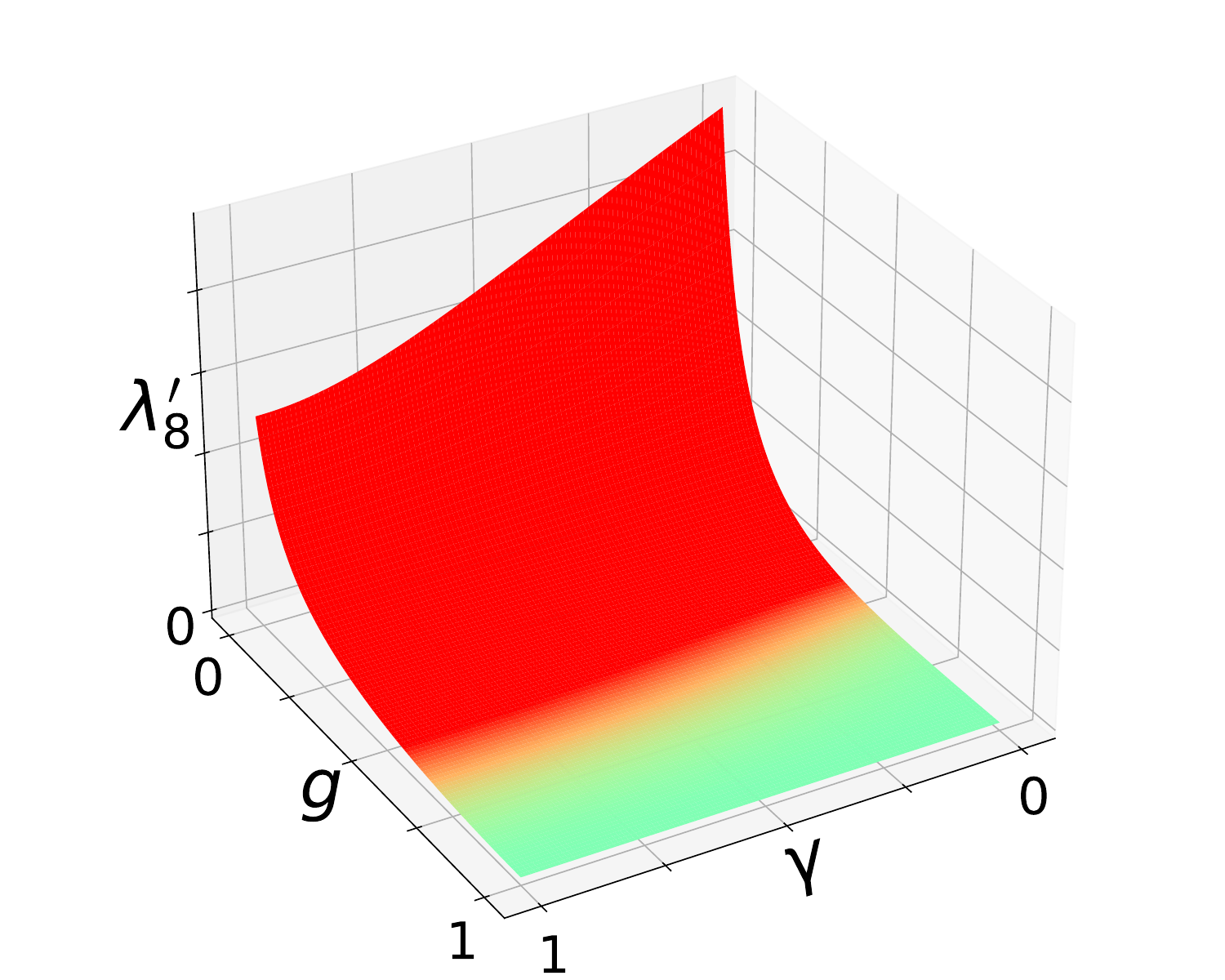}
			\caption{$\lambda_8^\prime$}
			\label{fig:lambda-prime8}
		\end{subfigure}
		\caption{Plot of the eigenvalues of the approximate mollified Hessian in \eqref{ee-barrier-hessian}: 
			Eigenvalues of the first term  ( \figref{lambda-alpha1} and \figref{lambda-alpha23});
			Eigenvalues of the second term   ( \figref{lambda-beta1} and \figref{lambda-beta23});
			Eigenvalues due to the third and forth term (\figref{lambda-prime7} and \figref{lambda-prime8}). }
	\end{figure}
	
	\paragraph{Eigen values and vectors of the first two terms} The eigenvalues of the first two terms   in \eqref{ee-barrier-hessian} 
	\textit{each} have the form 
	$$\lambda_{(\ast)1}=2\left(\frac{\partial \Tilde{b}}{\partial (\ast)}+2(\ast)\frac{\partial^2 \Tilde{b}}{\partial (\ast)^2}\right),$$ and $\lambda_{(\ast)2}=\lambda_{(\ast)3}=2{\partial \Tilde{b}}/{\partial (\ast)},$
	where $(\ast)$ is a substitute for either $\gamma$ or $g$ in the whole equation.
	The eigenvalues $\lambda_{\gamma1}$, $\lambda_{\gamma2}$ and $\lambda_{\gamma3}$ are plotted in \figref{lambda-alpha1} and \figref{lambda-alpha23}, which show that $\lambda_{\gamma1}$ is greater-than zero when $\gamma \lessapprox 1/3\epsilon_x$ while $\lambda_{\gamma2}$ and $\lambda_{\gamma3}$ are always greater-than or equal-to zero. 
	Likewise, the eigenvalues $\lambda_{g1}$, $\lambda_{g2}$ and $\lambda_{g3}$ are plotted in \figref{lambda-beta1} and \figref{lambda-beta23}, where $\lambda_{g1}$ is always greater-than or equal-to zero, and $\lambda_{g2}$ and $\lambda_{g3}$ are always less-than or equal-to zero.

	The corresponding eigenmatrices of $\lambda_{(\ast){i}}$  are (\cf~Eq. (12), (15) and (16) respectively in \cite{10.1145/3306346.3323014})
	\begin{align}\label{eq:qalphabeta-eigenmat}
		\mbQ_{(\ast)1} &=\mbN_{(\ast)}, &
		\mbQ_{\gamma{3}} =\begin{bmatrix}0&1&0\\-1&0&0\\0&0&0\end{bmatrix}\mbN_\gamma,\nonumber\\
		\mbQ_{(\ast)2} &=\begin{bmatrix}0&0&0\\0&0&1\\0&-1&0\end{bmatrix}\mbN_{(\ast)}, &
		\mbQ_{g{3}} =\begin{bmatrix}0&0&1\\0&0&0\\-1&0&0\end{bmatrix}\mbN_g,
	\end{align}
	where $\mbN_{(\ast)} = \mbn_{(\ast)}\mbn_{(\ast)}^T$. We also have $\mbQ_{\gamma{i}} \perp \mbQ_{\gamma{j}}$, and $\mbQ_{g{i}} \perp \mbQ_{g{j}}, i \ne j$ with  $i,j = 1,2,3$, and $\forall_{i}\,\mbQ_{g{i}} \perp \mbQ_{\gamma{i}}$, which means that these eigenmatrices are orthogonal. Readers are referred to our technical supplement for a more detailed derivation.

	\paragraph{Third and forth term}  
	Analysis of the last two terms of \eqref{ee-barrier-hessian} will give two eigenpairs
	\begin{equation}\label{eq:q7q8}
	\lambda_7=2t,\,\mbQ_{7}=\begin{bmatrix}0&0&0\\&\mbn_\gamma^T&\\&\mbn_g^T&\end{bmatrix}\, \text{and }
	\lambda_8=-2t,\,\mbQ_{8}=\begin{bmatrix}0&0&0\\&-\mbn_\gamma^T&\\&\mbn_g^T&\end{bmatrix},
	\end{equation}
	where $t=\frac{\partial^2 \Tilde{b}}{\partial\gamma\partial g}\cdot 
	\sqrt{c}f
	$. However, $\mbQ_{7}$ and $\mbQ_{8}$ are not orthogonal to $\mbQ_{\gamma1}$ and $\mbQ_{g1}$, which means these four bases are not eigenvectors of the complete mollified barrier Hessian expression in \eqref{ee-barrier-hessian}.
	
	We resolve this by replacing with (\cf~\figref{lambda-prime7} and \figref{lambda-prime8})
	\begin{align}
		\lambda_7^\prime&=\frac{\lambda_{\gamma1}}{2}+\frac{\lambda_{g1}}{2}-p,\, \mbQ_7^\prime=\begin{bmatrix}0&0&0\\0&k_1&0\\&\mbn_g^T&\\\end{bmatrix}\, \mathrm{and}\, \label{eq:Q7-l7-prime}\\ \lambda_8^\prime&=\frac{\lambda_{\gamma1}}{2}+\frac{\lambda_{g1}}{2}+p,\, \mbQ_8^\prime=\begin{bmatrix}0&0&0\\0&k_2&0\\&\mbn_g^T&\\\end{bmatrix}\label{eq:Q8-l8-prime},
	\end{align}
	where $k_1=\frac{\lambda_{\gamma1}-\lambda_{g1}-2p}{8t}$, $k_2=\frac{\lambda_{\gamma1}-\lambda_{g1}+2p}{8t}$, and $$p=\frac{\sqrt{\lambda_{\gamma1}^2-2\lambda_{\gamma1}\lambda_{g1}+\lambda_{g1}^2+64t^2}}{2}.$$ 
	The subspace spanned by $\mbQ_7^\prime$ and $\mbQ_8^\prime$ now contains $\mbQ_{7}$, $\mbQ_{8}$, $\mbQ_{\gamma1}$ and $\mbQ_{g1}$.
	This subspace is also orthogonal to $\mbQ_{\gamma2}$, $\mbQ_{\gamma3}$, $\mbQ_{g2}$ and $\mbQ_{g3}$, which is necessary to form a valid eigenspace of the complete mollified barrier Hessian in \eqref{ee-barrier-hessian}.
	The procedure for how we arrive at \eqref{Q7-l7-prime} and \eqref{Q8-l8-prime} are summarized in 
	our technical supplement. 
	
	The final analytically-projected mollified Hessian is thus
	\begin{equation} \label{eq:ee-barrier-psd-hess}
	\frac{\partial^2 \Tilde{b}}{\partial \mbJ^2} = \lambda_{\gamma2}\mbQ_{\gamma2}\otimes\mbQ_{\gamma2}+
	\lambda_{\gamma3}\mbQ_{\gamma3}\otimes\mbQ_{\gamma3} 
	+
	\lambda_{8}^\prime \mbQ_{8}^\prime \otimes  \mbQ_{8}^\prime.
	\end{equation} 
	We have found a total of six eigenpairs but only three are useful since the eigenvalues $\lambda_{g2}$, $\lambda_{g3}$ and $\lambda_7^\prime$ are always less-than or equal-to zero (\cf~\figref{lambda-beta23} and \figref{lambda-prime7}).

	\section{Implementation}
	\label{sec:implementation}
	This section will summarise our implementation of \eqref{energy-min} using a Gauss-Newton solver. We also summarise details of our GPU implementation. 
	
	\subsection{Overview} The simulation pipeline is summarised in \algref{gpuipc}.
	Given an input triangulated surface mesh, it is first pre-processed (\eg~tetrahedralization, BVH construction \etc) during initialization, which is then followed by actual simulation (\ie~timestepping via implicit integration). 
	For each timestep, we determine contact pairs by traversing the scene BVHs, querying with all points (to the triangle-BVH) and edges (to the edge-BVH). 
	A pair arises if the distance between the respective primitives (\eg~a point and a triangle) is less than the computational accuracy target $\hat{d}$. We also set boundary conditions (prescribed positions, velocities and/or force \etc, including gravity) before time-stepping the simulation. 
	We follow the standard architecture of a PN solver, with line search and simultaneously constructing and projecting each approximate barrier Hessian from our analysis.

	\begin{algorithm}[t]
		\caption{\label{alg:gpuipc}Implementation overview}
		\SetAlgoLined
		\LinesNumbered
		\SetKwFunction{initSim}{init}
		\SetKwFunction{updateBVHs}{updateBVHs}
		\SetKwFunction{findPrimitivePairs}{findContactPairs}
		\SetKwFunction{updateFrictionState}{updateFrictionState}
		\SetKwFunction{projectHessian}{getPSDForceJacobian}
		\SetKwFunction{updateBoundaryConditions}{updateBoundaryConditions}
		\SetKwFunction{updateBarrierStiffnessParams}{setBarrierStiffParams}
		
		\SetKwRepeat{Do}{do}{while}
		\initSim() \tcp{Load data, initial conditions ${\mbx}^{t=0}$ \etc}  
		\For{each timestep $t$ }
		{
			\updateBVHs($\mbx^{t}$)\;
			\findPrimitivePairs($\mbx^{t}$) \tcp{point-triangle pairs \etc}
			$\mbx \leftarrow \mbx^{t}$\; 
			$\mathcal{I}_{\mathrm{prev}} \leftarrow \mathcal{I}(\mbx)$\; 
			$\mbx_{\mathrm{prev}} \leftarrow \mbx^t$\;
			\Do{$\frac{\|\mbd\|_{\infty}}{\Delta{t}} \le \varepsilon_d$}  {
				$\mbg \leftarrow \frac{\partial\mathcal{I}(\mbx)}{\partial \mbx} $ \tcp{Energy gradient} 
				$\mathbb{H} \leftarrow \mbM + \mathrm{\projectHessian}\left(\frac{\partial^2\mathcal{I}(\mbx)}{\partial \mbx^2}\right)$ \label{lin:psd-hess-proj}\;
				$\mathbb{H} \mbd = - \mbg$ \label{lin:solve-lin-sys} \tcp{PCG solve}
				\updateBVHs() \tcp{Enlarge BVs with $\mbd$}
				$\alpha \leftarrow\min(1, \mathrm{maxFeasibleStepSize})$ \tcp{ACCD}
				\Do{$\mathcal{I}(\mbx) > \mathcal{I}_{\mathrm{prev}}$}  {
					$\mbx \leftarrow \mbx_{\mathrm{prev}} + \alpha \mbd$\; \findPrimitivePairs($\mbx$)\;
					$\mbx_{\mathrm{prev}}$) \;
					$\alpha \leftarrow \frac{\alpha}{2}$\;
				}
				$\mbx_{\mathrm{prev}} \leftarrow \mbx$\;
				$\mathcal{I}_{\mathrm{prev}} \leftarrow \mathcal{I}(\mbx)$\;
				Update $\kappa$, boundary conditions \etc\;
			}
			\Return $\mbx$\;
		}
	\end{algorithm}
	
	\begin{algorithm}[t]
		\SetAlgoLined
		\LinesNumbered
		\DontPrintSemicolon
		\SetKwFunction{getDeformationGradient}{getGapFunctionTerms}
		\SetKwFunction{getVertices}{getLocalContactVertices}
		\SetKwFunction{getForceJacobian}{buildApproxForceJacobian}
		\SetKwFunction{getEigenPairs}{EvalEigenPairs}
		\tcp{Similar steps per contact-pair}
		\For{each pair $p$ }
		{
			$\mbx_p\leftarrow \getVertices(\mbx, p)$\;
			$\lbrace\lambda_i, \mbQ_{i}\rbrace \leftarrow \getEigenPairs(\mbx_p)$ \tcp{\eg~\eqref{lambda1}}
			$\partial^2 b/\partial \mbJ^2 \leftarrow \sum_{i}{\max(\lambda_i^\text{thr}, \lambda_i) \mbQ_i\otimes\mbQ_i^T}$ \tcp{\eg~\eqref{psd-gen-hess2}} 
		}
		\caption{\label{alg:proj-hess}Steps to evaluate our approximation of \eqref{hess-tens}}
	\end{algorithm}
	
	\subsection{GPU implementation}
	Existing IPC methods are designed to run either on CPUs (\eg~\citet{RigidIPC,10.1145/3386569.3392425}), or via a hybrid combination of the CPU and GPU (\eg~\citet{MedialIPC,DBLP:journals/corr/abs-2201-10022}). 
	Hybrid methods are highly efficient but will require marshalling data portions (per Newton iteration) between the CPU and GPU. 
	Our eigenanalysis permits a \textit{completely} GPU based implementation of IPC when combined with recent elastic deformation energies (\cite{10.1145/3241041,10.1111/cgf.14111,10.1145/3306346.3323014,OurArapPaper} since constructing our barrier Hessian is reduced to evaluating a scaled tensor/outer-product  (see ~\eqref{psd-gen-hess} and \algref{proj-hess}). {For friction, \citet{10.1145/3386569.3392425-ts} describe a method to PSD project the respective friction Hessian, using a $2 \times 2$ matrix projection that we also use. The method is relatively straight-forward as it involves simply solving a quadratic problem, which is also efficiently parallelized on the GPU. { For a more comprehensive examination of the SPD project of friction Hessian and its impact, readers are referred to the supplemental material.}}
	
	\paragraph{Data structures} 
	Our BVH representation follows \citet{Karras12} (although any other suitable method like \citet{https://doi.org/10.1111/cgf.13948} can be used), which offers a depth-first search (DFS) layout that is ideal for cache-efficient parallel queries using points and edges.
	BVH traversal will return contact pairs between each query-object (a point or an edge) and the primitive(s) stored in the BVH (triangles or edges).
	The returned pair(s) for a single query may be further reduced to secondary cases (like a \textit{point-edge} case from \textit{point-triangle}) depending on the number of constraints in the active set (\cf~\eqref{pt-dist-orig} and  \eqref{ee-dist-orig}).
	BVHs are particularly attractive for collision detection on the GPU as they allow for efficient caching with temporal coherence via front-tracking \cite{WangTMT18}.
	
	Contact pairs are stored as tuples of vertex indices
	$\mathbf{t} = \left(i_0, i_1, i_2, i_3\right)$ 
	where $i_{0\ldots 3}$ are these indices.
	The \textit{type} is then encoded in the sign information of the indices as $\begin{bmatrix} + ,{+} , + , + \end{bmatrix}$ for \textit{edge-edge}; $\begin{bmatrix} - , {+} , + , + \end{bmatrix}$ for \textit{point-triangle}; $\begin{bmatrix} - , {+} , + , \# \end{bmatrix}$ for \textit{point-edge}; and $\begin{bmatrix} - , {+} , \# , \# \end{bmatrix}$ for \textit{point-point}.  
	The symbol $\#$ is simply a placeholder denoting nullity (\ie~a prescribed value that we can evaluate and compare with). 
	The close-to-parallel \textit{edge-edge} case is encoded with $\begin{bmatrix} + , {-} , + , + \end{bmatrix}$ for \textit{edge-edge}; $\begin{bmatrix} - , {-} , + , - \end{bmatrix}$ for \textit{point-edge}; and $\begin{bmatrix} - , {-} , - , - \end{bmatrix}$ for \textit{point-point}.
	These tuples are distinguished from normal cases based on the sign information of second index $i_{1}$, which is now negative.

	\paragraph{Parallel solver optimizations} 
	Our linear system (\cf~\alglref{solve-lin-sys} in \algref{gpuipc}) is solved using modified preconditioned conjugate gradients (modified-PCG)~\cite{largeStepCloth}.
	We design the steps of each PCG iteration to bypass full construction of the linear system, using local matrix-vector multiplications to then accumulate results in a global output vector.
	Briefly, threads (in a CUDA warp) are each mapped to one scalar entry in the local system matrix
	to then perform multiplication between this scalar and the corresponding component of the input vector (\eg~the vector $\mbc$ as in Line (5) of the modified-PCG algorithm in \cite{largeStepCloth}). 
	The multiplication result computed by each thread is stored in private 
	register memory, which is then accumulated (locally within each CUDA warp) before adding to the global output vector.
	We provide further implementation detail in our technical supplement, where we also summarise a lock-free `warp reduction' strategy for synchronising register memory access between threads. 
	
	{\paragraph{Preconditioning of the CG solver} Although our optimized matrix-vector multiplications can significantly reduce the time required for each CG iteration, the total number of CG iterations can still have significant impact on overall performance, especially when dealing with highly stiff elastic materials (\eg~with Young's Modulus $E=1e7$). 
		We address this issue by incorporating the state-of-the-art Multilevel Additive Schwarz (MAS) Preconditioner~\cite{wu2022gpu} for further improving the convergence rate of our matrix-free GPU-CG solver. Notably, this MAS preconditioner will only provide minor improvements in convergence rate when simulating with soft materials due to low conditioning of the system matrix. 
		Executing the MAS preconditioner will also incur additional overhead, which can reduce overall efficiency when compared to a simple block diagonal preconditioner~\cite{Tissot2019IterativeMF} (each block size is $3\times3$). Therefore, we trial the CG solver with both preconditioners per simulation and choose the one that is most efficient to take advantage of the benefits of both types of preconditioners.
	}

	\section{Results and Discussion}
	\label{sec:results}

	We present our evaluation and results in this section, which are produced on an Ubuntu system with a 16-core Intel Core i9 12900k CPU with 32GB of RAM, and an NVIDIA RTX 4090 GPU with 24GB of RAM. Simulations are updated with the modified-PCG solver \cite{largeStepCloth}, where we have used a  hybrid preconditioner with relative error of 1e-4 (\ie~to give $\delta_{new} < $ 1e-4$\delta_{0}$ as the termination condition of PCG).  {Higher-accuracy thresholds were tested, which we found to have little-to-no further impact on convergence as shown in our technical supplement.} The matrix of our linear system (arising at each Newton iteration) is positive-definite due to mass terms added to elastic/barrier Hessian(s). We also compare our system against the CPU implementation of IPC by \citet{10.1145/3386569.3392425} with CHOLMOD compiled with intel MKL LAPACK, BLAS and multi-threading for further acceleration. Results are produced with double-precision floating-point numbers, and deformation dynamics use the Stable neo-Hookean energy of \citet{10.1145/3180491}, as well as the Baraff-Witkin cloth of \citet{10.1111/cgf.14111} (\figref{cloth-sim}).

	\paragraph{Omitting the second term of the force Jacobian}
	
\begin{figure}[t]
		\centering
		\begin{subfigure}[b]{0.9\columnwidth}
			\centering
			\includegraphics[width=\columnwidth]{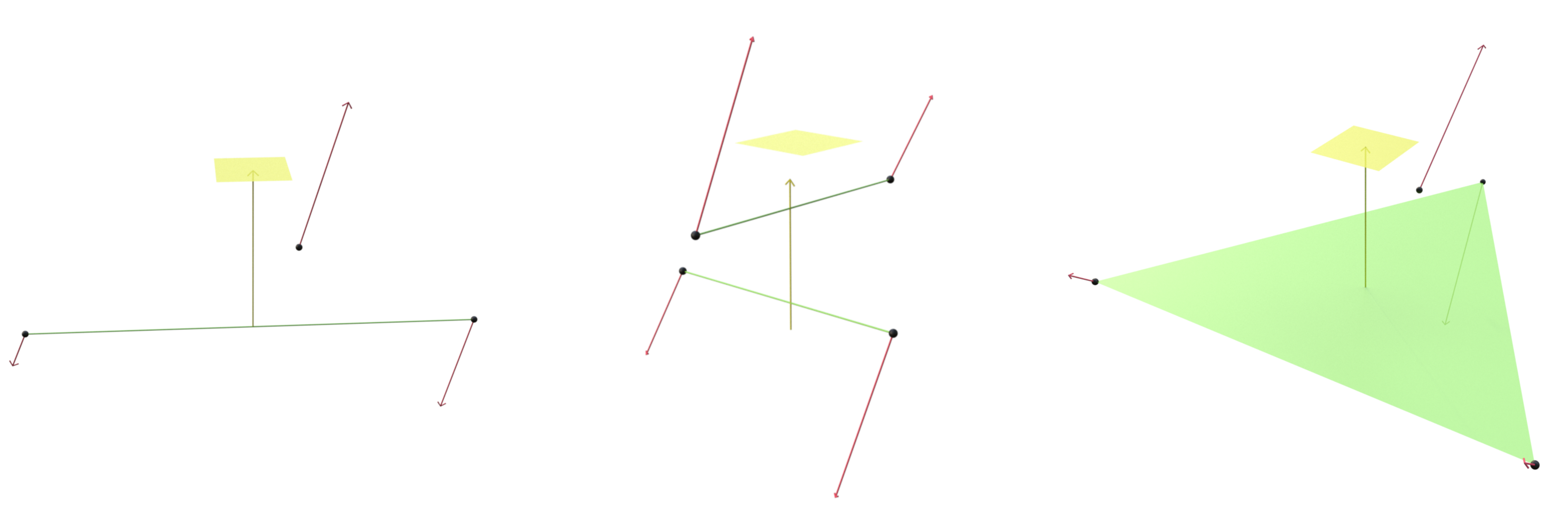}
			\subcaption{\label{fig:search-dir-li}\citet{10.1145/3386569.3392425}}
		\end{subfigure}
		\begin{subfigure}[b]{0.9\columnwidth}
			\centering
			\includegraphics[width=\columnwidth]{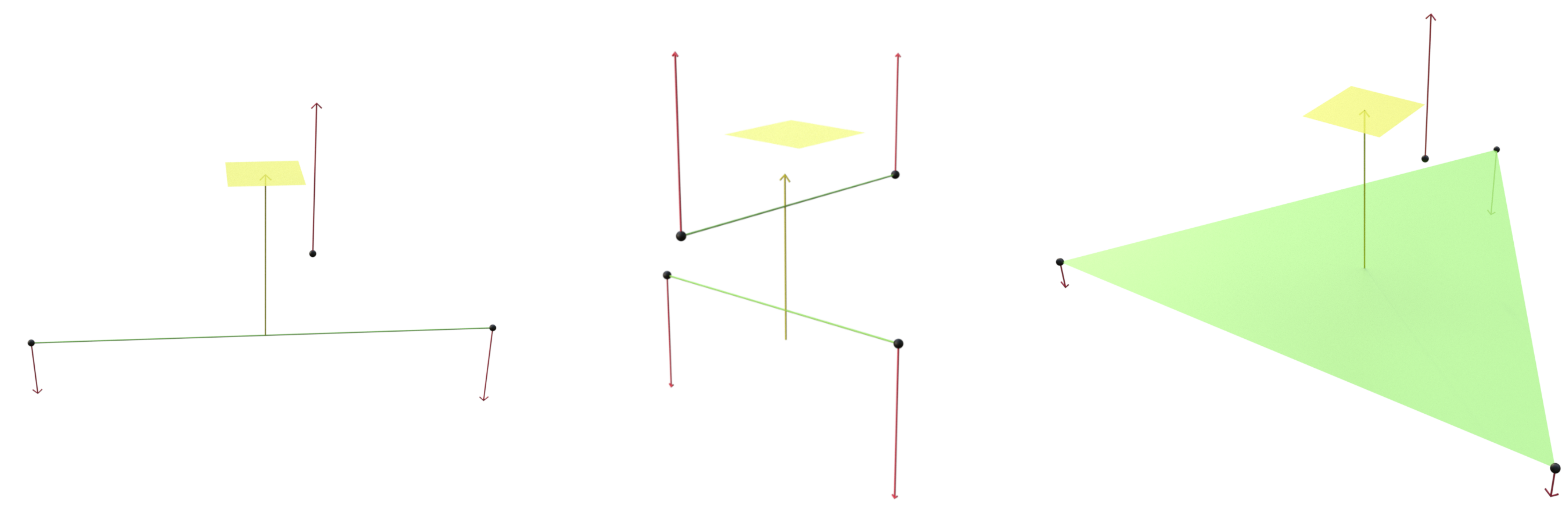}
			\subcaption{\label{fig:search-dir-ours}Ours}
		\end{subfigure}
		
		\caption{\label{fig:search-dir} Evaluating the entire force-Jacobian in \eqref{hess-tens} versus evaluating only the first term. The latter (\figref{search-dir-ours}) restricts the solution vector $\mbd$ (visualized by the red arrows) to the contact normal direction, and the former (\figref{search-dir-li}) is affected by higher order variations between contact pairs due to the nonlinearity of Euclidean distance. The influence of these high-order variations on $\mbd$ are similarly observed with the distance based formulation of \citet{10.1145/3386569.3392425}. Refer to our supplementary video. }
	\end{figure}

	\figref{search-dir} shows the effect of evaluating the entire force-Jacobian in \eqref{hess-tens} versus evaluating only the first term.
	A truncation of force-Jacobian terms gives approximate solutions $\mbd$, where these solutions remain congruent with the optimal search direction of `steepest descent':  
	Each local approximate solution is aligned with the contact normal vector $\mbn$, and barrier energies are minimized precisely by components of the exact solution (computed with the full Hessian) that are parallel-to this normal. {Utilizing a normal-aligned search direction can lead to a slightly faster convergence (up to 25\%) in nonlinear optimization. However, it does not always guarantee faster convergence, as the local optimal search direction may not necessarily result in the globally optimal convergence rate. This is evident in the results presented in rows Fig. \ref{fig:cloth-twisting} and Fig. \ref{fig:rods-twisting} of Tab. \ref{tab:stats}, where the full Hessian method \cite{10.1145/3386569.3392425} achieved a slightly faster convergence.}
	
\begin{figure}[htbp]
		\centering
		\begin{subfigure}[b]{\columnwidth}
			\centering
			\includegraphics[width=\columnwidth, trim=0 30 0 120, clip]{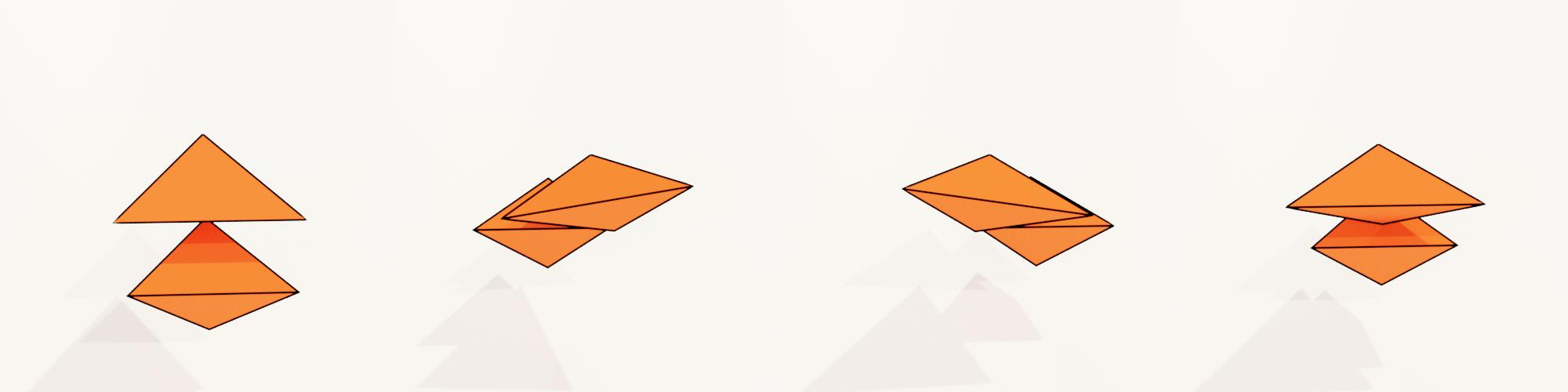}
			\subcaption{\label{fig:tet-ne-li}\citet{10.1145/3386569.3392425}}
		\end{subfigure}
		\begin{subfigure}[b]{\columnwidth}
			\centering
			\includegraphics[width=\columnwidth, trim=0 30 0 120, clip]{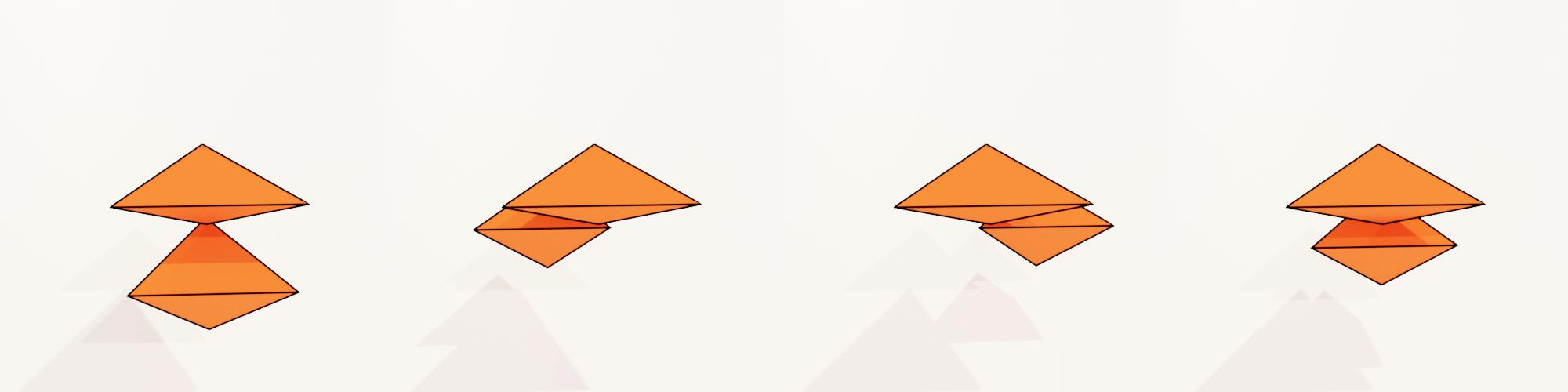}
			\subcaption{\label{fig:tet-ne-ours}Ours}
		\end{subfigure}
		\caption{\label{fig:tet-topple}Validation: Our resting contact is exact. Conversely, using the full Hessian can induce minute variations in motion trajectories subject to the relative positioning between contact pair vertices (point and triangle in this case). Using the full Hessian is only able to achieve artefact-free motion when the local point of contact lies at the barycenter of the test triangle.}
	\end{figure}

\begin{figure}[htbp]
		\centering
		\includegraphics[width=\columnwidth]{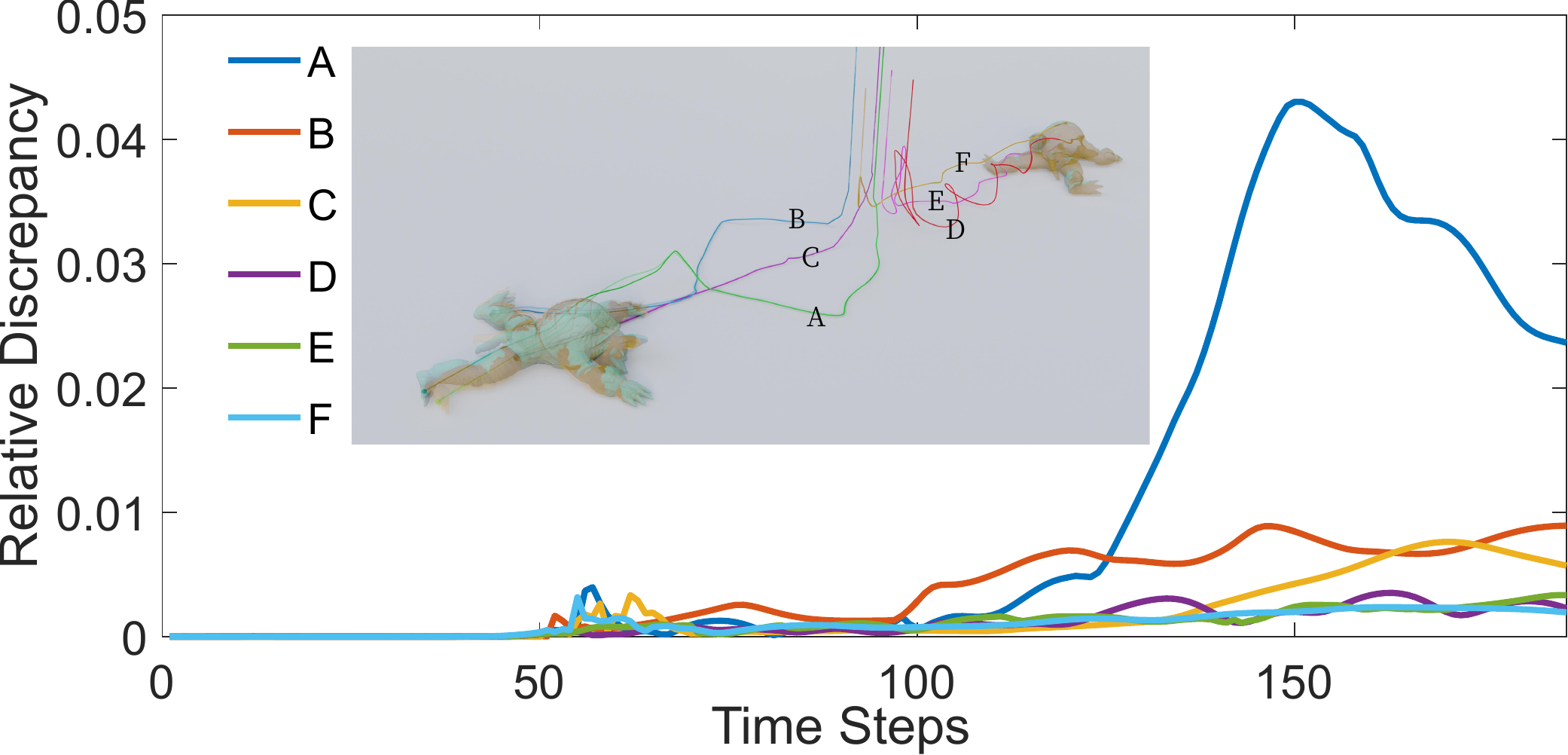}
		\caption{\label{fig:numerical-comp}Comparison of methods: motion discrepancy graphs which are presented comparing a simulation (see inset figure) using the full barrier Hessian versus our approach using only the first term. The curves refer to the discrepancy (measured as mean-squared error) of labelled vertices (A, B, C, D, E and F) on the surface of two Armadillo models. Simulations are run 10 times and averaged per time-step followed by normalization w.r.t. the scene bounding box diagonal of the first time-step. The armadillo models shaded in green denote simulation results evaluated with the full Hessian (\ie~equivalent to \citet{10.1145/3386569.3392425}), the orange models represent our results. Transparent curves are thus reference trajectories, while those which are solid are our computed paths.}
	\end{figure}

	Approximating the Hessian has further advantages. Using a simple unit-test setup shown in \figref{tet-topple}, we also find that solutions computed by evaluating the full Hessian can give inconsistent motion over time, 
	introducing numerical drift-like perturbation of simulation trajectories
	in the absence of any other external loading besides gravity. Our approximations do not suffer from such drifting.
	\figref{numerical-comp} also shows that our GN approximation generally produces the same visual quality as the reference implementation using the full barrier Hessian in complex simulations.
	Further demonstrations of the robustness of our method with similar unit tests as \citet{10.1145/3386569.3392425}  are shown in our supplementary material.

	\begin{figure}[htbp]
		\centering
		\includegraphics[width=0.9\columnwidth]{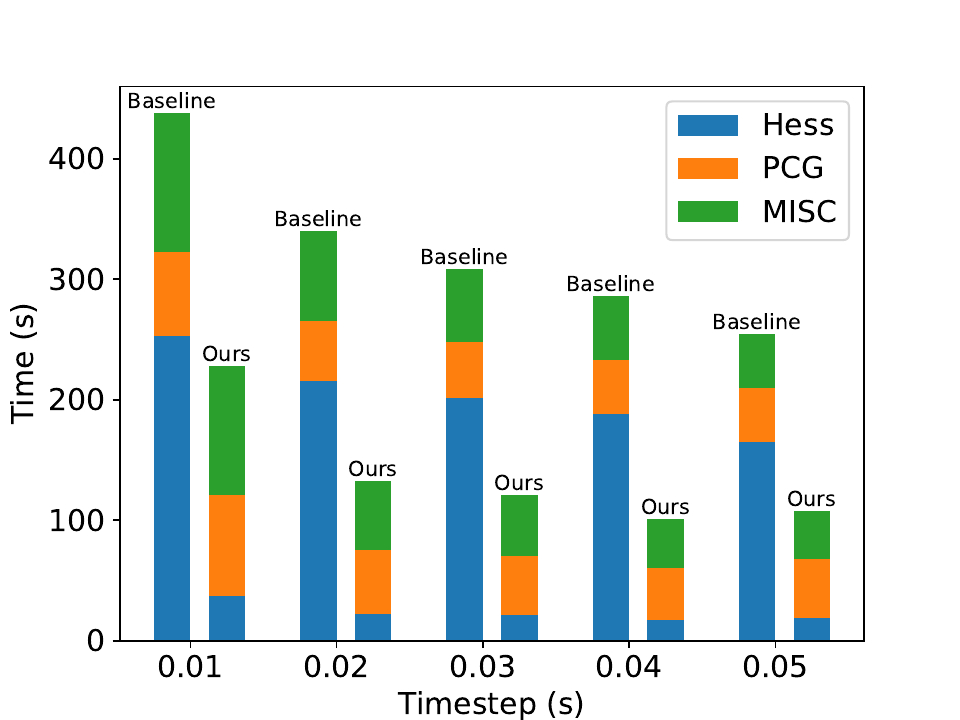}
		\caption{\label{fig:spd-proj} Impact of eigendecomposition: We use the funnel dolphin example shown in \figref{funnel} to demonstrate the time consumption of our IPC barrier Hessian projections on the GPU versus another GPU implementation strategy~\cite{MedialIPC} (baseline) that relies on CPU-side numerical eigendecomposition with the Eigen library. `PCG' is the linear solver and `Misc' are all other remaining tasks, which are shown here to highlight the time-significance of Hessian construction and projection. }
	\end{figure}
	
	\paragraph{Performance of solver and our analytic eigensystems} 
	\figref{spd-proj} summarises the advantage of our simultaneous construction and projection of barrier Hessians on the GPU when deployed into a full simulation pipeline. 
	We compare with a simulator (`baseline'), which performs the same operations as ours except that the barrier Hessians are constructed on the GPU with CPU-side numerical eigendecomposition with Eigen following Medial-IPC\footnote{{The heterogeneous CPU-GPU implementation in Medial-IPC is driven by the size of the barrier Hessian, where the $12\times 12$ Hessian plays a significant role. Achieving accurate numerical eigendecomposition for such a large Hessian requires numerous iterations with numerical decomposition methods. Unfortunately, these methods are not well-suited for parallelization on GPU. Lower-accuracy decomposition results can lead to substantial deviations and convergence failures in Newton iterations. In contrast, the Eigen library provides reliable numerical results and performs well on CPU, making it more suitable for handling this particular task.}}\cite{MedialIPC}.
	Our simultaneous construction and projection of barrier Hessians is over 10$\times$ faster than the baseline case using the funnel dolphin example (\figref{funnel}) to give approximately 3$\times$ speedup in overall simulation time. Thus, the ease and flexibility with which our analytic eigensystems translate into actual code is also beneficial for improving overall data-parallel performance {especial when the number of collision pairs is larger (\eg~ by $4\times$) than the number of finite elements}. 
	
	We additionally perform a focused comparison of the time to evaluate our analytic eigensystems versus the time to compute the traditional procedure of constructing and then numerically projecting barrier Hessians with Eigen \cite{eigenweb}. Elapsed time is measured 1k times to estimate averages. On the CPU we achieve at least {4.0}$\times$ and up-to 7.8$\times$ speedup over Eigen. The GPU implementation of our analytic eigensystems up-to 7.2$\times$ faster than our CPU implementation. Refer to \tabref{ana-ea-speedup}. {It is worth noting that such an SPD performance problem only arises in a well-optimized GPU IPC framework, where the linear solver and collision detection components are well optimized.}
	
	\begin{table}
		\caption{\label{tab:ana-ea-speedup} Comparison of time (milliseconds) to construct and project barrier Hessian (\eqref{4th-ord-tens-vec}), which has dimensions $\realNum^{3n\times3n}$, where $n$ is the number of vertices comprising the contact-pair. Thus, we have $\realNum^{12\times12}$ for \textit{point-triangle} and \textit{edge-edge} cases;  $\realNum^{9\times9}$ for \textit{point-edge}; and $\realNum^{6\times6}$ for \textit{point-point}. The first column shows the number of barrier Hessians. Our analytic eigensystems are faster with upto 7.8$\times$ speedup on CPU. 
		}

		\resizebox{\columnwidth{}}{!}{
			\begin{tabular}{cl|cc >{\columncolor{LightCyan}}c|c >{\columncolor{LightCyan}}c}
				\toprule
				\multicolumn{2}{c}{\texttt{${\partial^2 b}/{\partial \mbJ^2}$}} & \multicolumn{3}{c}{CPU} & \multicolumn{2}{c}{GPU} \\ 
				\midrule
				Count & Dims & Eigen & Ours & Speedup & Ours & Speedup\\
				\midrule
				\multirow{3}{*}{$10^7$} & $\realNum^{12\times12}$ &2.42e3   & 4.95e2      &    4.88$\times$         &9.01e1 &5.50$\times$   \\
				&$\realNum^{9\times9}$ &1.45e3   &2.29e2      &     6.33$\times$            &4.10e1 &5.58$\times$ \\
				&$\realNum^{6\times6}$ &4.79e2   &9.68e1    &    4.94$\times$           &1.32e1 & 7.28$\times$ \\
				\midrule
				\multirow{3}{*}{$10^6$} & $\realNum^{12\times12}$ &2.42e2   & 5.01e1      &    4.82$\times$         &9.03 &5.54$\times$   \\
				&$\realNum^{9\times9}$ &1.49e2   &2.21e1      &     6.73$\times$            &4.05 &5.46$\times$ \\
				&$\realNum^{6\times6}$ &5.02e1   &9.71    &    5.17$\times$           &1.37 & 7.05$\times$ \\
				\midrule
				\multirow{3}{*}{$10^5$}&$\realNum^{12\times12}$ &2.36e1   & 4.83      &    4.47$\times$         &1.08 &5.16$\times$   \\
				&$\realNum^{9\times9}$ &1.44e1   &2.22      &     6.48$\times$            &4.23e-1 &5.25$\times$ \\
				&$\realNum^{6\times6}$ &5.01   &6.38e-1    &    7.86$\times$           &1.62e-1 & 3.92$\times$ \\
				\midrule
				\multirow{3}{*}{$10^4$}&$\realNum^{12\times12}$ &2.33   & 5.22e-1      &    4.46$\times$         &1.76e-1 &3.23$\times$   \\
				&$\realNum^{9\times9}$ &1.50   &2.25e-1      &     6.65$\times$            &6.28e-2 &3.58$\times$ \\
				&$\realNum^{6\times6}$ &5.04e-1   &7.55e-2    &    6.67$\times$           &2.62e-2 & 2.87$\times$ \\
				\midrule
				\multirow{3}{*}{$10^3$}&$\realNum^{12\times12}$ &2.38e-1   & 5.57e-2      &    4.27$\times$         &1.65e-1 &3.37e-1$\times$   \\
				& $\realNum^{9\times9}$ &1.55e-1   &3.05e-2      &     5.09$\times$            &5.69e-2 & 5.37e-1$\times$ \\
				& $\realNum^{6\times6}$ &5.82e-2   &1.42e-2    &    4.08$\times$           &2.23e-2 & 6.37e-1$\times$ \\
				\bottomrule
			\end{tabular}
		}
	\end{table}
	
	\begin{table}
		\caption{\label{tab:par-solver-speedup} Comparison of parallel PCG solver time (in seconds).  We compare against an implementation where each local/per-Barrier term matrix-vector multiplication (during PCG) is computed by one GPU thread. Table columns are as follows: Total time spent computing linear system (PCG); Miscellaneous simulation tasks (Misc); Total simulation time (TimeTot); total number of conjugate gradient iterations (\#cg); Total number of Newton solver iterations (\#Newton); Average number of conjugate gradient iterations per Newton solve iteration (\#cg per iter). { Both methods are tested with the same PCG threshold of $1\mathrm{e}{-4}$.}
		}
		\resizebox{\columnwidth{}}{!}{
			\begin{tabular}{rr|cccccc}
				\toprule
				&  & PCG & Misc & TimeTot & \#cg & \#Newton & avg \#cg per iter \\ 
				\midrule
				\multirow{3}{*}{\figref{rods-twisting}} & Ours  & 1.18e3 & 5.83e2  & 1.76e3 & 2.66e6 & 2.19e4 & 121.1  \\
				& Baseline  & 4.69e3
				& 5.83e2  & 5.27e3 & 2.65e6 & 2.19e4 & 121.0  \\
				& \textbf{Speedup}  & \textbf{3.98}$\times$ & -  & \textbf{2.99}$\times$ & - & - & -  \\
				\midrule
				\multirow{3}{*}{\figref{cloth-twisting}} & Ours  & 1.12e3 & 1.26e3  & 2.38e3 & 3.42e6 & 4.24e4 & 80.6  \\
				& Baseline  &4.28e3 & 1.25e3  & 5.53e3 & 3.43e6 &4.23e4 & 81.0    \\
				& \textbf{Speedup}  & \textbf{3.82}$\times$ & -  & \textbf{2.32}$\times$ & - & - & -  \\
				\midrule
				\multirow{3}{*}{\figref{bunnies-dropping}} & Ours  & 3.68e3 & 2.37e3  & 6.05e3 & 9.50e5 &1.41e4 & 67.4  \\
				& Baseline  &1.51e4 & 2.37e3  & 1.75e4 & 9.54e5 &1.41e4 & 67.7    \\
				& \textbf{Speedup}  & \textbf{4.11}$\times$ & -  & \textbf{2.89}$\times$ & - & - & -  \\
				\midrule
				\multirow{3}{*}{\figref{balls-dropping}} & Ours  & 1.25e3 & 8.80e2  & 2.13e3 & 2.10e6 &1.72e4 & 122.1  \\
				& Baseline  &5.01e3 & 9.00e3  & 5.91e3 & 2.13e6 &1.74e4 & 122.3    \\
				& \textbf{Speedup}  & \textbf{4.00}$\times$ & -  & \textbf{2.77}$\times$ & - & - & -  \\
				\bottomrule
			\end{tabular}
		}
	\end{table}
	
	To evaluate the performance improvement due to our fine-grain strategy for parallelizing local matrix-vector multiplications during PCG, we compare with an alternative implementation that directly scales the number of GPU threads by the number of contact pairs and finite elements. That is, computing one local matrix-vector multiplication per GPU thread like \citet{gpumpm}. Our results are shown in \tabref{par-solver-speedup}, which are based on evaluations using four scenes. Our PCG solver is up-to $4\times$ faster than the baseline reference (averaging $3.97\times$), which leads to an overall $2\times$ speedup in simulation time.

	\begin{figure}[htbp]
		\begin{tabular}{@{}cccc@{}}
			\includegraphics[width=0.225\textwidth, trim=350 130 380 550, clip]{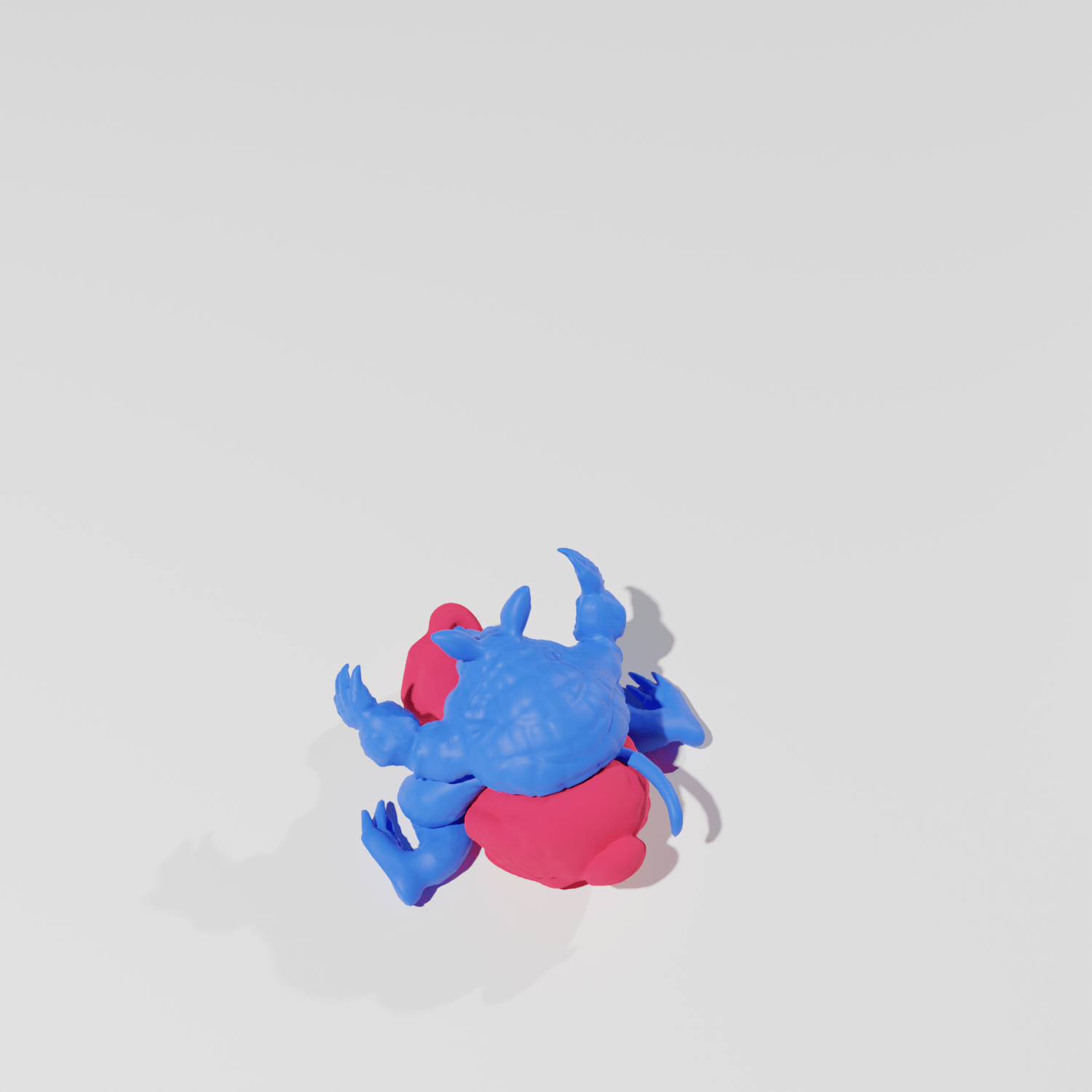} & 
			\includegraphics[width=0.225\textwidth, trim=350 130 380 550, clip]{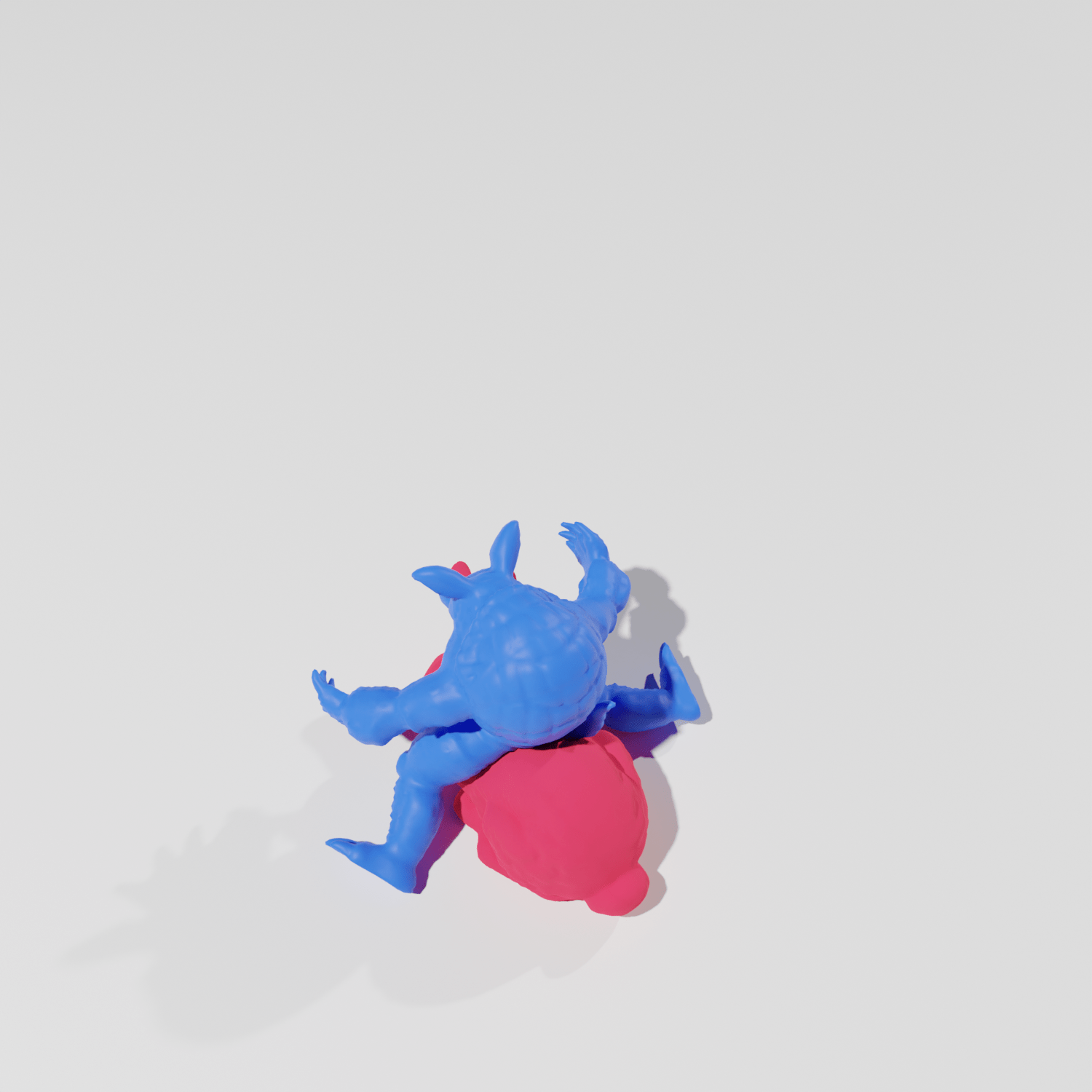} \\
			(a) 1e4  & (b) 1e5  \\
			\includegraphics[width=0.225\textwidth, trim=350 130 380 550, clip]{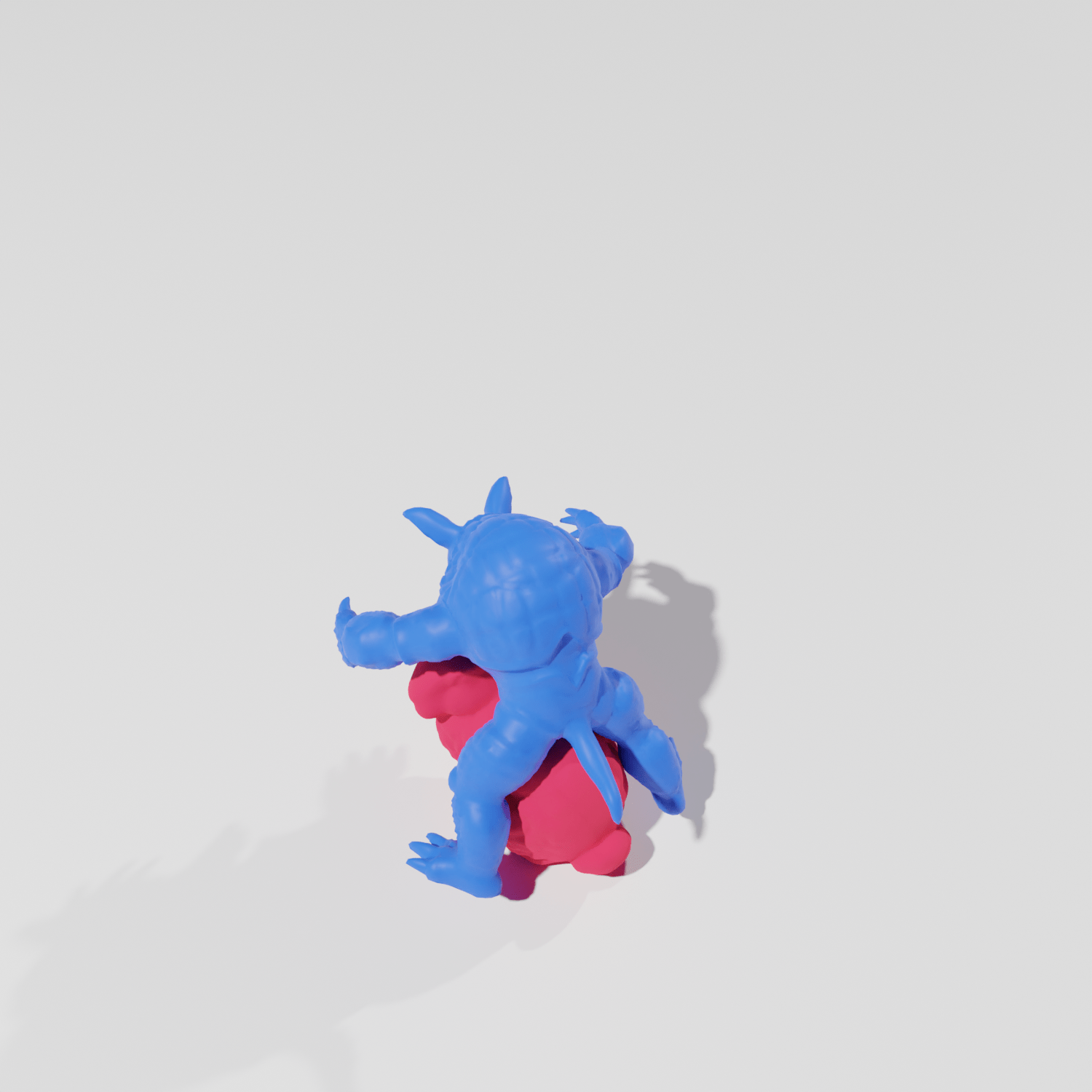} & 
			\includegraphics[width=0.225\textwidth, trim=350 130 380 550, clip]{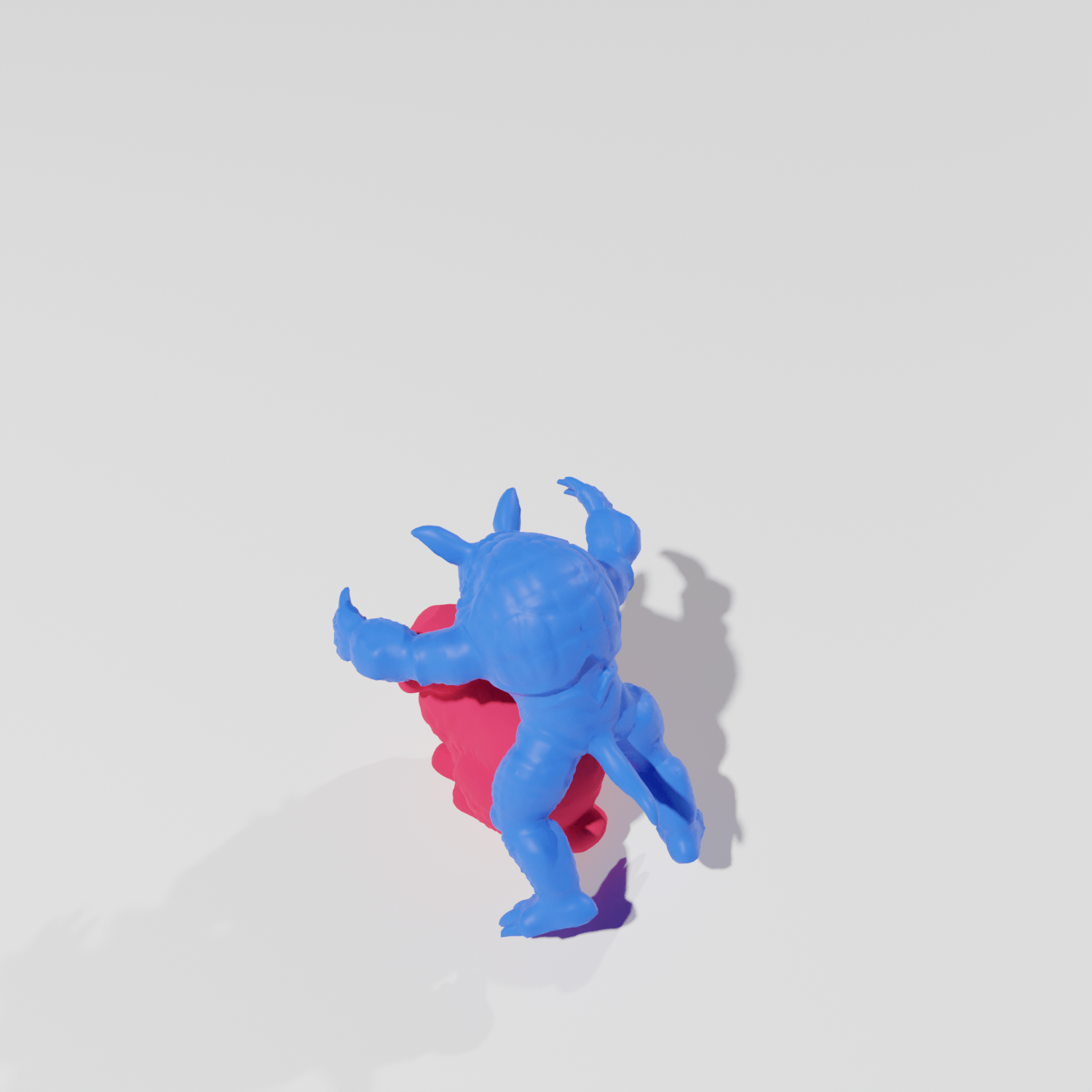} \\
			(c) 1e6 & (d) 1e7  \\
		\end{tabular}
		\caption{\label{fig:large_stiffness}Simulating with different values for material stiffness (Youngs modulus): Our framework can handle extremely large stiffness simulations efficiently.}
	\end{figure}
	
	\begin{figure}[htbp]
		\centering
		\includegraphics[width=\columnwidth]{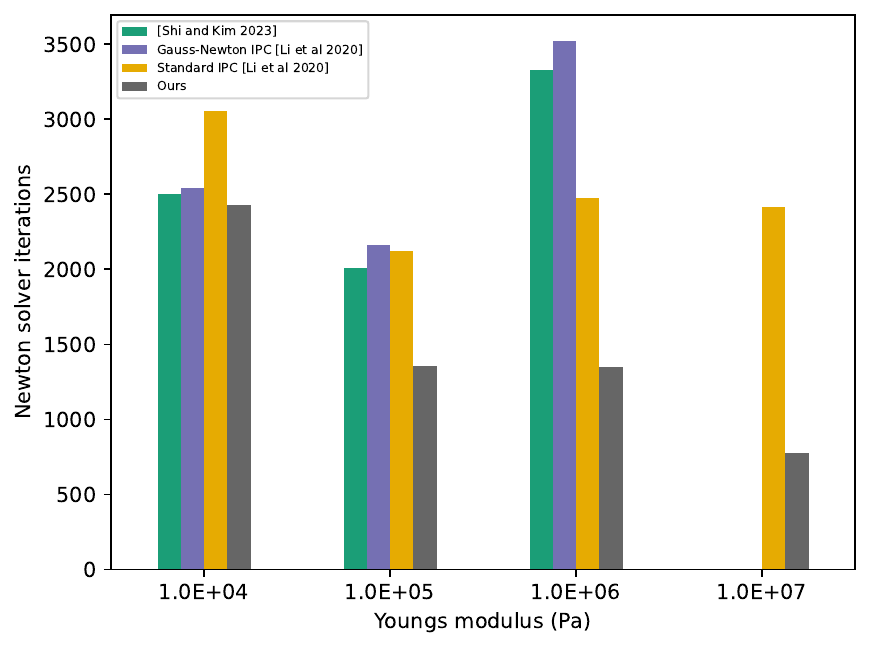}
		\caption{
			A plot comparing the number of Newton solver iterations versus material stiffness (GPa). The plotted data {(\cf~\tabref{youngs-mod-to-iterations})} is obtained from the demos shown in \figref{large_stiffness} (a) (b) (c) and (d), respectively. Our implementation requires the least number of iterations for the values of Young modulus tested (up-to 3$\times$ less). 
			Thus, our method is more efficient than \citet{10.1145/3386569.3392425} {as well as \citet{10.1145/3606934}} by overcoming the drawbacks typically associated with a Gauss-Newton approximation of the barrier Hessian like significantly low convergence rates or even failure to converge with highly stiff materials (\eg~ $E=1\mathrm{e}{7}$).
		}
		\label{fig:youngs-mod-to-iterations}
	\end{figure}
	
	\begin{table}[!ht]
		\centering
		\caption{{A comparison based on the number of Newton solver iterations w.r.t material stiffness (Youngs modulus). See also \figref{youngs-mod-to-iterations}}.}
		\resizebox{\columnwidth}{!}{
			\begin{tabular}{ccccc}
				\toprule
				Youngs modulus (Pa) &  \multicolumn{4}{c}{Method} \\
				\cmidrule{2-5}
				& \citet{10.1145/3606934} & Gauss-Newton IPC & Standard IPC & Ours \\ 
				\midrule
				$1\mathrm{E}{4}$ & 2501 & 2543 & 3053 & 2427 \\ 
				$1\mathrm{E}{5}$ & 2006 & 2162 & 2123 & 1358 \\ 
				$1\mathrm{E}{6}$ & 3326 & 3519 & 2475 & 1352 \\ 
				$1\mathrm{E}{7}$ & \textbf{FAIL} & \textbf{FAIL} & 2412 & 779 \\ 
				\bottomrule
		\end{tabular}}
		\label{tab:youngs-mod-to-iterations}
	\end{table}
	
	\begin{figure}[!ht]
		\centering
		\includegraphics[width=\columnwidth]{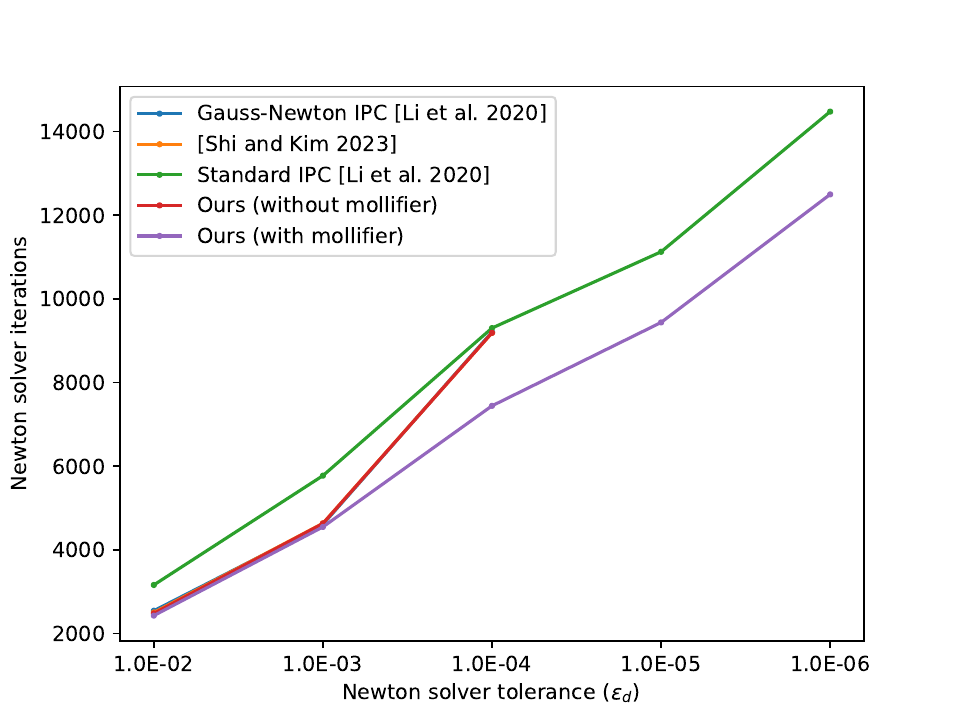}
		\caption{{Comparison of PN solver iteration count versus convergence threshold $\varepsilon_d$ (scaled by scene bounding box diagonal), where we have used the demo in \figref{large_stiffness} (a) to obtain our data for a fair comparison since the Gauss-Newton IPC and the method of \citet{10.1145/3606934} are susceptible to failure (\eg~when $\varepsilon_d=10^{-5}$ or material is stiff). We compare our analytically projected mollified Hessian in \secref{moll-bar-Hess-ana} with four alternative implementations to show the practical advantage of our method in improving convergence rates even with small thresholds $\varepsilon_d$. The data shown here is also provided in \tabref{proj-newt-iters} for reference}.  
		}
		\label{fig:par-ee-iterations-comp}
	\end{figure}
	
	\begin{table}[htbp]
		\centering
		\caption{{A comparison based on the behaviour of the projected-Newton solver, where we measure the number of iterations versus tolerance. See also \figref{par-ee-iterations-comp} for a visualisation}.}
		\resizebox{\columnwidth}{!}{
			\begin{tabular}{cccccc}
				\toprule
				Tolerance &  \multicolumn{5}{c}{Method} \\
				\cmidrule{2-6}
				& Gauss-Newton IPC & \citet{10.1145/3606934} & Standard IPC & Ours (\textit{without} mollifier) & Ours \\ 
				\midrule
				$1\mathrm{E}{-2}$ & 2543 & 2501 & 3159 & 2478 & 2427 \\ 
				$1\mathrm{E}{-3}$ & 4606 & 4633 & 5772 & 4628 & 4546 \\ 
				$1\mathrm{E}{-4}$ & 9206 & 9186 & 9300 & 9186 & 7441 \\ 
				$1\mathrm{E}{-5}$ & \textbf{FAIL} & \textbf{FAIL} & 11123 & \textbf{FAIL} & 9434 \\ 
				$1\mathrm{E}{-6}$ & \textbf{FAIL} & \textbf{FAIL} & 14474 & \textbf{FAIL} & 12495 \\ 
				\bottomrule
			\end{tabular}
		}
		\label{tab:proj-newt-iters}
	\end{table}

	\paragraph{Material stiffness and performance} \figref{youngs-mod-to-iterations} and \tabref{youngs-mod-to-iterations} show the impact of material stiffness on the number of Newton solver iterations, where we compare our method with  \citet{10.1145/3386569.3392425} and their discussed Gauss-Newton method (\cf~\eqref{bar-hess-orig}), as well as the recent method of \citet{10.1145/3606934}. {The detailed experimental settings are provided in row Fig. \ref{fig:large_stiffness} of Tab. \ref{tab:stats}, which were utilized for all comparison methods.} For the values tested, we find that our method requires the least amount of iterations to converge in every case (up to 3$\times$ less), which demonstrates that our combination of projected Hessian approximations with mollification can lead to a more efficient simulator. {The method of \citet{10.1145/3606934} requires a consistently higher number of iterations (up to 2.4$\times$ more in our tests) and we found it unable to simulate stiff materials \eg~with $E=1\mathrm{e}{7}$.}

	\begin{figure}[htbp]
		\centering
		\includegraphics[width=\columnwidth]{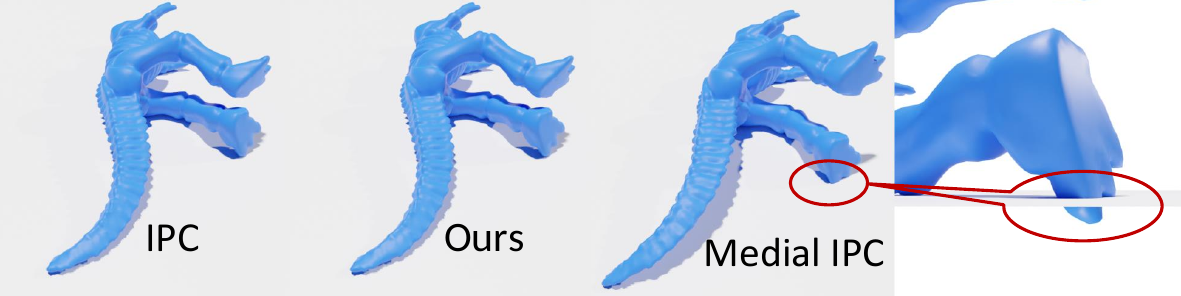}
		\caption{We compare our GIPC framework with Medial-IPC  \cite{MedialIPC}. We use the scene from Medial-IPC's publicly available source code, which we configure as follows: $\Delta t = 0.02, E = 1e3, \upsilon = 0.35, \rho = 2$, which aligns with the default setting of Medial IPC. The solver threshold is set to $\varepsilon_d$ = 1e-2l and the distance threshold to $\hat{d}$ = 1e{-3}l, which aligns with the default setting of standard IPC.  }
		\label{fig:medialIPCCP}
	\end{figure}

	\paragraph{Nearly-parallel edges} The significance of our novel analytically projected mollified barrier Hessian is demonstrated with comparison against alternative strategies as shown in \figref{par-ee-iterations-comp} {and \tabref{proj-newt-iters}}. The impact of our approach to handling nearly-parallel \textit{edge-edge} contacts is most observable when simulating with extremely small solver thresholds $\varepsilon_d$. 
	Comparing across the spectrum from $\varepsilon_d = 10^{-2}$ to $\varepsilon_d = 10^{-6}$, we find that our mollifier yields the lowest number of iterations in most settings evaluated. 
	Our improvement is most significant when using extremely small thresholds {(\eg~$\varepsilon_d = 10^{-5}$ and above), where Gauss-Newton IPC, our approach \textit{without} mollification and the method of \citet{10.1145/3606934} fail}.  
	
	\begin{figure}[htbp]
		\centering
		\begin{subfigure}[b]{0.42\textwidth}
			\centering
			\includegraphics[width=\columnwidth]{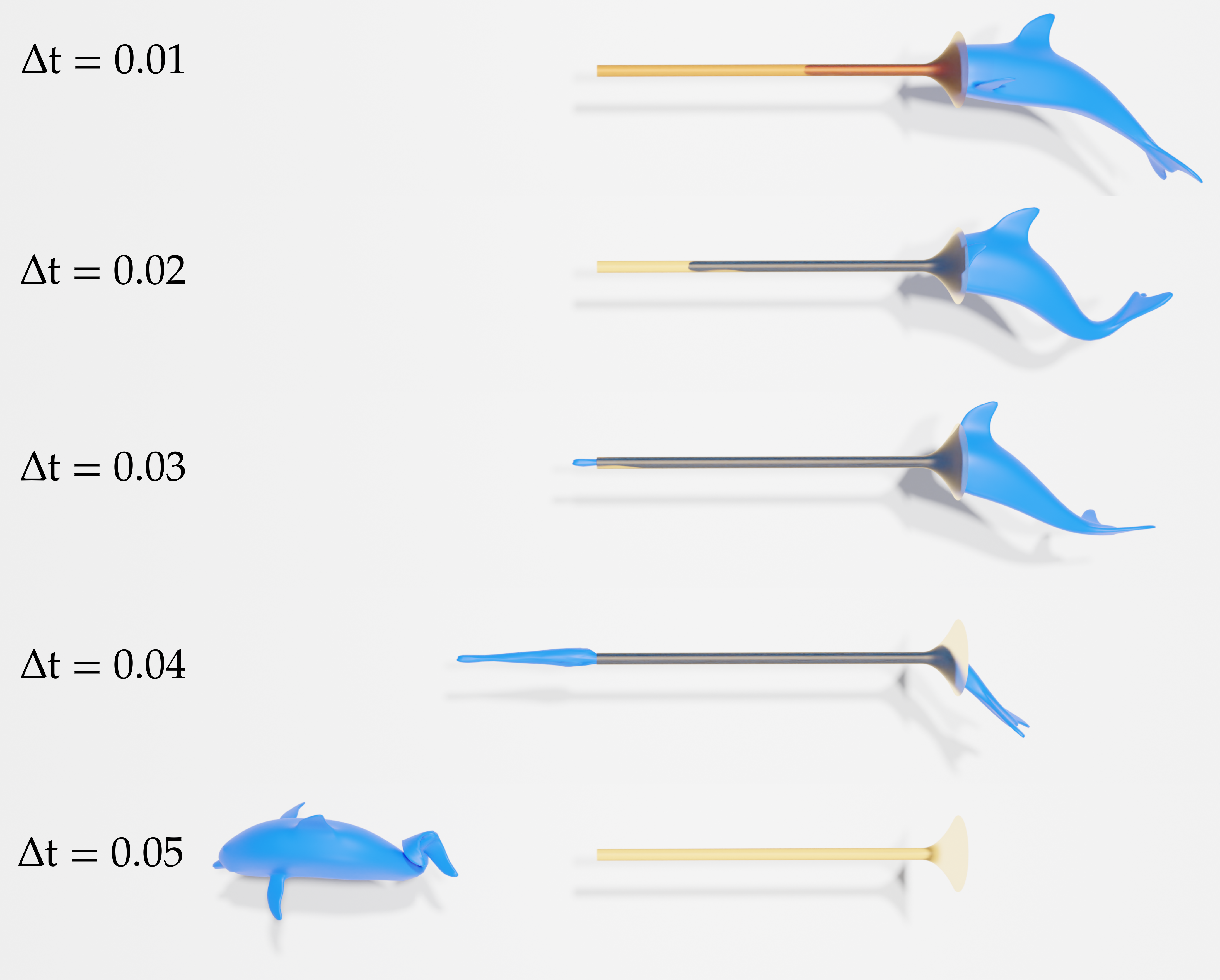}
		\end{subfigure}
		\caption{\label{fig:funnel} Funnel test: From top to bottom we show simulation with varying timestep sizes. The simulation remains both intersection- and inversion-free throughout all time steps.}
	\end{figure}
	
	\paragraph{Funnel dolphin} We also demonstrate the robustness of our contact resolution with elastic material undergoing severe compression and strong boundary conditions. Repeating \citet{10.1145/3386569.3392425}'s funnel test, we pull a elastic material dolphin model through a co-dimensional funnel mesh. Several time step sizes are evaluated ranging from 0.01s to 0.05s with the results shown in \figref{funnel}. Our simulations remain intersection- and inversion-free across all time-steps, demonstrating the robustness of our method even for relatively large time step sizes. 
	
	\paragraph{Medial-IPC}
	\figref{medialIPCCP} shows an experimental setup, which we use to compare results with Medial-IPC \cite{MedialIPC} (using their publicly available source code) to evaluate visual quality and performance.
	Medial-IPC offers a fast reduced simulation framework with GPU optimization but does not resolve contacts exactly due to an inconsistency of medial primitives
	and deformed vertices positions 
	(see \figref{medialIPCCP}, right).
	{ We conducted our performance comparisons with Medial IPC on Windows because the available implementation of Medial-IPC source code is based on this operating system. It is worth noting too that standard IPC \cite{10.1145/3386569.3392425} runs significantly faster on Linux than on Windows for several reasons (including better support for Intel's MKL), some of which are unknown. Thus, to ensure a fair comparison with IPC for this experiment, we still run it on Linux with the same the optimizations mentioned for CPU IPC.}
	For performance, we have found our method (average 0.38s/frame) to be $5.2\times$ faster than Medial-IPC \textit{in this setup} (average 2.0s/frame), where we have also found Medial-IPC to be 2.2$\times$ faster than standard IPC \cite{10.1145/3386569.3392425} (average 4.3s/frame). {\color{red}
	}
	Our method is 11.3$\times$ faster than standard IPC here, which is generally lower than other results shown in \tabref{stats}. Such a lower speedup of our method over standard IPC is to be expected however due the relatively low amount of work for GPU parallelism, while the relative stiffness is large.

	\paragraph{Overall performance} The overall performance summary of our method is provided in \tabref{stats}, where we report both CPU \cite{10.1145/3386569.3392425} and GPU (ours) running time. For reference dynamics please see our supplemental video. Our method achieves significant speedup in all cases evaluated. Our GPU method is on average 46.7$\times$ faster than the reference CPU implementation. Our lowest speedup is 11.3$\times$, and in the best case we achieve up-to 95.7$\times$ faster performance. For reference, the rod twist simulation in \figref{rods-twisting} (twisted for 50s) took over 20.3 hours with the CPU implementation, compared to a mere 29.3 minutes with our simulator. { To further validate the enhancements brought about by our barrier method, we replicated the scenarios depicted in \figref{large_stiffness} and \figref{funnel} using CPU-IPC \cite{10.1145/3386569.3392425} alongside our barrier method. Additional details and results can be found in our technical supplement.}
	
	The supplementary dressed-actor animation shown in \figref{cloth-sim} is modelled with {160k} triangles and simulated with the finite element formulation of the Baraff-Witkin model \cite{10.1111/cgf.14111}. The simulation took 0.09s per Newton iteration using a timestep size of 0.01s, with an average of 25 iterations per timestep to solve fully converged deformation and frictional contact. The whole animation sequence contains 2678 timesteps and is computed within 1.8 hours.
	The human body motion is from the AMASS dataset \cite{AMASS:ICCV:2019}

	\paragraph{Differentiable simulation} 
	IPC and its variants~\cite{10.1145/3386569.3392425, Li2021CIPC, MedialIPC} have introduced differentiable contact handling methods. 
	{Building upon this foundation, \citet{huang2022differentiable} have developed a differentiable solver based on IPC. However, its performance is hindered by the bottleneck of CPU-IPC, resulting in slow computation times.}
	We here show a material parameter optimization demo by making use of the differentiability of our GPU simulator with a cloth example (see \figref{diffcloth}). The apparel in the middle row (blue) shows the forward simulation with a random initial stretch stiffness parameter; the lower row (purple) shows the simulation with optimized stretch stiffness solved with the inverse method of \citet{face2005} and our barrier gradients, such that it matches the target deformation shown in \figref{diffcloth}, top row. Our optimized simulation closely matches the target trajectory. 
	Note that the lower part of the apparel is composed of two layers and thus the deformation due to self-contacts must be also differentiable for the optimization to converge. 
	
	{While there is still room for further exploration of differentiable simulation techniques, our method brings notable performance advantages to the field. Specifically, our approach introduces a highly efficient GPU IPC framework that ensures intersection-free simulations. As evidenced by our results, this inverse simulation (the cloth is modeled with approximately 15,000 triangles.) was successfully completed in just 200 seconds, encompassing 300 time steps, showcasing the speed and effectiveness of our method.}
	
	\begin{figure}[t]
		\centering
		
		\begin{subfigure}[b]{0.44\columnwidth}
			\centering
			\includegraphics[width=\columnwidth, trim=400 200 400 200, clip]{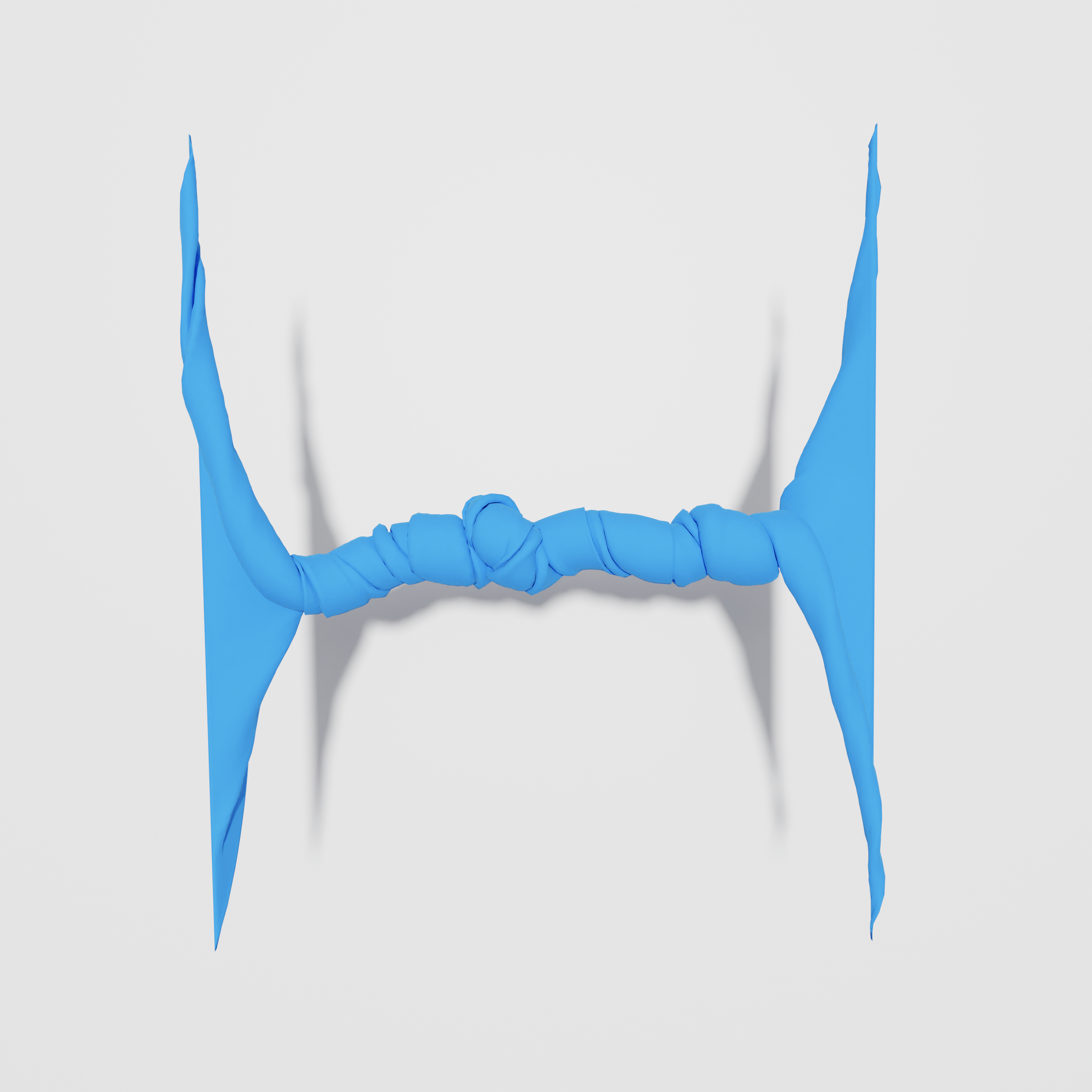}
		\end{subfigure}
		\begin{subfigure}[b]{0.44\columnwidth}
			\centering
			\includegraphics[width=\columnwidth, trim=400 200 400 200, clip]{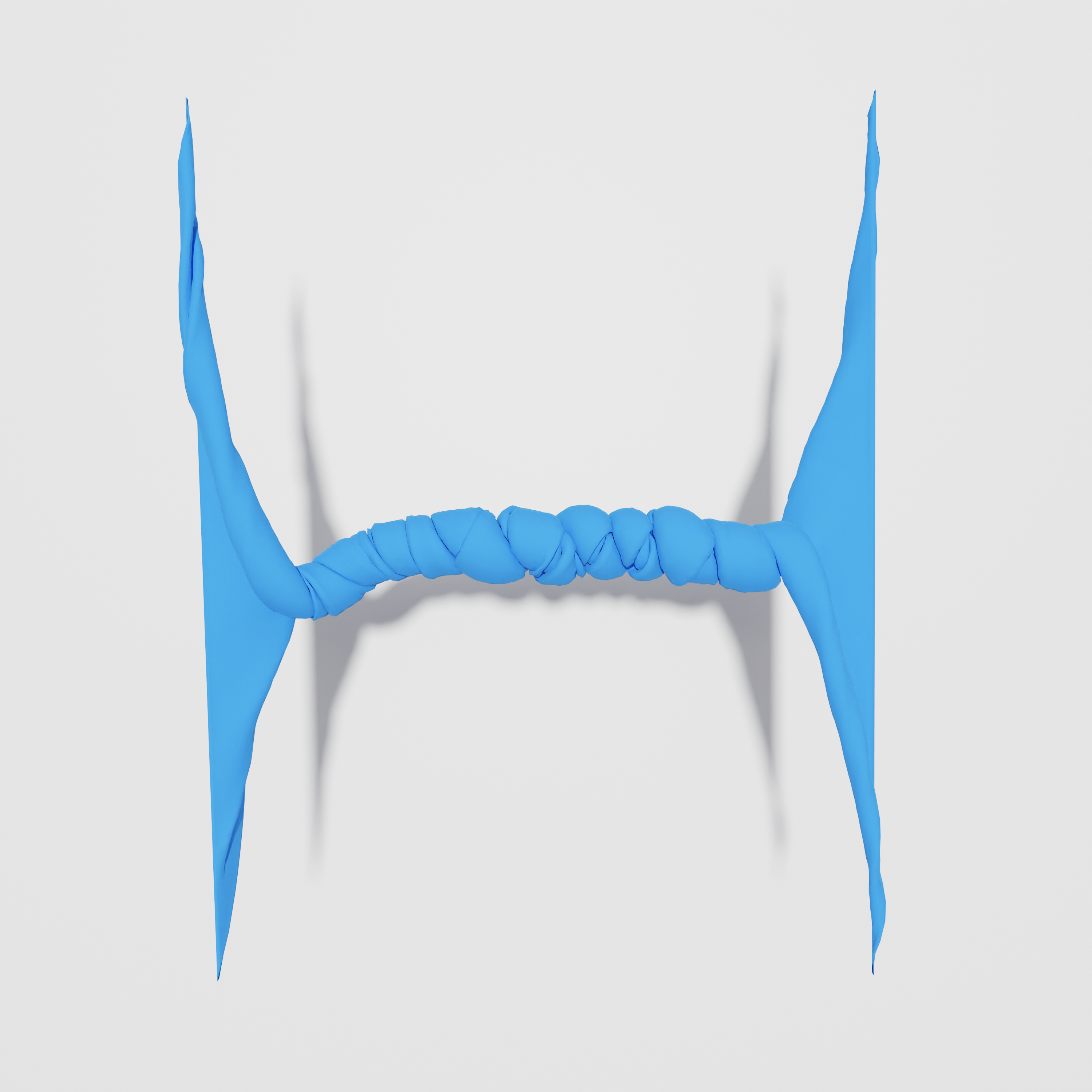}
		\end{subfigure}
		\caption{\label{fig:cloth-twisting}Twisting cloth: Here we reproduce \citet{10.1145/3386569.3392425}'s cloth stress test with extreme twisting of a volumetric mat for 100s.
		}
	\end{figure}

	\begin{figure}[htbp]
		\centering
		\begin{subfigure}[b]{0.42\textwidth}
			\centering
			\includegraphics[width=\columnwidth]{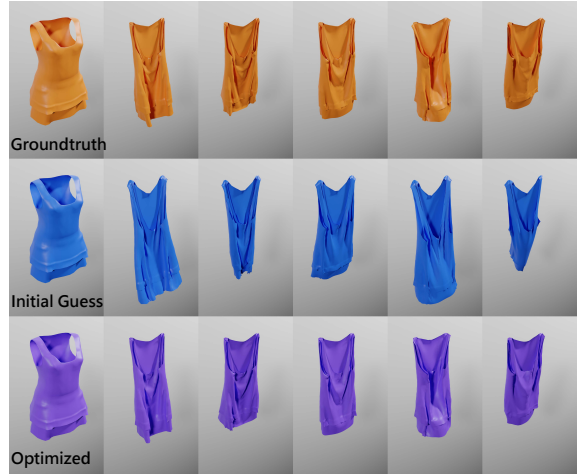}
		\end{subfigure}
		
		\caption{\label{fig:diffcloth} Differentiable test: We show a simulation with optimized stretch stiffness to match the target trajectory.}
	\end{figure}
	
	\begin{figure}[htbp]
		\centering
		\begin{subfigure}[b]{0.42\textwidth}
			\centering
			\includegraphics[width=\columnwidth, trim=0 0 0 0, clip]{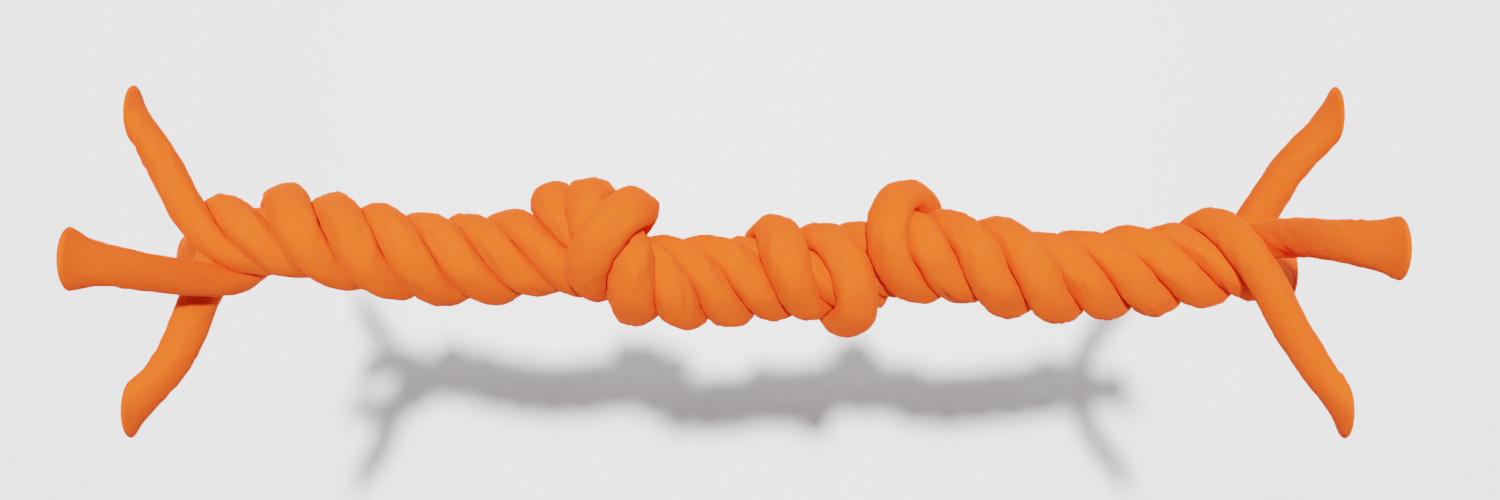}
		\end{subfigure}
		\begin{subfigure}[b]{0.42\textwidth}
			\centering
			\includegraphics[width=\columnwidth, trim=0 0 0 0, clip]{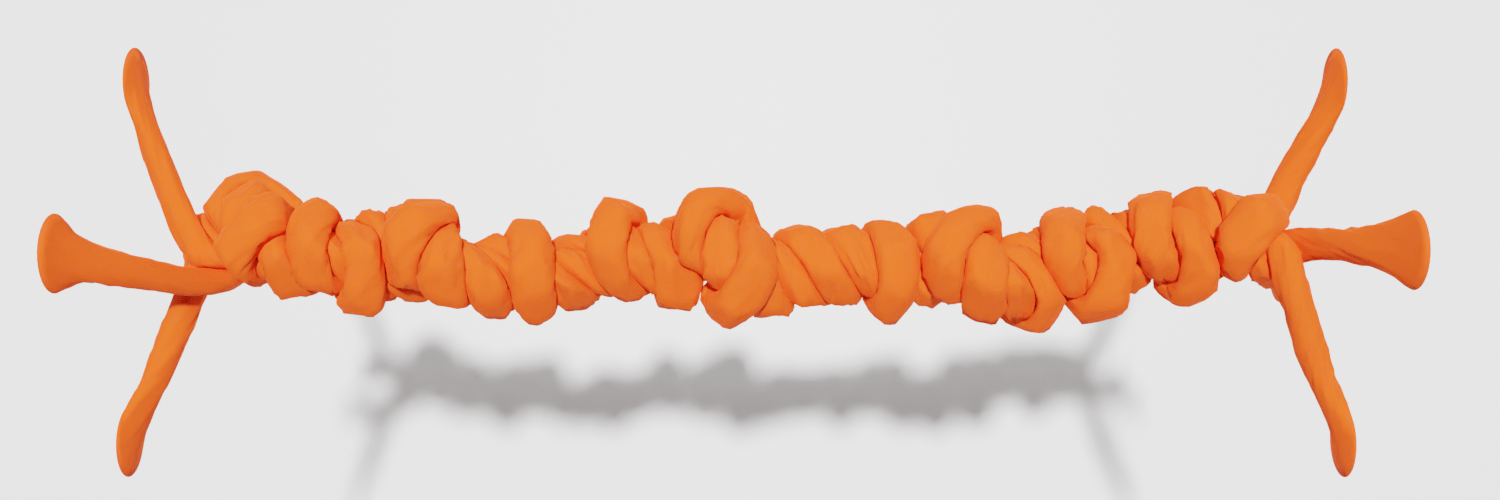}
		\end{subfigure}
		\caption{\label{fig:rods-twisting}Rod twist: In this example, we reproduce the rod twisting test. 
		}
	\end{figure}
	\begin{figure}[htbp]
		\centering
		\begin{subfigure}[b]{0.4\textwidth}
			\centering
			\includegraphics[width=\columnwidth, trim=400 260 400 230, clip]{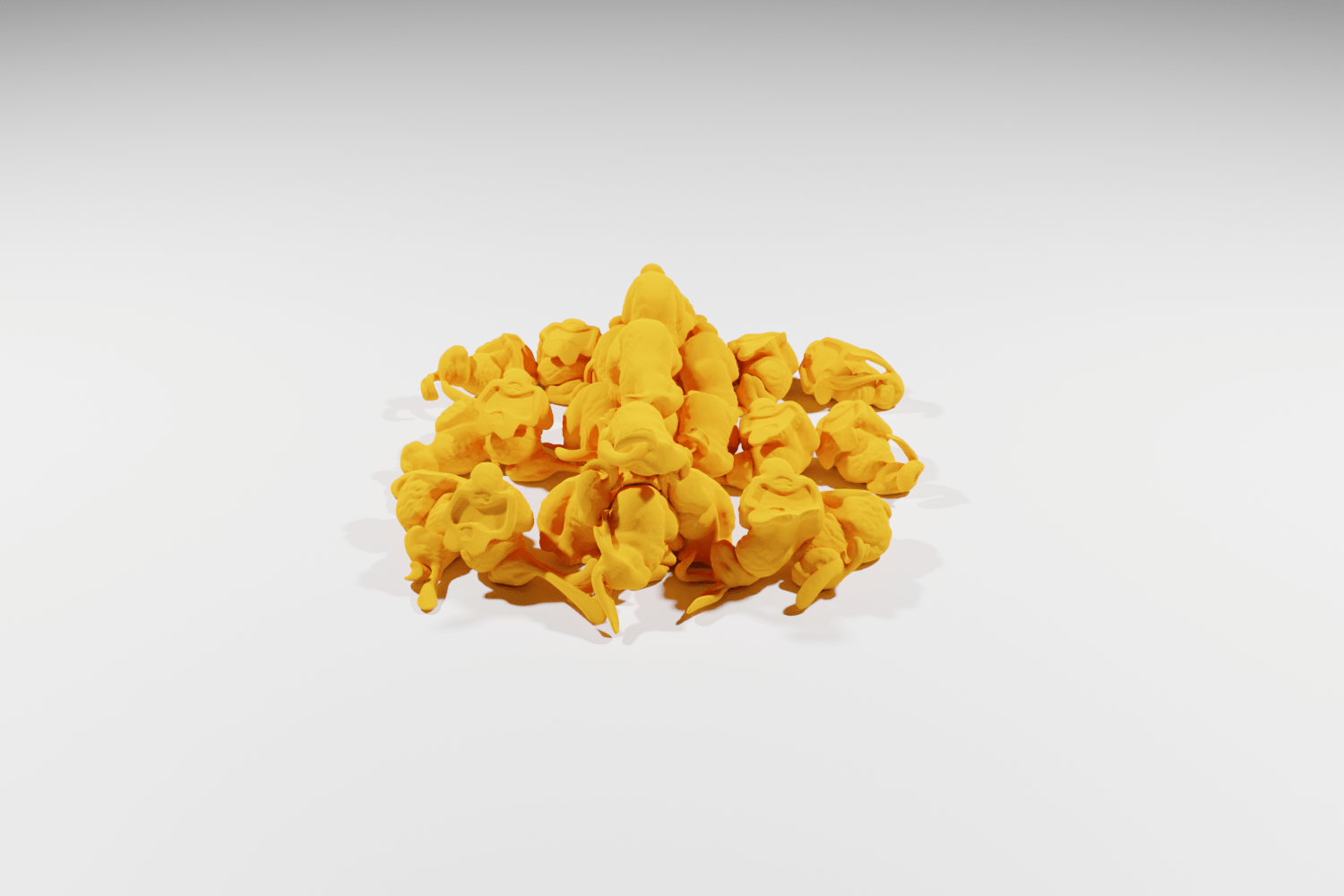}
		\end{subfigure}
		\caption{\label{fig:bunnies-dropping}Bunny drop: We simulate 27 Stanford Bunny models containing $2158k$ tetrahedra and a total of $1686k$ degrees of freedom. Our GPU simlation is over 95$\times$ than standard IPC \cite{10.1145/3386569.3392425} on the CPU.  
		}
		
	\end{figure}
	\begin{figure}[htbp]
		\centering
		\centering
		\begin{subfigure}[b]{0.22\textwidth}
			\centering
			\includegraphics[width=\columnwidth, trim=250 200 350 350, clip]{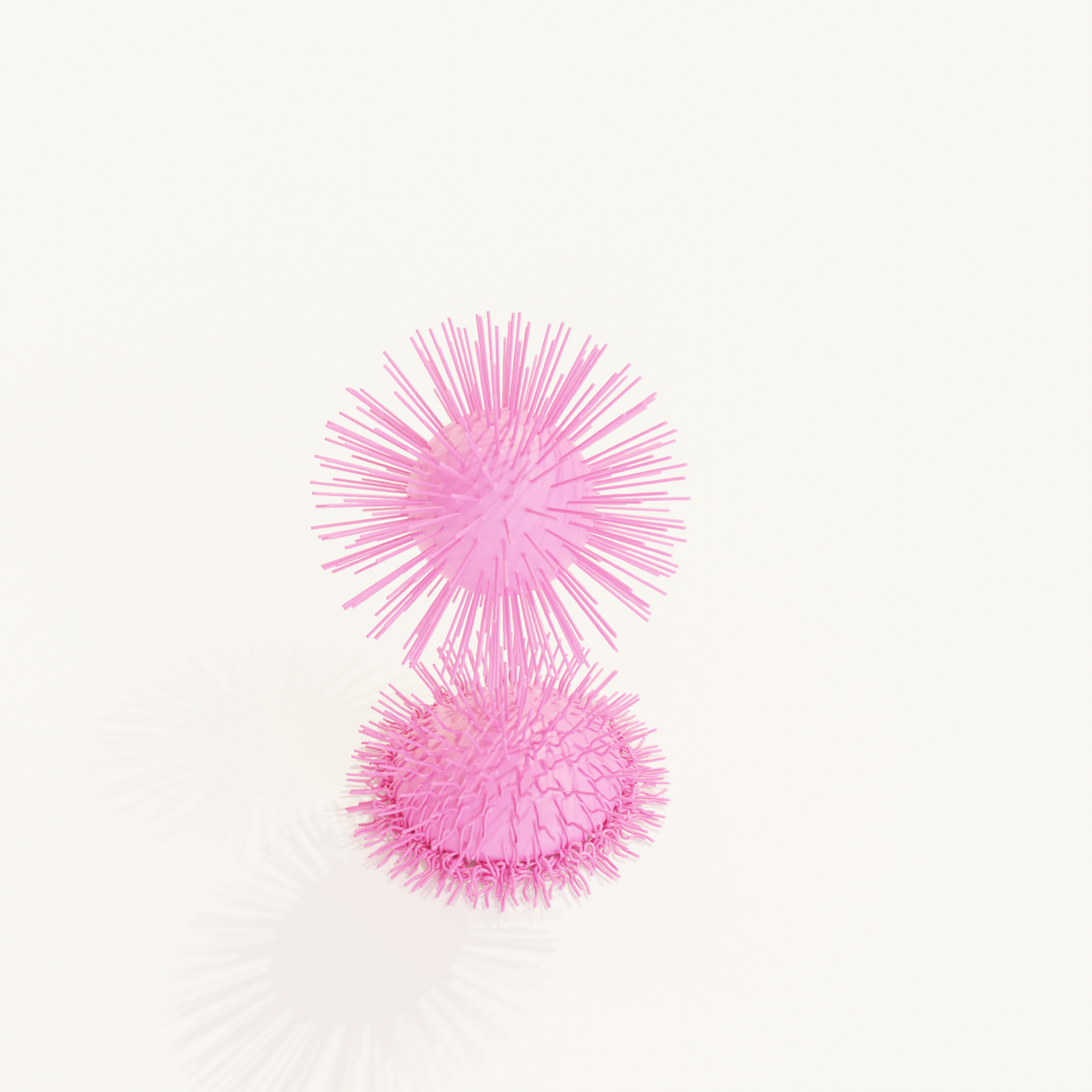}
		\end{subfigure}
		\begin{subfigure}[b]{0.22\textwidth}
			\centering
			\includegraphics[width=\columnwidth, trim=250 000 350 550, clip]{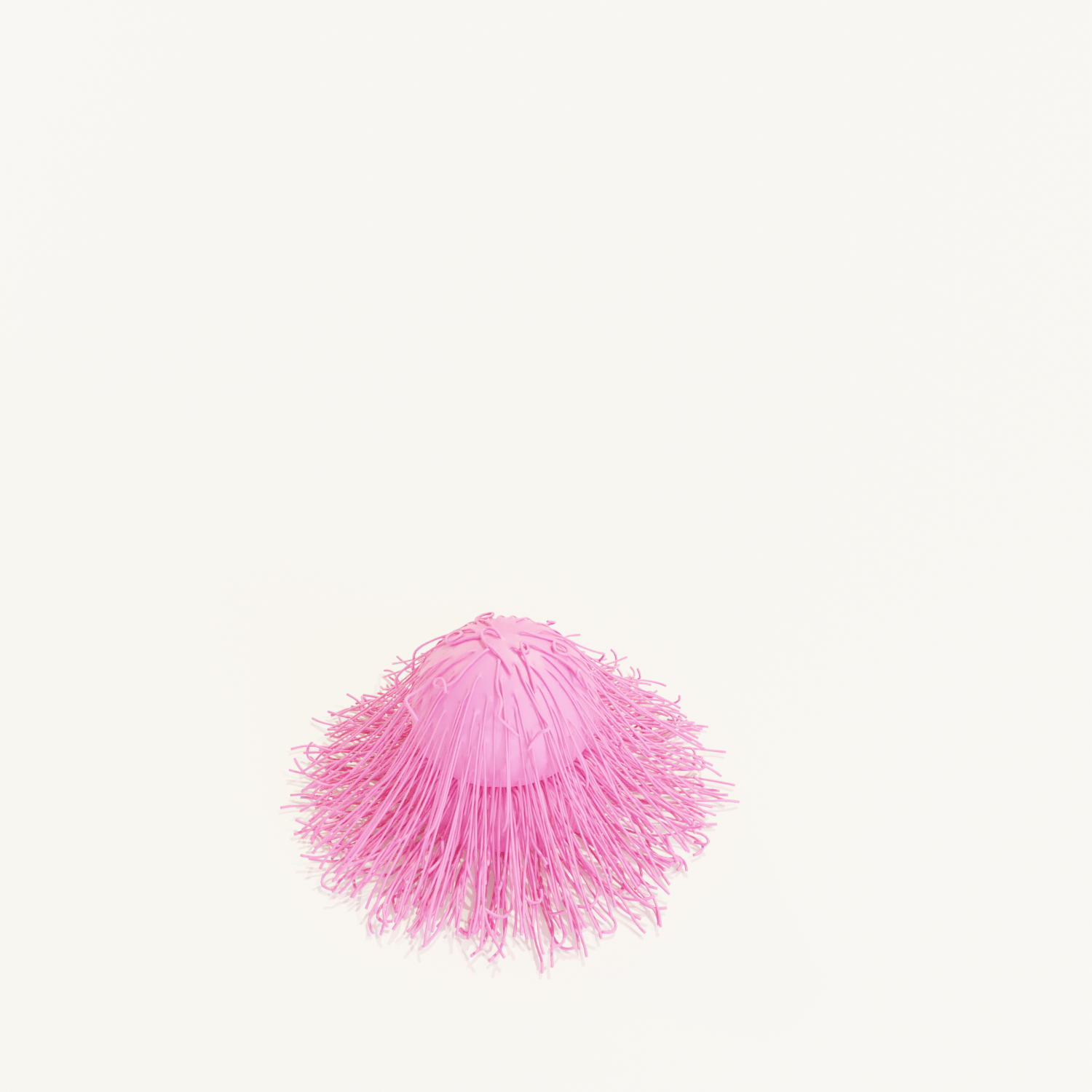}
		\end{subfigure}
		\caption{\label{fig:balls-dropping}Furry balls: Two soft spheres embellished with protruding soft spikes are dropped to the ground, resulting in intricate collisions between their strands with deformation.
		}
	\end{figure}
	
	\begin{figure}[htbp]
		\centering
		
		\begin{subfigure}[b]{0.2\textwidth}
			\centering
			\includegraphics[width=\columnwidth, trim=0 50 200 200, clip]{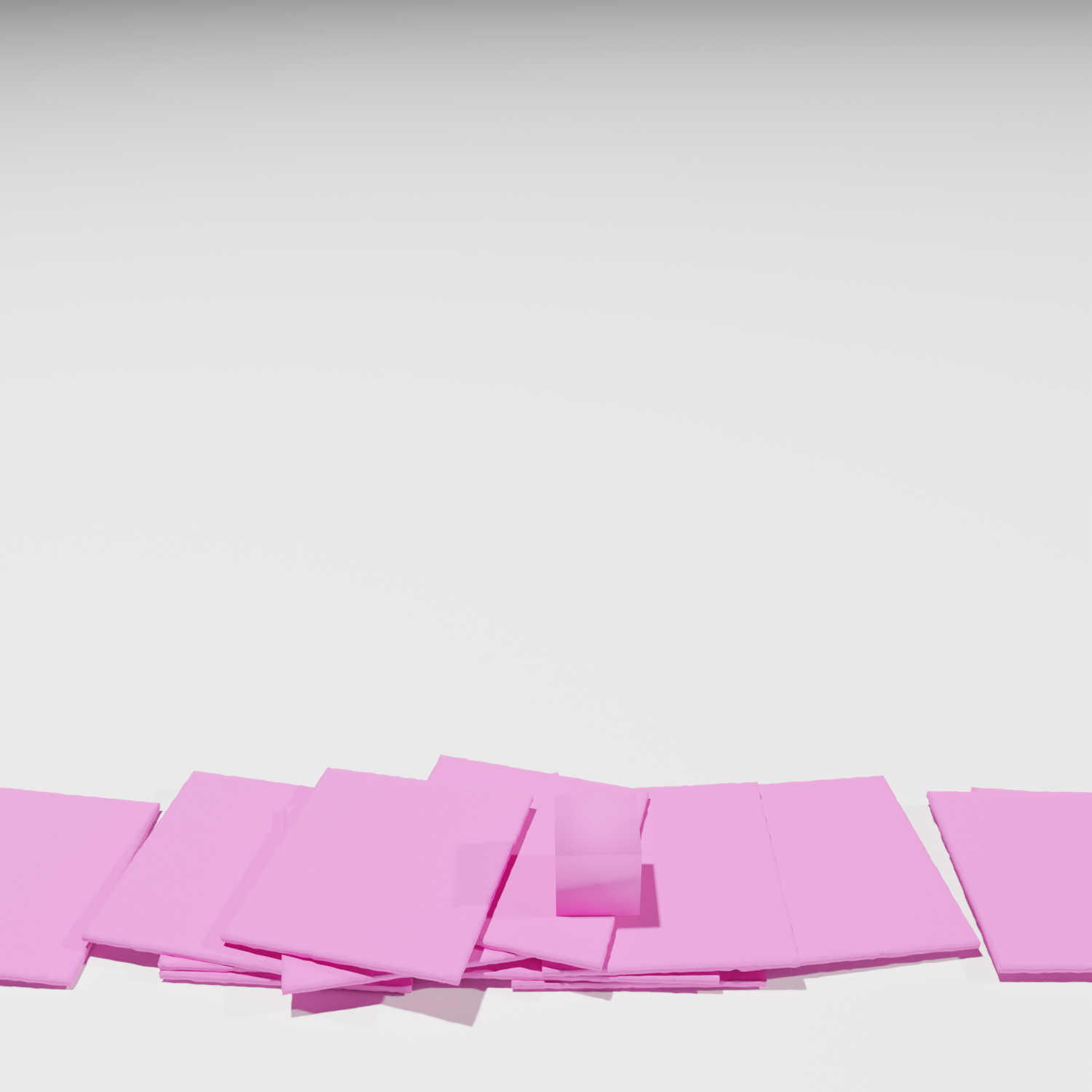}
		\end{subfigure}
		\begin{subfigure}[b]{0.2\textwidth}
			\centering
			\includegraphics[width=\columnwidth, trim=0 50 200 200, clip]{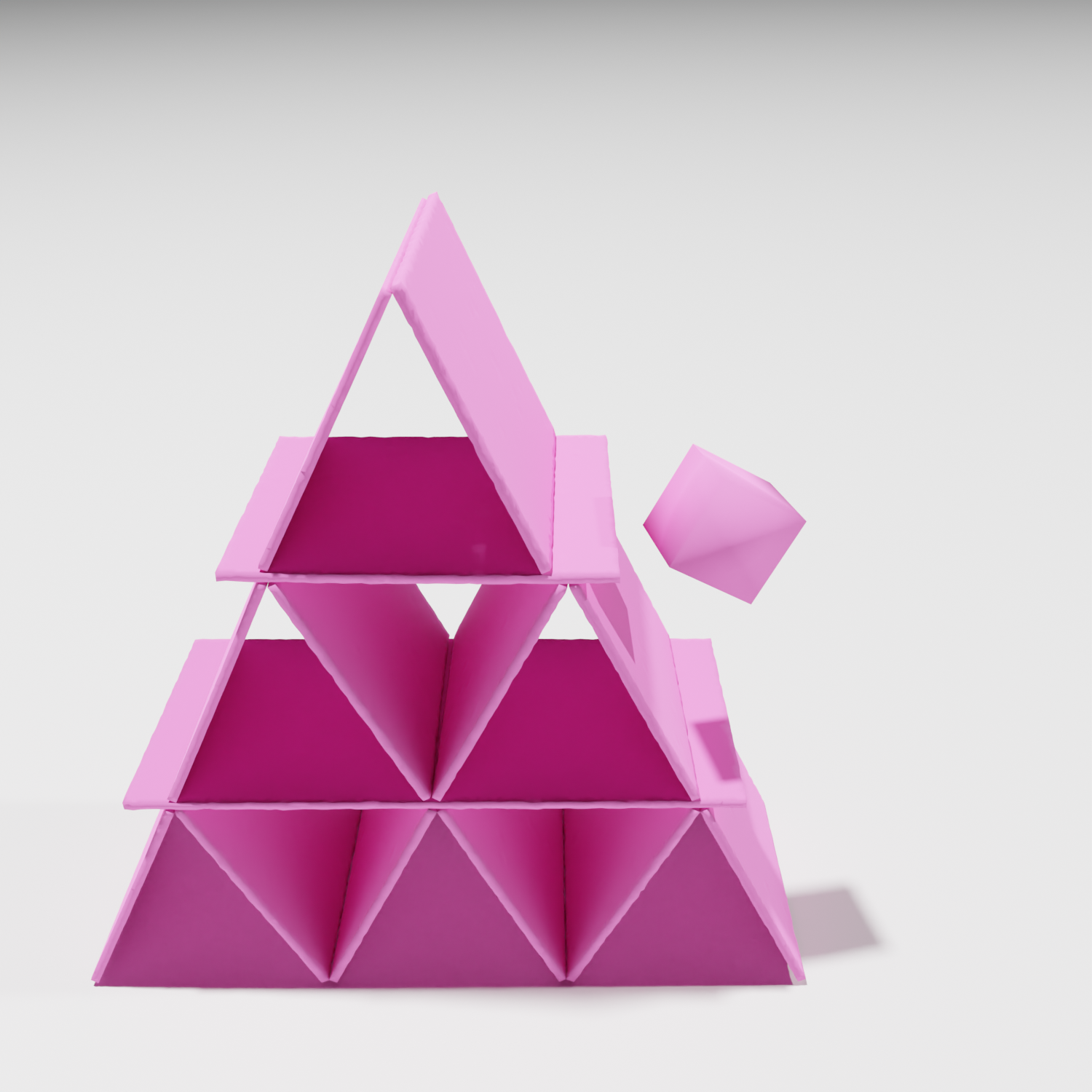}
		\end{subfigure}
		
		\caption{\label{fig:house}Stiff card house: we simulate a card house with stiff boards (E = 0.1GPa) and impact the house with a dropping block. Left: the house is fail to keep its shape as we omit the friction between boards. Right: we keep the house shape with frictional contact between the boards.}
	\end{figure}

	\begin{figure}[htbp]
		\centering
		\begin{subfigure}[b]{0.22\textwidth}
			\centering
			\includegraphics[width=\columnwidth, trim=350 0 180 500, clip]{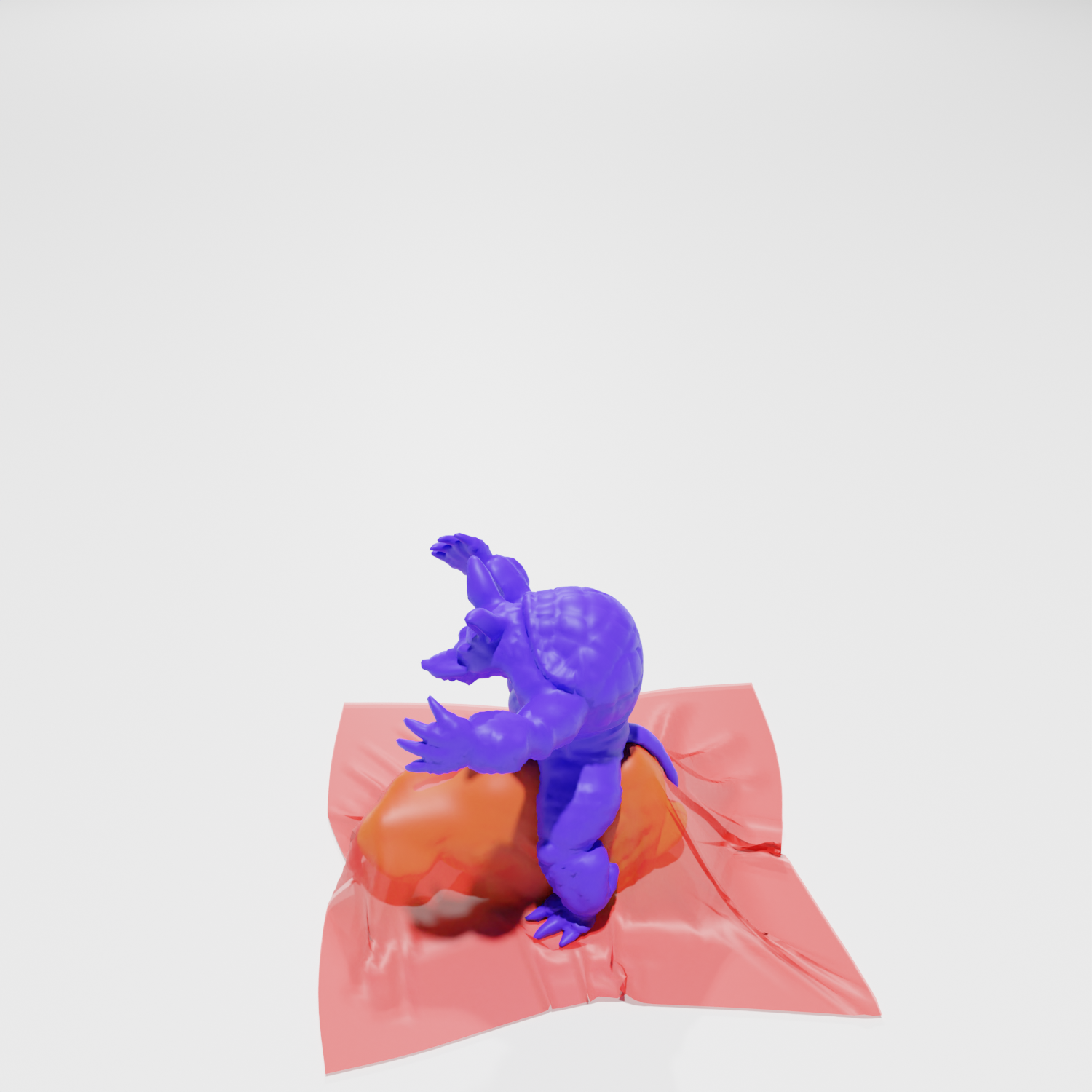}
		\end{subfigure}
		\begin{subfigure}[b]{0.22\textwidth}
			\centering
			\includegraphics[width=\columnwidth, trim=350 0 180 500, clip]{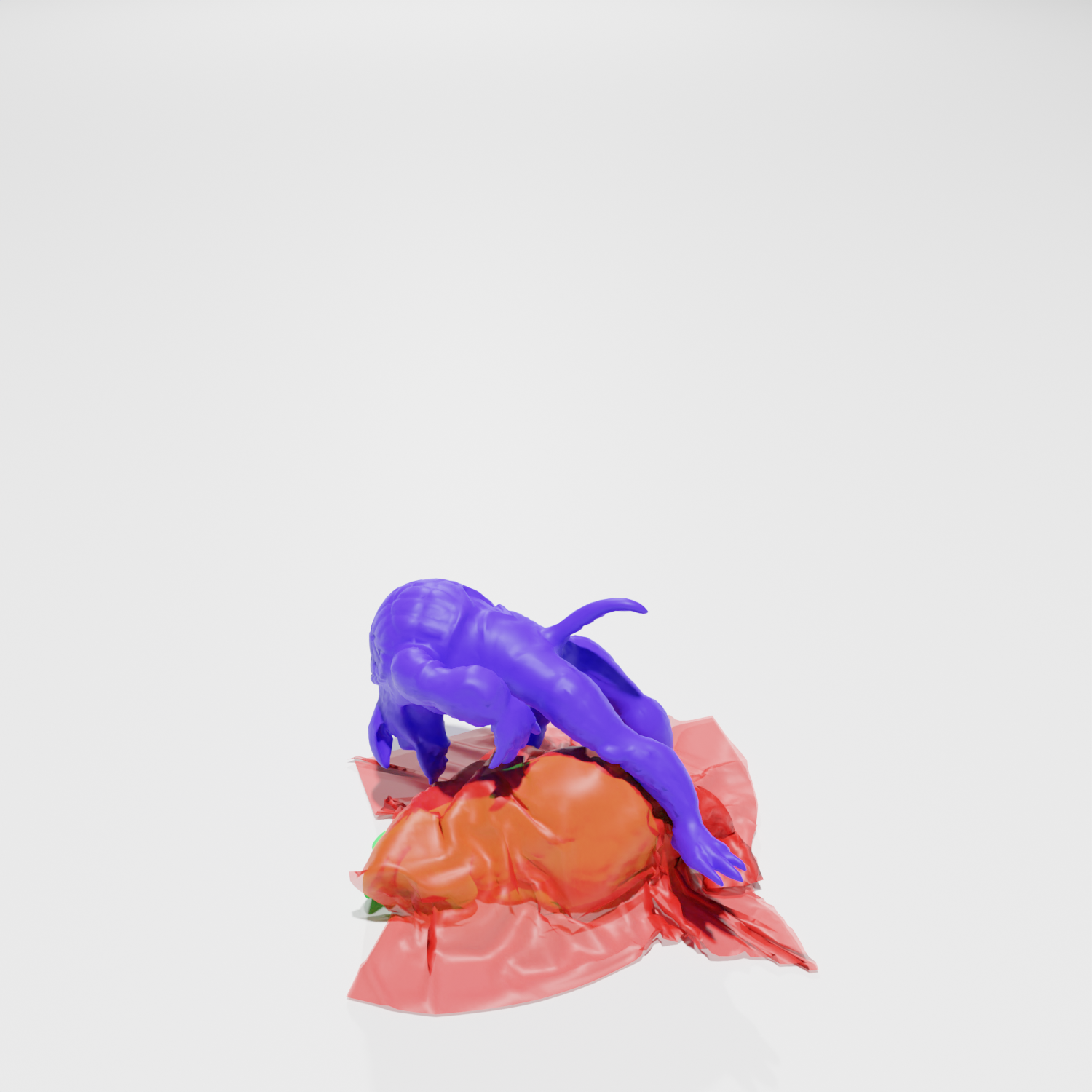}
		\end{subfigure}
		\caption{\label{fig:bunny-cloth}Here we show the ability to likewise simulate multiple objects of different codimension simultaneously with frictional contact.}
	\end{figure}
	
	\begin{table*}[t]
		\centering
		\resizebox{\textwidth}{!}{
			\begin{tabular}{r|ccccc|cccccccccccccc|c}
				\toprule
				& \texttt{v},\texttt{t},\texttt{f}  & $\rho$,$E$,$\upsilon$ & $\hat{d}$, $\varepsilon_d$ & $\mu$, $\epsilon_{v}$ & $\Delta{t}$, \#$\Delta{t}$& \multicolumn{2}{c}{\texttt{buildCP}} &  \multicolumn{2}{c}{\texttt{buildGH}} & \multicolumn{2}{c}{\texttt{solve}} & \multicolumn{2}{c}{\texttt{CCD}}   & \multicolumn{2}{c}{\texttt{\#i}} &   \multicolumn{2}{c}{\texttt{misc}} &\multicolumn{2}{c}{\texttt{timeTot}} & \textbf{speedup} \\ 
				& & & & & & cpu&gpu & cpu&gpu & cpu&gpu & cpu&gpu & cpu&gpu & cpu&gpu & cpu&gpu & (\textit{gpu vs. cpu}) \\
				\midrule
				\figref{cloth-sim} & 80k, -, 160k  & 1e2,1e2,0.49 & 3e-4l, 4e-2l & 0.4, 1e-3l  & 0.01, 2678 & - & 9.96e2 & - & 2.70e2 & - & 4.78e3 & - & 3.02e2 & - & 25.1 & - & 44 & - & 6.40e3 & - \\ 
				\rowcolor{LightCyan}
				\figref{cloth-twisting} & 45k, 133k, 89.9k  & 1e3,1e4,0.48 & 1e-3l, 1e-2l & -  & 0.01, 5k & 2.63e4 & 6.31e2 & 4.91e3 & 3.79e2 & 7.23e4 & 1.12e3 & 2.08e4 & 1.19e2 & 8.33 & 8.48 & 6.27e3 & 131 & 1.31e5 & 2.38e3 & \textbf{55.0}$\times$ \\ 
				\figref{rods-twisting} & 53.3k, 202k, 79.8k  & 1e3,1e4,0.49 &1e-3l, 1e-2l & -  & 0.01, 5k & 1.21e4 & 1.62e2 & 3.55e3 & 2.78e2 & 4.16e4 & 1.18e3 & 9.90e2 & 3.35e1 & 4.27 & 4.39  & 5.83e3& 106 & 7.30e4 & 1.76e3 & \textbf{41.5}$\times$ \\
				\rowcolor{LightCyan}
				\figref{bunnies-dropping} & 518k, 2158k, 562k  & 1e3,1e4,0.49 & 3e-4l, 1e-2l & 0.2, 1e-3l  & 0.01, 224 & 1.17e5 & 6.70e2 & 2.75e4 & 1.56e3 & 3.75e5 & 3.68e3 & 5.77e4 & 1.01e2 & 79.4 & 63.2 & 1.44e3& 39.7& 5.79e5& 6.05e3& \textbf{95.7}$\times$ \\ 
				\figref{balls-dropping} & 123k, 330k, 211k  & 1e3,1e4,0.49 & 4e-4l, 2e-2l & -  & 0.01, 200 & 2.66e4 & 3.94e2 & 4.16e3 & 3.93e2 & 5.98e4 & 1.25e3 & 2.02e4 & 8.97e1 & 89.9 & 85.9 & 1.95e2 & 3.30 & 1.11e5 & 2.13e3 & \textbf{52.1}$\times$ \\
				\rowcolor{LightCyan}
				\figref{bunny-cloth} & 36k, 135k, 46k  & \makecell{1e3,1e4,0.49(body)\\1e2,1e2,0.49(cloth)} & 3e-4l, 1e-2l & 0.5, 1e-3l  & 0.01, 350 & 6.65e3 & 1.10e2 & 1.63e3 & 1.27e2 & 1.83e4 & 2.67e2 & 2.83e3 & 1.89e1 & 48.1 & 43.2 & 1.15e2 & 3.86 & 2.96e4 & 5.28e2 & \textbf{56.1}$\times$ \\

				\figref{large_stiffness} & \makecell{32k, 135k, 38k}  &  \makecell{1e3,1e4,0.49\\1e3,1e5,0.49\\1e3,1e6,0.49\\1e3,1e7,0.49}&1e-3l, 1e-2l & - & \makecell{0.01, 95\\0.01, 80\\0.01, 100\\0.01, 81} & \makecell{6.33e2\\4.09e2\\4.66e2\\4.62e2} & \makecell{8.76\\3.95\\3.70\\2.07} & \makecell{2.97e2\\2.05e2\\2.39e2\\2.33e2} & \makecell{2.28e1\\1.25e1\\1.22e1\\6.97} & \makecell{3.39e3\\2.32e3\\2.74e3\\2.62e3} & \makecell{3.60e1\\3.92e1\\7.63e1\\8.49e1} & \makecell{5.41e2\\3.58e2\\4.04e2\\3.94e2} & \makecell{2.86\\1.35\\1.24\\0.73} & \makecell{33.2\\26.5\\24.8\\29.8} & \makecell{25.5\\17.0\\13.5\\9.61} & \makecell{1.93e1\\1.81e1\\2.14e1\\3.11e1} & \makecell{0.68\\0.40\\0.66\\0.43} & \makecell{4.88e3\\3.31e3\\3.87e3\\3.74e3} & \makecell{7.11e1\\5.74e1\\9.41e1\\9.51e1} & \makecell{\textbf{68.6}$\times$\\\textbf{57.7}$\times$\\\textbf{41.1}$\times$\\\textbf{39.3}$\times$} \\
				
				\rowcolor{LightCyan}
				\figref{house} & 77k, 290k, 108k  & 1e3,1e8,0.40 & 1e-4l, 1e-2l & 1.0, 1e-3l  & 0.01, 250 & 6.74e2 & 5.58 & 2.28e2  & 1.67e1 & 1.98e3 & 1.15e2 & 4.35e2 & 2.08 & 4.78 & 3.97 & 5.30e1 & 4.64 & 3.37e3 & 1.44e2 & \textbf{23.4}$\times$ \\
				
				\figref{medialIPCCP} & 17k, 56k, 34k  & 2,1e3,0.35 & 1e-3l, 1e-2l & -  & 0.02, 200 & 2.20e2 & 3.90 & 48.1 & 6.37 & 3.73e2 & 5.96e1 & 2.03e2 & 2.94 & 7.61 & 7.60 & 1.59e1& 3.29& 8.60e2& 7.61e1& \textbf{11.3}$\times$ \\

				\rowcolor{LightCyan}
				\figref{funnel} & \makecell{8k, 36k, 10k(dolphin)\\30k, -, 60k(funnel)}  & 1e3,1e4,0.40&1e-3l, 1e-2l & - & \makecell{0.01, 300\\0.02, 150\\0.03, 100\\0.04, 75\\0.05, 60} & \makecell{2.64e3\\1.62e3\\1.24e3\\9.69e2\\8.86e2} & \makecell{7.91e1\\4.27e1\\3.67e1\\2.82e1\\2.85e1} & \makecell{5.47e2\\4.27e2\\3.62e2\\3.28e2\\3.13e2} & \makecell{2.94e1\\2.23e1\\2.14e1\\1.75e1\\1.49e1} & \makecell{3.46e3\\2.55e3\\2.13e3\\1.72e3\\1.59e3} & \makecell{6.71e1\\5.31e1\\4.88e1\\4.30e1\\4.56e1} & \makecell{2.06e3\\1.36e3\\1.09e3\\8.93e2\\8.28e2} & \makecell{1.79e1\\1.25e1\\1.17e1\\1.01e1\\1.05e1} & \makecell{22.7\\29.6\\36.8\\38.4\\43.6} & \makecell{20.9\\23.4\\29.6\\30.5\\38.3} & \makecell{1.84e2\\1.03e2\\6.82e1\\5.38e1\\4.47e1} & \makecell{4.51\\3.02\\2.31\\1.83\\1.52} & \makecell{8.89e3\\6.06e3\\4.89e3\\3.97e3\\3.66e3} & \makecell{1.98e2\\1.33e2\\1.21e2\\1.01e2\\1.01e2} & \makecell{\textbf{44.9}$\times$\\\textbf{45.3}$\times$\\\textbf{40.4}$\times$\\\textbf{39.4}$\times$\\\textbf{36.2}$\times$} \\

				\bottomrule
		\end{tabular}}
		\caption{\label{tab:stats}Performance summary using comparison \citet{10.1145/3386569.3392425}. The columns are as follows: number of vertices, including the interior for tetrahedral meshes (\texttt{v}); number of tetrahedra (\texttt{t}); number of surface triangles (\texttt{f}); time step size in seconds ($\Delta{t}$); material density($\rho$), Young's modulus ($E$) in units of pascals Pa, and Poisson's ratio ($\upsilon$); computational accuracy target in meters ($\hat{d}$) which is set w.r.t. to the scene bounding box diagonal length l; Newton Solver tolerance threshold ($\epsilon_d$); friction coefficient ($\mu$) and velocity magnitude bound ($\epsilon_{v}$);Total number of time steps (\#$\Delta{t}$); Total time to build/find contact pairs (\texttt{buildCP}); Total time to build energy gradients and Hessians for all types (\texttt{buildGH}); Total linear solver time (\texttt{solve}); Total CCD time (\texttt{CCD});  Average number of Newton iterations per time step (\texttt{\#i}); Total time for remaining miscellaneous tasks (\texttt{misc}); Total simulation compute time (\texttt{timeTot}); Estimated speedup of GPU implementation (ours) versus CPU (\citet{10.1145/3386569.3392425}). All time measurements are presented in seconds. 
		}
	\end{table*}
	
	\section{Conclusion and Discussion}
	\label{sec:conclusion}
	We have presented a new  barrier function that requires minimal effort to ensure positive semi-definiteness of its approximate Hessian for efficient Gauss-Newton optimization of IPC.
	Proximal contact is represented as compressive distortion of a simplex, giving rise to the {constraint Jacobian} from the positions of vertices in the stencil of a contact pair. 
	This constraint Jacobian is the source of our novel derivation and analysis of the barrier function, enabling closed-form expressions for the eigendecomposition of the approximate Hessian. 
	With the exception of the near-parallel \textit{edge-edge} case, all configurations of contact are found to have the same expressions (three) for their eigensystem. One eigenvalue is always positive (and others are negative) which we use to PSD project the approximate Hessian and define a filtering strategy for improving Gauss-Newton convergence rates.
	
	{
		As an `oriented-simplex' method, our approach is inspired by the idea of constructing deformation gradients \cite{10.1145/2343483.2343501} with matrices composed of vectors along the edges of a simplex, then inverting one of the matrices and multiplying. 
		This idea is remarkably powerful, enabling stress-like measurement of contact where the gradient of the barrier potential w.r.t the Jacobian $\partial b / \partial \mbJ$ is the linear operator for mapping a normal vector $\mbn$ in the local space (of a contact) to the force vector $\mbf_{c} = -\partial b/\partial \mbx$ in world space across the surface perpendicular to this $\mbn$. 
		
		Our constraint Jacobian is related but differs from the deformation gradient in that it is always diagonal, which permits efficient and elegant evaluations of the barrier energy. 
	}
	The resulting construction is free of explicit assemblage (\ie~without actual use of vectors along simplex edges, matrix inversions nor multiplications) and naturally extends to our mollification of the barrier function in situations of degenerate near-parallel edges.
	
	\paragraph{Limitations and future work}
	{Despite integrating the state-of-the-art MAS preconditioner ~\cite{wu2022gpu} into our framework, our approach can still exhibit reduced speedup as the stiffness of the problem increases (see also \tabref{stats}, rows of \figref{large_stiffness}, \figref{funnel} and \figref{house}). Hence, it remains beneficial to explore more efficient preconditioners in future work for further improvement of the convergence rate.}

	Our method still uses the friction potential of \citet{10.1145/3386569.3392425} with a lagged sliding basis and magnitude of the normal force {(refer to our technical supplement for more details)}.
	This implies that we suffer from the same limitations that affect accuracy. 
	Specifically, we find that friction-forces are not guaranteed to lie within the local contact plane, which implies that their normal component tends to influence the net barrier force acting in this direction to affect convergence.
	This can be remedied by using a larger threshold for the friction mollifier (see \S 5.3 in \cite{10.1145/3386569.3392425}) but this is done to the detriment of static friction accuracy, as larger thresholds induce viscous drag when objects (\ie~contact primitives) should be at standstill.
	These problematic normal friction-force components can likewise be ameliorated by reducing the respective mollifier threshold but this will deteriorate the solver convergence rate. For these reasons, having a friction model that that guarantees strictly in-plane force components with moderate thresholds remains an open problem, which will be meaningful to address in future works.

	\begin{acks}
		This work was partially funded by the Research Grant Council of Hong Kong (GRF 17210222). This work was also supported by the Innovation and Technology Commission of the HKSAR Government under the InnoHK initiative (TransGP project),
		and the JC STEM Lab of Robotics for Soft Materials funded by The Hong Kong Jockey Club Charities Trust.
		Finally, we thank the reviewers for their detailed and insightful feedback.
		
	\end{acks}

	\bibliographystyle{ACM-Reference-Format}
	\bibliography{bibliography}
	
	\appendix
	
	\begin{figure}[t]
		\centering
		\includegraphics[width=0.6\columnwidth]{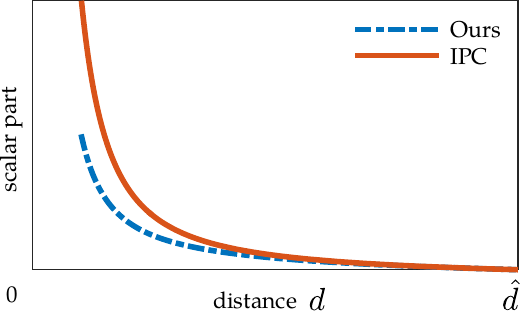}
		\caption{\label{fig:gauss_compare}Comparison with the Gauss-Newton method discussed in \cite{10.1145/3386569.3392425}. We plot the scalar part of the approximate Hessian as a function of distance $d$.}
	\end{figure}
	\section{Gauss-Newton approximation comparison}
	\label{app:gn-appr-comp}
	This section provides a simplified demonstration to show the difference between our Gauss-Newton approximation of the barrier Hessian (\cf~\secref{bar-fn-and-derivs}) and the method discussed in Section 4.3 of \cite{10.1145/3386569.3392425}.
	We will use the \textit{point-point} test case for the comparison. 
	
	Let the determinant (\cf~\eqref{sigma-sum}) be $f = \sqrt{d}$, for the purpose of matching the Euclidean distance based formulation of \citet{10.1145/3386569.3392425}. This implies that we have (\cf~\eqref{reduced-F}) $\mbJ = \begin{bmatrix} \sqrt{d}\end{bmatrix} \in \realNum^{1\times1}$ from which we can evaluate our gap function as $g = \bar{g} = d$ that we will then substitute into the smoothly-clamped barrier function (\cf~\eqref{ipc-barrier-orig}) of \citet{10.1145/3386569.3392425} to align our method with this original formulation for comparison. 
	
	Revisiting our approximation of the force Jacobian, we have  
	\begin{align}\label{eq:gauss1}
		\frac{\partial^2 b}{\partial \mbx^2}  &=  {\frac{\partial \mbJ}{\partial \mbx}}: \frac{\partial^2 b}{\partial \mbJ^2} :\frac{\partial \mbJ}{\partial \mbx} + \cancel{\frac{\partial^2 \mbJ}{\partial \mbx^2}\colon \frac{\partial b}{\partial \mbJ}},\nonumber\\
		&={\frac{\partial \sqrt{d}}{\partial \mbx}}:\frac{\partial^2 b}{\partial \mbJ^2}:{\frac{\partial \sqrt{d}}{\partial \mbx}}\nonumber\\
		&={\frac{1}{2\sqrt{d}}\frac{\partial d}{\partial \mbx}} :\frac{\partial^2 b}{\partial \mbJ^2}:{\frac{1}{2\sqrt{d}}\frac{\partial d}{\partial \mbx}},\nonumber\\
		&={\frac{\partial d}{\partial \mbx}} :\left(\frac{1}{4d} \frac{\partial^2 b}{\partial \mbJ^2}\right):{\frac{\partial d}{\partial \mbx}},
	\end{align}
	where the barrier Hessian is given by (\cf~ \eqref{psd-gen-hess}) $$\frac{\partial^2 b}{\partial \mbJ^2} = \left(4g\frac{\partial^2 b}{\partial {g}^2}+2\frac{\partial b}{\partial {g}}\right)\mbQ_1 \otimes \mbQ_1$$ according to our analytic eigensystems. 
	
	Since the eigenmatrix $\|\mbQ_1\|_F^2 = 1$ is normalized, we arrive at the form
	\begin{align}\label{eq:gauss2}
		\frac{\partial^2 b}{\partial \mbx^2}  &=
		{\frac{\partial d}{\partial \mbx}} :\left(\frac{1}{4d} \frac{\partial^2 b}{\partial \mbJ^2}\right):{\frac{\partial d}{\partial \mbx}}\nonumber\\
		&= \frac{1}{4d}\left(4g\frac{\partial^2 b}{\partial {g}^2}+2\frac{\partial b}{\partial {g}}\right){\frac{\partial d}{\partial \mbx}}\otimes\frac{\partial d}{\partial \mbx} \nonumber\\
		&=\frac{1}{4d}\left(4d\frac{\partial^2 b}{\partial {d}^2}+2\frac{\partial b}{\partial {d}}\right){\frac{\partial d}{\partial \mbx}}\otimes \frac{\partial d}{\partial \mbx}\nonumber\\
		&=\left(\frac{\partial^2 b}{\partial {d}^2}+\frac{1}{2d}\frac{\partial b}{\partial {d}}\right)\nabla_{\mbx} d  \otimes\nabla_{\mbx} d,
	\end{align}
	which resembles the first term in \eqref{bar-hess-orig}. 
	The primary difference is that \citet{10.1145/3386569.3392425}'s proposed Gauss-Newton approximation will only use ${\partial^2 b}/{\partial {d}^2}$ as the scalar part, whereas we use $$\frac{\partial^2 b}{\partial {d}^2}+\frac{1}{2d}\frac{\partial b}{\partial {d}}.$$ A visual comparison is also provided in \figref{gauss_compare}, which compares the two expressions to reveal that our approximated Hessian produces smaller values than the truncated Hessian proposed in \cite{10.1145/3386569.3392425}. The effect is that our method will produce search directions $\mbd=({{\partial^2 b}/{\partial \mbx^2}})^{-1}{{\partial b}/{\partial \mbx}}$ that have larger magnitude. 
	We found that the truncated Hessian of \citet{10.1145/3386569.3392425} is effective when simulating soft materials with a relatively large threshold in the iterative solver. 
	However, increasing stiffness (or encountering nearly-parallel \textit{edge-edge} cases) will significantly deteriorate the convergence of this method with potential for failure. 
	In contrast, our proposed analytically projected Gauss-Newton method performs well even under these conditions.
	
	\section{The norms of $\frac{\partial b}{\partial \mbx}$ and $\frac{\partial^2 b}{\partial \mbx^2}$}
	\label{app:magnitude_evaluation}
	This section is a supplement to \figref{gradient-mag-diff}, where we present derivations showing the scalar expressions for $\|{\partial b}/{\partial \mbx}\|_2$ and $\|{\partial^2 b}/{\partial \mbx^2}\|_F$ that are used to plot the lines shown. 
	
	\paragraph{Expression of the barrier gradient norm} A closed-form expression for the magnitude of the barrier gradient can be obtained directly from \eqref{barrier-contact-force}, which we rewrite here in vectorized form as
	\begin{equation}\label{eq:vec-grad}
	\frac{\partial b}{\partial \mathbf{x}} = \frac{\partial b}{\partial g}\vecOp\left(\frac{\partial \mathbf{J}}{\partial \mathbf{x}}\right)\vecOp\left(\frac{\partial g}{\partial \mathbf{J}}\right),
	\end{equation}
	where the expression for ${\partial g}/{\partial \mathbf{J}}$ is given in \eqref{dg-dJ}. Only the terms ${\partial b}/{\partial g}$  and ${\partial g}/{\partial \mathbf{J}}$ describe the sought magnitude since ${\partial \mathbf{J}}/{\partial \mathbf{x}}$ is just a change-of-basis tensor.
	
	Using the expression for the eigenmatrix $\mathbf{Q}_1$ in \eqref{q1}, we can rewrite ${\partial g}/{\partial \mathbf{J}}$ as
	\begin{equation}\label{eq:new-dg-dJ}
	\frac{\partial g}{\partial \mathbf{J}} = 2\sqrt{g}\mathbf{Q}_1.
	\end{equation}
	Substituting \eqref{new-dg-dJ} into \eqref{vec-grad}, we have
	\begin{equation}
	\frac{\partial b}{\partial \mathbf{x}} = 2\sqrt{g}\frac{\partial b}{\partial g}\vecOp\left(\frac{\partial \mathbf{J}}{\partial \mathbf{x}}\right)\vecOp\left(\mathbf{Q}_1\right),
	\end{equation}
	to give $-2\sqrt{g}\frac{\partial b}{\partial g} = \|{\partial b}/{\partial \mbx}\|_2$ as the expression for the magnitude of the barrier gradient since the eigenmatrix $\|\mbQ_1\|_F^2 = 1$ is normalized. 
	
	\paragraph{Expression of the barrier force-Jacobian norm} 
	The norm $\|{\partial^2 b}/{\partial \mbx^2}\|_F$ of our approximate barrier force-Jacobian  is the primary eigenvalue (\eqref{lambda1}) of the barrier Hessian (\eqref{4th-ord-tens}).
	To show this, we use \eqref{4th-ord-tens-vec} by
	\begin{align}\label{eq:gauss3}
		\frac{\partial^2 b}{\partial \mathbf{x}^2} &= \vecOp\left(\frac{\partial \mathbf{J}}{\partial \mathbf{x}}\right) \vecOp\left(\frac{\partial^2 b}{\partial \mathbf{J}^2}\right) \vecOp\left(\frac{\partial \mathbf{J}}{\partial \mathbf{x}}\right)^T,\nonumber\\
		&=\vecOp\left(\frac{\partial \mathbf{J}}{\partial \mathbf{x}}\right) \left(\left(4g\frac{\partial^2 b}{\partial {g}^2}+2\frac{\partial b}{\partial {g}}\right)\vecOp\left(\mathbf{Q}_1\right)\vecOp\left(\mathbf{Q}_1\right)^T\right) \vecOp\left(\frac{\partial \mathbf{J}}{\partial \mathbf{x}}\right)^T,\nonumber\\
		&=\left(4g\frac{\partial^2 b}{\partial {g}^2}+2\frac{\partial b}{\partial {g}}\right)\vecOp\left(\frac{\partial \mathbf{J}}{\partial \mathbf{x}}\right) \vecOp\left(\mathbf{Q}_1\right)\vecOp\left(\mathbf{Q}_1\right)^T \vecOp\left(\frac{\partial \mathbf{J}}{\partial \mathbf{x}}\right)^T,
	\end{align}
	to give $4g\frac{\partial^2 b}{\partial {g}^2}+2\frac{\partial b}{\partial {g}} = \|{\partial^2 b}/{\partial \mbx^2}\|_F$ as the expression for the norm of our approximate force-Jacobian.
	
\end{document}


	
	\title{Technical Supplement to ``GIPC: Fast and stable Gauss-Newton optimization of IPC barrier energy''}
	
	\author{Kemeng Huang}
	\orcid{0000-0001-9147-2289}
	\email{kmhuang@connect.hku.hk}
	\email{kmhuang819@gmail.com}
	\affiliation{%
		\institution{The University of Hong Kong, TransGP}
		\country{Hong Kong}
	}

	\author{Floyd~M.~Chitalu}
	\authornote{Currently unaffiliated. Was at The University of Hong Kong while contributing to this work.}
	\orcid{0000-0001-9489-8592}
	\email{floyd.m.chitalu@gmail.com}
	\affiliation{%
		\institution{The University of Hong Kong}
		\country{Hong Kong}
	}
	
	\author{Huancheng Lin}
	\orcid{0000-0003-4446-1442}
	\email{lamws@connect.hku.hk}
	\affiliation{%
		\institution{TransGP, The University of Hong Kong}
		\country{Hong Kong}
	}
	
	\author{Taku Komura}
	\orcid{0000-0002-2729-5860}
	\email{taku@cs.hku.hk}
	\affiliation{%
		\institution{The University of Hong Kong, TransGP}
		\country{Hong Kong}}
	
	\renewcommand\shortauthors{Huang\etal}
	
	\begin{abstract}
		{This document provides supplementary details about our derivations and implementation of the main paper. \secref{cp-dg} describes the reference (explicit) constructions of the constraint Jacobian, together with the normal vector. We then use these descriptions to detail our derivations (and simplifications) of the change-of-basis tensor in \secref{cob-tensor}.
			We also prove in \secref{eigenanalysis} that the first eigenvalue (Eq. (18) in the paper) is always positive, and that remaining eigenvalues (Eq. (19) in the paper) are always negative. 
			\secref{deriv-details} describes the intermediate steps we use to arrive at the eigenpairs of our mollified barrier Hessian.
			In \secref{friction}, we provide a summary of the analysis conducted on the friction Hessian, which is followed by supplementary implementation details of our global matrix-free PCG solver in \secref{par-solver}. Supplementary unit test results are provided in \secref{unit-tests}. We also provide a comparison of our barrier method and the original formulation on the same hardware (CPU) in \secref{barrier-tests}. Further supplemental results concerning different linear solvers and the impact of friction-Hessian projection on overall performance are provided in \secref{linear-solver-tests} and \secref{friction-impact}, respectively.}
	\end{abstract}

	%
	%

	\ccsdesc[500]{Computing methodologies~Physical simulation}
	\ccsdesc[500]{Computing methodologies~Collision detection}
	\ccsdesc[500]{Computing methodologies~Massively parallel algorithms}
	
	%
	%

	\keywords{IPC, Barrier Hessian, Eigen Analysis, 
		GPU}

	\maketitle

	\section{Explicitly Constructing Jacobians}\label{sec:cp-dg}
	
	\begin{figure*}[t]
		\centering
		\includegraphics[width=\textwidth]{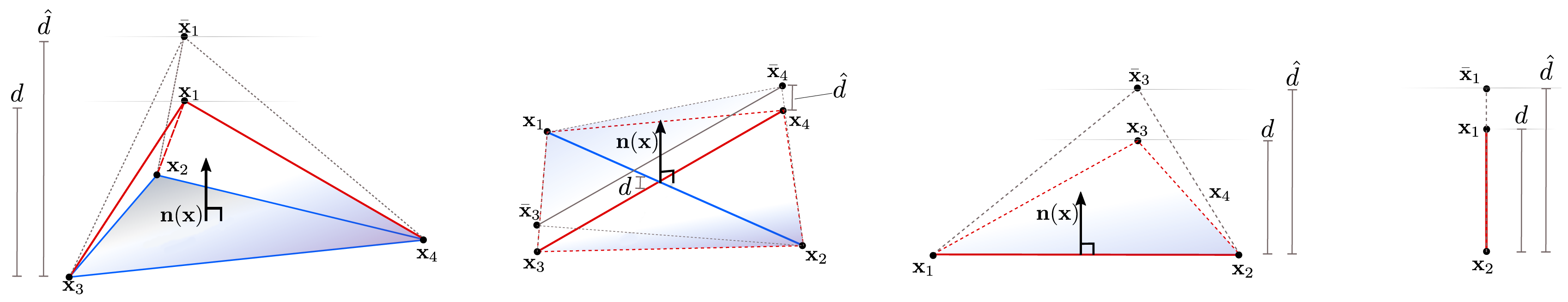}
		\caption{\label{fig:contact-pair-dg}The primitive contact pairs that are tested for intersection. From left to right, we have \textit{point-triangle} (tetrahedron); \textit{edge-edge} (tetrahedron); \textit{point-edge} (triangle); and \textit{point-point} (line-segment). In each case, the vertices can be seen as forming a simplex with which we formulate an impermeable barrier energy using a Jacobian (matrix) constructed from these vertices.}
	\end{figure*}
	
	In this section, we describe the reference method for explicitly computing the constraint Jacobian $\mbJ(\mbx, \hat{d})$ 
	and the normal vector $\mbn(\mbx)$ for a given contact pair (see also \figref{contact-pair-dg}).
	Several of the constructions we will describe can be found throughout literature (see \eg~\cite{muller2015air,10.1145/2343483.2343501,KANE19991}) but we provide them here for self-containment.
	
	Vertices of a contact pair form a simplex, with which we measure distance $d$ via a matrix $\mbF(\mbx, \hat{d}) = \mbE(\mbx)\bar{\mbE}(\mbx, \hat{d})^{-1}$, where the terms defining this $\mbF$ denote the ideal- $\bar{\mbE}$ and current-shape $\mbE$ matrices of the simplex, respectively.
	{In what follows, we will show how to compute $\mbE$ and $\bar{\mbE}$ for each possible simplex/contact-pair. 
	}
	
	\subsection{Point-triangle} The \textit{point-triangle} case represents a tetrahedron, where the variables constituting the matrix $\mbF$ are
	\begin{align*}
	\mbE &= \begin{bmatrix} \mbx_2-\mbx_1|\mbx_3-\mbx_1|\mbx_4-\mbx_1\end{bmatrix}\in \realNum^{3\times3}\nonumber\\
	\bar{\mbE} &= \begin{bmatrix} \bar{\mbx}_2-\bar{\mbx}_1|\bar{\mbx}_3-\bar{\mbx}_1|\bar{\mbx}_4-\bar{\mbx}_1\end{bmatrix}\in \realNum^{3\times3}.
	\end{align*}
	We construct $\bar{\mbE}$ with
	\begin{align*}
	{ \bar{\mbx}_1} ={\mbx}_1+\left(\hat{d} - d\right)\mbn,\quad
	{ \bar{\mbx}_i} ={\mbx}_{i},\quad i = 2, 3, 4
	\end{align*} 
	where $\mbn$ is the triangle normal
	\begin{equation}\label{eq:pt-normal}
	\mbn = \frac{(\mbx_{3}-\mbx_{2})\times(\mbx_{4}-\mbx_{2})}{\|(\mbx_{3}-\mbx_{2})\times(\mbx_{4}-\mbx_{2})\|}\in \realNum^{3\times 1},
	\end{equation}
	and $d = \mbv \cdot \mbn$ is the distance between the point $\mbx_1$ and triangle $\mbx_2$, $\mbx_{3}$, $\mbx_{4}$ using $\mbv = \mbx_1-\mbx_{2}$.
	
	\subsection{Edge-edge} The \textit{edge-edge} case also represents a tetrahedron, where $\mbE$ and $\bar{\mbE}$ are constructed like the \textit{point-triangle} case. However, the ideal shape positions are now
	\begin{align*}
	{ \bar{\mbx}_1} &={\mbx}_{1}, \quad { \bar{\mbx}_2} ={\mbx}_{2},\nonumber \\
	{ \bar{\mbx}_3} &={\mbx}_3+\left(\hat{d} - d\right)\mbn,\quad
	{ \bar{\mbx}_4} ={\mbx}_4+\left(\hat{d} - d\right)\mbn,
	\end{align*}
	where the normal is computed as in \eqref{pt-normal} but using the edge vectors, and with $d=\mbv\cdot \mbn$ representing the distance between these edges using $\mbv = \mbx_3-\mbx_1$.
	
	\subsection{Point-edge} The \textit{point-edge} case represents a triangle, with a non-square $\mbF$ constructed with
	\begin{align*}
	\mbE &= \begin{bmatrix} \mbx_2-\mbx_1|\mbx_3-\mbx_1\end{bmatrix}\in \realNum^{3\times2}\\
	\bar{\mbE} &= \begin{bmatrix} \bar{\mbx}_2-\bar{\mbx}_1|\bar{\mbx}_3-\bar{\mbx}_1\end{bmatrix}\in \realNum^{2\times2},
	\end{align*}
	We can compute $\bar{\mbE}$ by projecting vertex components to the plane 
	\begin{align*}
	\bar{\mbx}_1 =\mbK\mbH\left(\mbx_1+\left(\hat{d}- d\right)\mbn_e\right),\quad
	\bar{\mbx}_2 =\mbK\mbH\mbx_2,\quad
	\bar{\mbx}_3 =\mbK\mbH\mbx_3,
	\end{align*}
	where the in-plane edge normal $\mbn_e$ ($\mbn_e\perp \mbx_2\mbx_3$) is given by 
	\begin{equation*}
	\mbn_e = \frac{(\mbx_2-\mbx_3)\times \mbn_t}{\|(\mbx_1 - \mbx_2)\times \mbn_t\|}\in \realNum^{3\times 1},
	\end{equation*}
	with $\mbn_t$ denoting triangle normal computed as in \eqref{pt-normal} but using $\mbx_1$, $\mbx_2$ and $\mbx_3$. The  matrix $\mbH \in \realNum^{3\times 3}$ represents a rotation to align the triangle to a canonical axis plane, \eg~ with normal $\mbn_t = (0, 1, 0)$. Several methods exist to determine this rotation with one example being Rodrigues' formula \cite{10.5555/561828} from which we get
	\begin{equation}\label{eq:rotation}
	\mbH= \begin{cases}
	2\frac{(\mbp+\mbb)(\mbp+\mbb)^T}{(\mbp+\mbb)^T(\mbp+\mbb)}-\mbI & \mbp\ne-\mbb\\
	\begin{bmatrix}
	-1&0&0\\0&-1&0\\0&0&1
	\end{bmatrix} & \mathrm{otherwise},
	\end{cases}
	\end{equation}
	where
	$
	\mbp\coloneqq \mbn_t$, $\mbb =\begin{bmatrix} 0 & 1 & 0\end{bmatrix}^T$
	and $\mbK\in \realNum^{2\times 3}$ is a matrix (subject to our choice of \mbb) to extract the $xz$ components of a 3D vector 
	$$
	\mbK=\begin{bmatrix}
	1&0&0\\
	0&0&1
	\end{bmatrix}.
	$$
	Finally, $d=\mbv\cdot \mbn_e$ is the distance between point $\mbx_1$ and edge $\mbx_2\mbx_3$  using $\mbv = \mbx_1-\mbx_2$.
	
	\subsection{Point-Point} The \textit{point-point} case reduces to a line segment and where the Jacobian is a vector. We have 
	\begin{align}
	\mbE = \mbv\in \realNum^{3\times1} \quad \mathrm{and} \quad
	\bar{\mbE} = \begin{bmatrix} \bar{\mbx}_2-\bar{\mbx}_1 \end{bmatrix}\in \realNum, \label{eq:pp-rest-shape}
	\end{align}
	where $\mbv = \mbx_2-\mbx_1$, and 
	\begin{align*}
	\bar{\mbx}_1 =\mbK\mbH\left({\mbx}_1+\left(\hat d- d\right) \mbn\right)\in \realNum \quad \mathrm{and} \quad
	\bar{\mbx}_2 =\mbK\mbH{\mbx}_2 \in \realNum,
	\end{align*}
	where  $\mbn = \frac{\mbv}{\|\mbv\|}$ is the normal, and the rotation $\mbH$ is constructed similarly to \eqref{rotation} \ie~ as one which aligns $\mbn$ to the canonical $y$-axis of $(0,1,0)$. Assuming the same choice for the vector $\mbb$ as above, the matrix $\mbK = \begin{bmatrix}
	0 & 1 & 0
	\end{bmatrix}$ now extracts the $y$-component of a 3D vector. Thus, $d = \mbv \cdot \mbn = \|\mbn\| = \|\mbv\|$.
	
	In all cases above, we have $\mbK\mbH\mbn$ as the normal vector used to evaluate the gap function, using $\mbn_e$ for the point-edge case.
	
	\section{Change-of-basis tensor} \label{sec:cob-tensor} 
	This section outlines how we compute the change-of-basis tensor ${\frac{\partial \mbF}{\partial \mbx} } \ \in \realNum^{\mathrm{dims}({\mbF})\times{3{s}}}$ from a contact pair comprised of ${s}$ vertices. 
	This tensor can be understood as a column vector with block entries of dimensions $\mathrm{dims}({\mbF}) \equiv \realNum^{3\times{m}}$ for an $m$-dimensional simplex.
	
	\paragraph{Definition} The general expression defining the change-of-basis tensor is given by
	\begin{equation}\label{eq:cobt-def}
	\frac{\partial {\mbF}}{\partial \mbx}=\frac{\partial \mbE}{\partial \mbx}\bar{\mbE}^{-1}+\mbE\frac{\partial \bar{\mbE}^{-1}}{\partial \mbx},
	\end{equation}
	where $\bar{\mbE}(\mbx(t))$ is our so-called `ideal' configuration matrix which varies unlike the case of hyperelastic materials where the analogous reference shape matrix is constant\footnote{See also the element rehabilitation scheme of \citet{10.1145/3306346.3323014}.}. This variation is evaluated with the identity
	\begin{equation*}\label{eq:dDmdx}
	\frac{\partial \bar{\mbE}^{-1}}{\partial \mbx_i}=-\bar{\mbE}^{-1}\frac{\partial \bar{\mbE}}{\partial \mbx_i}\bar{\mbE}^{-1},
	\end{equation*}
	and is computed w.r.t the $m$ vertices in the stencil of a contact pair. It is worth noting that the second term $\mbE\frac{\partial \bar{\mbE}^{-1}}{\partial \mbx_i}$ of Eq.~(\ref{eq:cobt-def}) has been observed to be negligible during the conducted empirical trials.
	
	\paragraph{Higher order derivatives} The second order derivative ${\partial^2 \left(\bar{\mbE}^{-1}\right)}/{\partial \mbx^2}$ exists too since the analytic expressions of some entries in $\bar{\mbE}$ are dependent on the normal vector $\mbn(\mbx)$ that is computed from the cross-product between edge-vectors. This is also the reason why our local force Jacobian (Eq. (15) in the paper) is dependent on ${\partial^2 {\mbF}}/{\partial \mbx^2}$ (${\partial^2 {\mbJ}}/{\partial \mbx^2}$ in the paper), which intuitively captures higher-order variations in $\bar{\mbE}$ due to a changing normal vector $\mbn(\mbx)$ w.r.t positions.

	{
		\subsection{Relationship between ${\mbF}$ and ${\mbJ}$}
		
		The contact constraint Jacobian $\mbJ \equiv \mbS$ that we use in the paper can be viewed as corresponding to the stretch factor of the matrix $\mbF = \mbR\mbS$ from polar decomposition\footnote{It is likewise possible  to take the perspective of singular value decomposition $\mbF = \mbU\Sigma\mbV^T$ from which $\mbJ = \Sigma$ since the method outlined in the paper may be viewed as working in the local principal stretch space of the deforming simplex along $-\mbn$.}. 
		This decomposition yields a rotation $\mbR$ and symmetric stretch $\mbS = \mbV \Sigma \mbV^T$, both of which will be diagonal in our method with $\mbR = \mbU = \mbV = \mbI$ due to the compressive distortion assumption of the simplex along the vector $-\mbn(\mbx)$ as outlined in Figure 2 of the paper. In general, either ${\mbF}$ or ${\mbJ}$ can be used to define the barrier function $b$, its gradient $\partial b/\partial \mbx$ and force Jacobian  $\partial^2 b/\partial \mbx^2$: The dimensions of the local force vector and force Jacobian matrix will be equal in either case, which can be demonstrated by applying tensor vectorization (\eg~ on either Eq. (13) or Eq. (15) in the paper) to show that tensor contraction in the derivatives will cancel out differences in the dimensions between  ${\mbF}$ and ${\mbJ}$ for a given $m$-dimensional simplex with $\mbx \in \realNum^{3s\times1}$. The change-of-basis tensor $\partial \mbJ/\partial \mbx$ is computed w.r.t the explicit entries in the diagonal structure of $\mbJ$.
	}
	
	\begin{algorithm}[htbp]
		\For{each thread, \texttt{thread\_id}}
		{
			\tcc{Variable initialization}
			\texttt{shared} \texttt{offset}\; 
			$\texttt{H\_dims} \leftarrow \texttt{m}^2$ \tcp{Total entries per Hessian $\in \realNum^{ \texttt{m}\times \texttt{m}}$}
			$\texttt{H\_id} \leftarrow \frac{\texttt{thread\_id}}{\texttt{H\_dims}}$ \tcp{Hessian index}
			$\texttt{R\_id} \leftarrow \frac{\texttt{thread\_id} \% \texttt{H\_dims}}{m}$ \tcp{Hessian row index}
			$\texttt{C\_id} \leftarrow (\texttt{thread\_id} \% \texttt{H\_dims})\% \texttt{m}$ \tcp{Hessian col index}
			$\texttt{v\_id} \leftarrow \frac{\texttt{C\_id}}{3}$ \tcp{Vertex index in Hessian}
			$\texttt{vc\_id} \leftarrow \texttt{C\_id}$ \% 3 \tcp{vertex component index}
			$\texttt{e\_id} \leftarrow \texttt{thread\_id} \% \texttt{m}$ \tcp{Entry index in Hessian row}
			
			$\texttt{h}\leftarrow \texttt{load}(\texttt{H\_id}, \texttt{R\_id},\texttt{C\_id}, \ldots)$ \tcp{Hessian entry value}
			\tcp{component of vector multiplied with Hessian}
			$\texttt{c} \leftarrow \texttt{load}(\texttt{v\_id}, \texttt{vc\_id}, \ldots)$ \;
			$\texttt{r} \leftarrow \texttt{c} \cdot \texttt{h}$ \tcp{scalar multiplication result}
			
			\If{$\texttt{thread\_id} == 0$}{
				$\texttt{offset} \leftarrow  (\texttt{m} - \texttt{e\_id}$);
			}
			
			\tcc{Wait for `\texttt{offset}' initialization}
			\texttt{barrier}()\;
			
			$\texttt{B\_id} \leftarrow \frac{\texttt{thread\_id} - \texttt{offset} + \texttt{m}}{\texttt{m}}$ \tcp{Hess row index in block}
			\tcp{Re-calibrated offset of 1st full Hessian row}
			$\texttt{l\_id} \leftarrow (\texttt{thread\_id} - \texttt{offset}) \% \texttt{m}$ \;
			\If{$\texttt{B\_id} == 0$} { 
				$\texttt{l\_id} \leftarrow \texttt{thread\_id}$\; 
			}
			
			$\texttt{w\_id} \leftarrow \texttt{thread\_id}\%32$ \tcp{Warp index}
			\tcp{is 1st thread after boundary}
			$\texttt{is\_boundary} \leftarrow (\texttt{l\_id} == 0) || (\texttt{w\_id} == 0)$ \;
			\tcp{set the bit value of mark according to boundary}
			$\texttt{mark} \leftarrow \texttt{cudaBrev}(\texttt{cudaBallot}(\texttt{is\_boundary}))$ \;
			\tcp{length of reduction range in warp}
			$\texttt{interval} \leftarrow \texttt{cudaClz}(\texttt{mark} \texttt{<<} (\texttt{w\_id} + 1))$  \;
			\tcp{clamp to \texttt{warp size}}
			$\texttt{interval} \leftarrow \min(\texttt{interval}, 31 - \texttt{w\_id})$ \;
			\tcc{\textbf{warp reduction} (accumulate 'r')}
			$\texttt{iter} \leftarrow 1$\;
			\While{$\texttt{iter} < m$}{
				\tcp{read value 'r' of neighbour(s) in warp}
				$\texttt{tmp} \leftarrow \texttt{cudaShfldown}(\texttt{r}, \texttt{iter})$ \;
				\If{$\texttt{interval} \ge \texttt{iter}$}{\texttt{r} += \texttt{tmp}}
				\texttt{iter} <<= 1 \tcp{bitshift}
			}
			\tcp{only 1st thread after boundary will write}
			\If{\texttt{is\_boundary}}{
				\texttt{addToOutputArray}(\texttt{r}) \tcp{write using atomics}
			}
		}
		\caption{\label{alg:matReduction}Warp reduction for matrix-vector multiplication. }
	\end{algorithm}
	
	\section{Eigenvalue properties}
	\label{sec:eigenanalysis}
	This section provides the proof showing that only the first eigenvalue $\lambda_1$ (\cf~Eq. (18) and Eq. (19) in the paper) is ever positive for the standard contact pairs \ie~those not representing the nearly-parallel case.
	We have used Eq. (12) in the main paper to obtain 
	\begin{align}\label{eq:lambda1}
	\lambda_1 &= \frac{2\hat d^4(6g+2g\ln(g)-7g^2-6g^2\ln(g)+1)}{g},\\
	\lambda_{2,3}&=\frac{-2\hat d^4\left(g-1\right)\left(g+2g\ln\left(g\right)-1\right)}{g^2},\label{eq:lambda23}
	\end{align} 
	as the expanded expressions that we use for our proof. {We would arrive at different expressions for $\lambda_{1,2,3}$ if we instead used Eq. (24) in the paper but the behaviour (plots) remains the same}.
	
\begin{figure}[t]
		\centering
		\begin{subfigure}[b]{0.23\textwidth}
			\centering
			\includegraphics[width=\columnwidth,trim=0 0 0 0, clip]{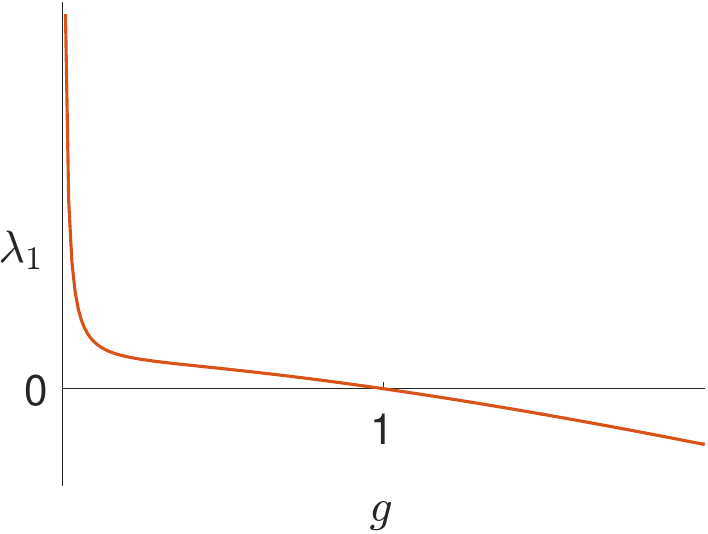}
			\caption{\label{fig:lambda0} Eigenvalue}
		\end{subfigure}
		\begin{subfigure}[b]{0.23\textwidth}
			\centering
			\includegraphics[width=\columnwidth,trim=0 0 0 0, clip]{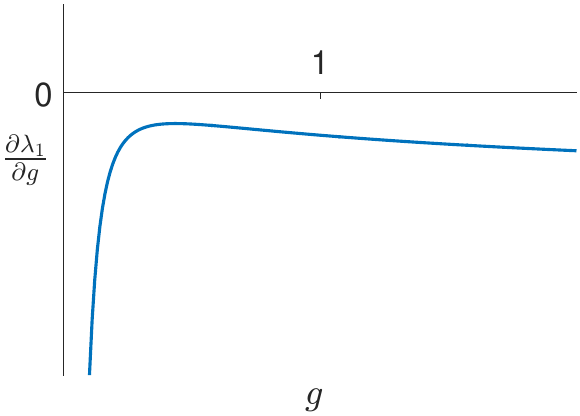}
			\caption{\label{fig:first_d_lambda0} First derivative}
		\end{subfigure}
		\caption{\label{fig:lambda-plot} A plot of \eqref{lambda1} in the paper and its derivatives, where $\hat d=1$.}
	\end{figure}

	\paragraph{$\lambda_1$ is always positive} 
	Since our gap function $g \in (0,1)$ has limited range, we have
	\begin{align}
	\lim_{g \to 0^+}\lambda_1 &= +\infty\label{eq:lambda-decr-inf}\\
	\lim_{g\to1}\lambda_1 &=0,\label{eq:lambda-decr}
	\end{align}
	where the condition $\lambda_1\ge0$ is always true (\figref{lambda0}). 
	
	\begin{proof}
		The first eigenvalue $\lambda_1$ 
		is monotonically decreasing, which we demonstrate using the 1st derivative of \eqref{lambda1} 
		\begin{align}\label{eq:lambda1-deriv}
		\frac{\partial \lambda_1}{\partial g}=\frac{-2\hat d^4(13g^2 - 2g + 6g^2\ln(g) + 1)}{g^2}.
		\end{align}
		\eqref{lambda1-deriv} is \emph{always} negative (\figref{first_d_lambda0}) which is inline with our observation in \eqref{lambda-decr} that $\lambda_1$ is ever decreasing from positive to zero as $g\to1$. 
		Therefore, $\frac{\partial \lambda_1}{\partial g} < 0$ is always true, which means $\lambda_1>0$ is also true when $g \in (0,1)$.
	\end{proof}
	
	\paragraph{$\lambda_{2,3}$ are always negative} 
	The second eigenvalue $\lambda_2$ and third eigenvalue $\lambda_3$ are always less than zero when $g \in (0,1)$. 
	\begin{figure}[t]
		\centering
		\includegraphics[width=0.5\columnwidth,trim=0 0 0 0, clip]{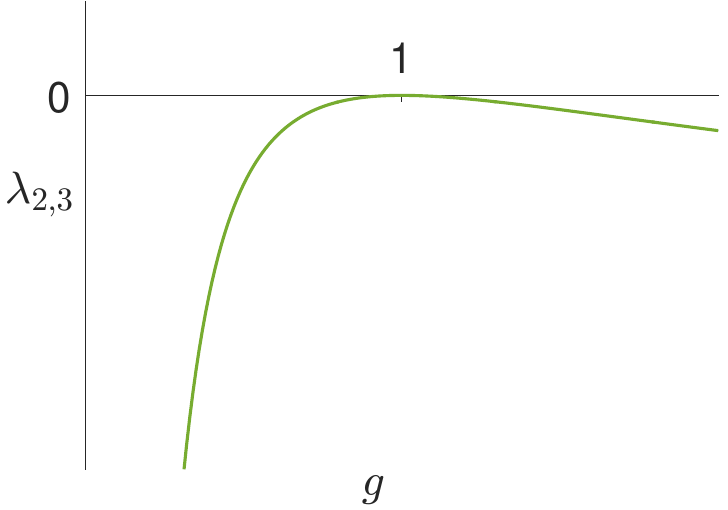}
		\caption{\label{fig:lambda23}A plot of \eqref{lambda23} in the paper, where $\hat d=1$.}
	\end{figure}
	\begin{proof}
		\eqref{lambda23} has an odd number of terms which are less-than or equal-to zero
		\begin{align}\label{eq:lambda23-proof}
		g-1&\le0,\nonumber\\g+2g\ln(g)-1&\le0, \nonumber\\-2\hat d^4&\le0.
		\end{align}
		Thus, we have $\lambda_{2,3}\le0$, which is also shown in \figref{lambda23}.
		
	\end{proof}
	
	\eqref{lambda-decr-inf}-\eqref{lambda23-proof} summarise why exactly one eigenpair is sufficient to guarantee positive semi-definiteness for minimizing our barrier energy.
	
	{
		\section{Approximate mollified Hessian terms} \label{sec:deriv-details}
		
		This section summarises of the steps to derive Eq. (32) in the paper (\secref{deriv-details-0}), and the steps we follow to arrive at the eigensystem of the last two terms of our mollified barrier Hessian (\secref{deriv-details-1}). 
		
		\subsection{Eigenvectors of the first two terms} \label{sec:deriv-details-0}
		
		We follow the approach outlined in \cite{10.1145/3306346.3323014}, to arrive at the eigenpairs
		$$\lambda_{\gamma 1}=2\left(\frac{\partial \Tilde{b}}{\partial \gamma}+2\gamma\frac{\partial^2 \Tilde{b}}{\partial \gamma^2}\right),\quad \mbQ_{\gamma 1} = \frac{1}{\sqrt{\gamma}}\mbJ\mbn_{\gamma}\mbn_{\gamma}^T.$$
		where
		$$\mbQ_{\gamma 1} =\frac{1}{\sqrt{c}}\begin{bmatrix}
		1&0&0\\
		0&\sqrt{c}&0\\
		0&0&f
		\end{bmatrix}\begin{bmatrix}0&0&0\\0&1&0\\0&0&0\end{bmatrix} = \mbn_{\gamma}\mbn_{\gamma}^T$$
		is the expansion from which we get the expression used in the paper.
		Similar steps are followed to obtain $\lambda_{g1}$ and $\mbQ_{g1} = \mbn_{g}\mbn_{g}^T$. 
		
		For the eigenpairs
		$$\lambda_{\gamma 2, \gamma3}=2\frac{\partial \Tilde{b}}{\partial \gamma}, \quad \mbQ_{\gamma2}, \mbQ_{\gamma3},$$
		we use the twist matrices from \citet{10.1145/3241041}
		$$\mbT_x = \begin{bmatrix}0&0&0\\0&0&1\\0&-1&0\end{bmatrix}, \quad \mbT_y = \begin{bmatrix}0&0&-1\\0&0&0\\1&0&0\end{bmatrix}, \quad \mbT_z = \begin{bmatrix}0&1&0\\-1&0&0\\0&0&0\end{bmatrix},$$
		to arrive at
		\begin{align}
		\mbQ_{\gamma 2} = normalize(\mbU\mbT_x\mathbf{\Sigma}\mbV^T\mbn_{\gamma}\mbn_{\gamma}^T), \\
		\mbQ_{\gamma 3} = normalize(
		(\sigma_y\hat{a}_y\mbU\mbT_z-\sigma_z\hat{a}_z\mbU\mbT_y)\mathbf{\Sigma}\mbV^T\mbn_{\gamma}\mbn_{\gamma}^T),\label{eq:eigenV23}
		\end{align}
		using SVD $\mbJ = \mbU\mathbf{\Sigma}\mbV^T$, where $\sigma_{y,z}$ are the second and third singular values from $\mathbf{\Sigma}$, and $\hat{a}_{y,z}$ are the entries of the vector $\mbV^T\mbn_{\gamma}$. We also have $\mbU = \mbV = \mbI$ due to the special diagonal structure of $\mbJ$ from which $\mbQ_{\gamma2}$ and $\mbQ_{\gamma3}$ can be simplified as
		$$\mbQ_{\gamma2} =\begin{bmatrix}0&0&0\\0&0&1\\0&-1&0\end{bmatrix}\mbn_{\gamma}\mbn_{\gamma}^T , \quad \mbQ_{\gamma3} =\begin{bmatrix}0&1&0\\-1&0&0\\0&0&0\end{bmatrix}\mbn_{\gamma}\mbn_{\gamma}^T.$$
		Similarly, we can get $\mbQ_{g2}$ and $\mbQ_{g3}$ by
		$$\mbQ_{g2} =\begin{bmatrix}0&0&0\\0&0&1\\0&-1&0\end{bmatrix}\mbn_{g}\mbn_{g}^T, \quad  \mbQ_{g3} =\begin{bmatrix}0&0&1\\0&0&0\\-1&0&0\end{bmatrix}\mbn_{g}\mbn_{g}^T.$$
	} 
	
	\subsection{Eigenpairs of the last two terms}\label{sec:deriv-details-1}
	We arrive at $\lambda_7\mbQ_{7}$ and $\lambda_8\mbQ_{8}$ in the paper with
	\begin{align}\label{eq:q7q8-orig-deriv}
	\frac{\partial^2 \Tilde{b}}{\partial \gamma\partial g}\left(\frac{\partial g}{\partial\mbJ} \otimes \frac{\partial \gamma}{\partial\mbJ} \right) +\frac{\partial^2 \Tilde{b}}{\partial g\partial \gamma}\left(\frac{\partial \gamma}{\partial\mbJ} \otimes \frac{\partial g}{\partial\mbJ} \right)
	=&
	\nonumber\\ \frac{\partial^2 \Tilde{b}}{\partial \gamma \partial  g}\mbg_g\mbg_\gamma^T+\frac{\partial^2 \Tilde{b}}{\partial g \partial  \gamma }\mbg_\gamma\mbg_g^T 
	=& \nonumber\\\begin{bmatrix}
	0&0&0&0&0&0&0&0&0\\
	0&0&0&0&0&0&0&0&0\\
	0&0&0&0&0&0&0&0&0\\
	0&0&0&0&0&0&0&0&0\\
	0&0&0&0&0&0&0&0&4t\\
	0&0&0&0&0&0&0&0&0\\
	0&0&0&0&0&0&0&0&0\\
	0&0&0&0&0&0&0&0&0\\
	0&0&0&0&4t&0&0&0&0
	\end{bmatrix}&,
	\end{align}
	which is an expansion of the last two terms of the mollified barrier Hessian, where $\mbg_\ast = \vecOp({\partial (\ast)}/{\partial \mbJ})$ with $\ast$ being a placeholder for $\gamma$ or $g$.
	We obtain the closed-form expressions of the eigenpairs by analysing this matrix in \eqref{q7q8-orig-deriv} as an eigenproblem.
	
	We arrive at our expressions for $\lambda_7^\prime\mbQ_7^\prime$ and $\lambda_8^\prime\mbQ_8^\prime$ in the paper
	by solving an eigenproblem on the auxiliary matrix 
	\begin{align} \label{eq:aux-mat}
	\mathbf{M} &= \lambda_{\gamma1}\vecOp\left(\mbQ_{\gamma1}\right)\vecOp\left(\mbQ_{\gamma1}\right)^T+\lambda_{g1}\vecOp\left(\mbQ_{g1}\right)\vecOp\left(\mbQ_{g1}\right)^T\nonumber\\
	&+\lambda_{7}\vecOp\left(\mbQ_{7}\right)\vecOp\left(\mbQ_{7}\right)^T+\lambda_{8}\vecOp\left(\mbQ_{8}\right)\vecOp\left(\mbQ_{8}\right)^T \nonumber\\
	&=
	\begin{bmatrix}
	0&0&0&0&0&0&0&0&0\\
	0&0&0&0&0&0&0&0&0\\
	0&0&0&0&0&0&0&0&0\\
	0&0&0&0&0&0&0&0&0\\
	0&0&0&0&\lambda_{\gamma1}&0&0&0&4t\\
	0&0&0&0&0&0&0&0&0\\
	0&0&0&0&0&0&0&0&0\\
	0&0&0&0&0&0&0&0&0\\
	0&0&0&0&4t&0&0&0&\lambda_{g1}\\
	\end{bmatrix},
	\end{align}
	based on the fact that the mollified Hessian is defined as 
	\begin{equation}
	\vecOp\left(\frac{\partial^2 \Tilde{b}}{\partial \mbJ^2}\right) \equiv \mathbf{M} + \sum_{i = \gamma2, \gamma3, g2, g3}\lambda_{i}\vecOp\left(\mbQ_{i}\right)\vecOp\left(\mbQ_{i}\right)^T,
	\end{equation}
	The non-orthogonal eigenmatrices $\mbQ_{\gamma1}$, $\mbQ_{g1}$, $\mbQ_{7}$, $\mbQ_{8}$ lie in the subspace represented by $\mbQ_7^\prime$ and $\mbQ_8^\prime$, which we determine by solving an eigenproblem on $\mathbf{M}$.
	{
		\section{Analysis of friction hessian}\label{sec:friction}
		
		This section provides a summary of the analysis conducted on the Hessian of the smooth friction model, as proposed by \cite{10.1145/3386569.3392425}. The model uses a `lagged' sliding basis, where contact force is computed using some quantities that are computed at the last time step.
		Specifically, the formulation of the local friction force is
		\begin{align}
		F_k(\mbx, \lambda_k^n, \boldsymbol{T}_k^n, \mu)=-\mu \lambda_k^n \boldsymbol{T}_k^n f_1(\|\mbu_k\|)\frac{\mbu_k}{\|\mbu_k\|},
		\end{align}
		where $\mu$ denotes the friction coefficient, $k$ here denotes the collision pair index, $\mbu_k = {\boldsymbol{T}_k^n}^T\mbx_k^r \in\realNum^{2\times1}$, $\mbx_k^r$ is the relative displacement of $k$-{th} collision pair, while $\lambda_k^n$ and $\boldsymbol{T}_k^n\in \realNum^{3s\times 2}$ represent the sliding basis and contact normal force derived from the previous time step that we refer to by the superscript $n$ here. The function denoted here by $f_1$ will provide a smooth and monotonic transition from 0 to 1 over a finite range
		\begin{equation}\label{eq:psd-gen-hess2}
		f_1(\|\mbu_k\|) = 
		\begin{cases}  
		-\frac{\|\mbu_k\|^2}{\epsilon_v^2 \Delta{t}^2}+\frac{2\|\mbu_k\|}{\epsilon_v \Delta{t}}, &    \|\mbu_k\|\in(0,\Delta{t}\epsilon_v) \\  
		1, &    \|\mbu_k\|>\Delta{t}\epsilon_v
		\end{cases},  
		\end{equation}
		which is also detailed in \cite{10.1145/3386569.3392425}. This leads to the corresponding friction potential 
		\begin{align}
		D_k(\mbx) = \mu\lambda_k^n f_0(\|\mbu_k\|).
		\end{align}
		Here, $f_0$ is defined such that $f_0'=f_1$ and $f_0(\epsilon_v h) = \epsilon_v \Delta{t}$ so that $F_k(\mbx)=-\nabla_{\mbx} D_k(x)$. The Hessian of $D_k(\mbx)$ can be simply given by:
		\begin{equation}\label{eq:FH}
		\begin{split}
		\nabla_{\mbx}^2D_k(\mbx) = \mu \lambda_k^n \boldsymbol{T}_k^n \left( \frac{f_1'(\|\mbu_k\|)\|\mbu_k\|-f_1(\|\mbu_k\|)}{\|\mbu_k\|^3}\mbu_k\mbu_k^T \right. \\
		\left. +\frac{f_1(\|\mbu_k\|)}{\|\mbu_k\|}\mbI_2
		\right)\boldsymbol{T}_k^{nT} \in \realNum^{3s\times3s}.
		\end{split}
		\end{equation}
		where $\mbI_2$ is a $2\times2$ identity matrix. Eigenanalysis of the Hessian in \eqref{FH} essentially boils down to the analysis of a simple $2\times2$ matrix, 
		\begin{align}
		\frac{f_1'(\|\mbu_k\|)\|\mbu_k\|-f_1(\|\mbu_k\|)}{\|\mbu_k\|^3}\mbu_k\mbu_k^T+\frac{f_1(\|\mbu_k\|)}{\|\mbu_k\|}\mbI_2,
		\end{align}
		which can be accomplished by analytically solving a quadratic characteristic polynomial.
	} 
	
	\section{Data-parallel solver}\label{sec:par-solver}
	This section provides the high level information (and algorithm) that we use to parallelise our GPU implementation of preconditioned conjugate gradients (PCG) \cite{10.5555/865018} {without} constructing the full system matrix.
	This scheme has been adopted previously, \eg~ by \citet{gpumpm} and \citet{10.1145/3386760}, where our novelty lies in the parallel optimization of local matrix-vector multiplications. The remaining vector-vector multiplications of PCG is optimized with the global reduction methods of \citet{ZhaoHLWQ20}.
	
	The traditional approach to solving the global linear system requires prior construction/assembly of a (sparse) system matrix, which involves incrementally \textit{adding} contributions into its block entries from the local Hessians~\cite{10.1145/2816795.2818081,largeStepCloth}. This is especially suitable for elastic deformation problems with finite elements or mass-spring systems where mesh topology is fixed to allow pre-computation of the non-zero entries/indices in the global matrix. 
	In our case the time-dependent nature of the number of IPC barrier energy terms precludes such pre-computation to affect the cost of booking-keeping for tracking dynamically-changing sparse matrix indexing offsets. 
	
\begin{figure}[t]
		\centering
		\includegraphics[width=0.6\columnwidth]{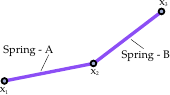}
		\caption{Mass spring system with two springs}
		\label{fig:mass-spring}
	\end{figure}

	\paragraph{Decomposing system matrix-vector multiplication} 
	It is possible to avoid such global construction by decomposing  
	system matrix back to the local form  
	using the distributive property of addition, which we use and can be demonstrated using a simple 2D mass-spring system with three nodes and two springs. 
	\figref{mass-spring} shows two springs endowed with a (local system) matrix each
	\begin{align}\label{eq:A-decom}
	\mbA &= \begin{bmatrix}
	\mbA_{11} & \mbA_{12}\\
	\mbA_{21} & \mbA_{22}
	\end{bmatrix} \equiv \frac{\partial^2\Psi_{A}}{\partial\mbx^2},\\
	\mbB &= \begin{bmatrix}
	\mbB_{11} & \mbB_{12}\\
	\mbB_{21} & \mbB_{22} 
	\end{bmatrix} \equiv \frac{\partial^2\Psi_{B}}{\partial\mbx^2},
	\end{align}
	which correspond to their energies $\Psi_{A}$ and $\Psi_{B}$, respectively, with block entries $\mbA_{ij}, \mbB_{ij} \in \realNum^{2\times2}$.
	The global system matrix may be constructed as follows:
	\begin{align*} \label{eq:mat-decomp}
	\underbrace{
		\begin{bmatrix}
		\mbA_{11} & \mbA_{12} & 0\\
		\mbA_{21} & \mbA_{22} & 0\\
		0 & 0 & 0
		\end{bmatrix}
	}_{\mathbb{A}}
	+
	\underbrace{
		\begin{bmatrix}
		0 & 0 & 0\\
		0 & \mbB_{11} & \mbB_{12}\\
		0 & \mbB_{21} & \mbB_{22}
		\end{bmatrix}
	}_{\mathbb{B}}
	=
	\underbrace{
		\begin{bmatrix}
		\mbA_{11} & \mbA_{12} & 0\\
		\mbA_{21} & \mbA_{22} + \mbB_{11}\\
		0 & \mbB_{12} & \mbB_{22}
		\end{bmatrix}}_{\mathbb{H}},
	\end{align*}
	with dimensions $\mathbb{H} \in \realNum^{6\times6}$ which is based on the connectivity of the system. 
	
	\begin{figure*}[htbp]
		\centering 
		\begin{subfigure}[b]{2\columnwidth}
			\centering
			\includegraphics[width=\textwidth]{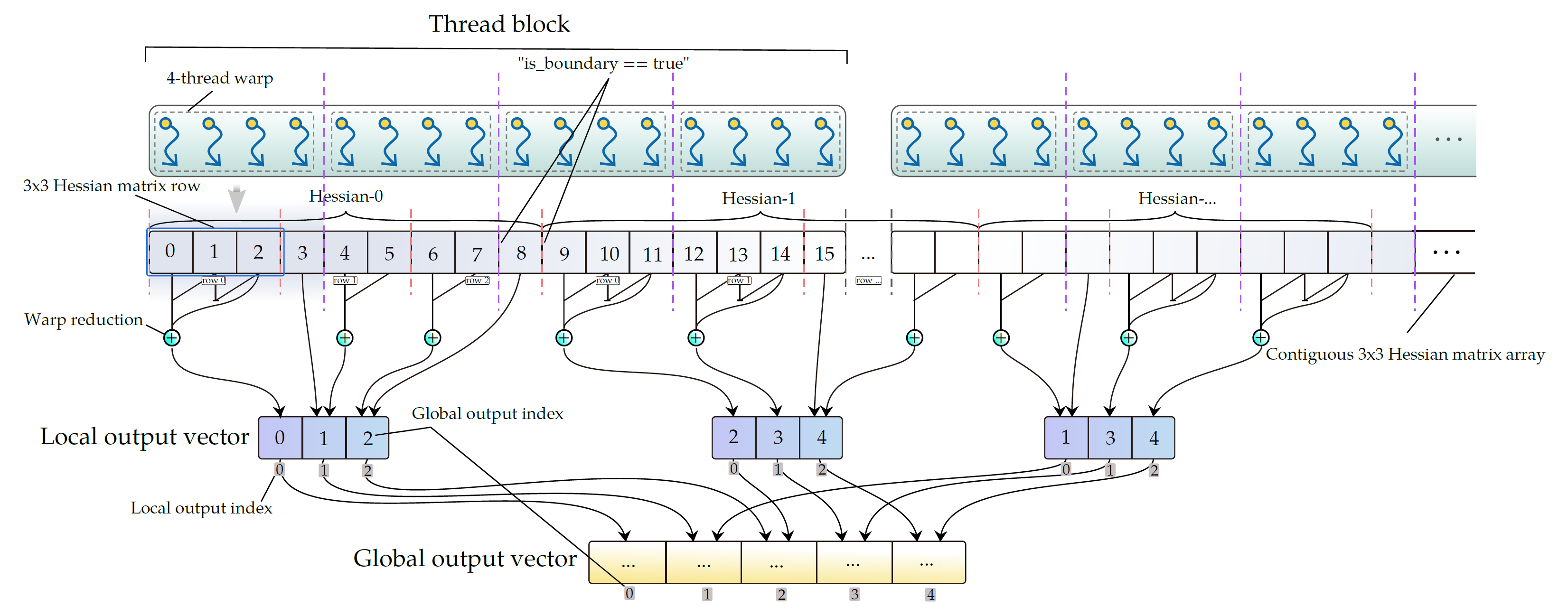}
		\end{subfigure}
		\caption{Warp reduction for parallel dense matrix-vector multiplication (see also \algref{matReduction}).}
		\label{fig:warp-reduction}
	\end{figure*}
	
	Two springs sharing one vertex will result in a summation of two specific block entries from $\mbA$ and $\mbB$, which are determined by a local-to-global index map. This map, which is constant, is based on the connectivity of the mass spring system, where `local' refers to node indices in a spring versus the `global' indices in the mass-spring system\footnote{Note that the same analogy carries through when dealing with contact using our IPC formulation: a collision between two surface boundary elements (\eg~a point and an triangle) will form simplex from which we construct our barrier function, its gradient and Hessian. A vertex will have a local index in the simplex and a global index in the respective mesh(es) in contact.}.  
	Thus, we have 
	\begin{align}
	\mathbb{H}\mbc &= ({\mathbb{A}}+{\mathbb{B}})\mbc,\nonumber\\
	&= {\mathbb{A}}\mbc + {\mathbb{B}}\mbc \in \realNum^{6\times1},
	\end{align} 
	with $\mbc = \begin{bmatrix}\mbc_1 & \mbc_2 & \mbc_3\end{bmatrix}^T$ and $\mbc_i \in \realNum^{2\times1}$ to give
	\begin{align*} 
	\begin{bmatrix}
	\mbA_{11}\mbc_1 +  \mbA_{12}\mbc_2\\
	\mbA_{21}\mbc_1 +  \left(\mbA_{22} + \mbB_{22}\right)\mbc_2 + \mbB_{23}\mbc_3\\
	\mbB_{32}\mbc_2 + \mbB_{33}\mbc_3
	\end{bmatrix} = \mathbb{H}\mbc = \mbc^\prime.
	\end{align*}
	where the vector $\mbc$ is synonymous with the variable defined in Line (5) of the `modified-pcg' Algorithm of \citet{largeStepCloth}.
	The global solution $\mbc^\prime$ can be decomposed into local solution vectors
	\begin{align}\label{eq:A-decom}
	\begin{bmatrix}
	\mbA_{11} & \mbA_{12}\\
	\mbA_{21} & \mbA_{22}
	\end{bmatrix}
	\begin{bmatrix}
	\mbc_{1} \\
	\mbc_{2}
	\end{bmatrix}
	= 
	\begin{bmatrix}
	\mbA_{11}\mbc_1 +  \mbA_{12}\mbc_2\\
	\mbA_{21}\mbc_1 +  \mbA_{22}\mbc_2\\
	\end{bmatrix} = \mbc_{\mba}^\prime,\in \realNum^{2\times1}
	\end{align}
	and 
	\begin{align}\label{eq:B-decom}
	\begin{bmatrix}
	\mbB_{11} & \mbB_{12}\\
	\mbB_{21} & \mbB_{22}\\
	\end{bmatrix}
	\begin{bmatrix}
	\mbc_{2} \\
	\mbc_{3}
	\end{bmatrix}
	= 
	\begin{bmatrix}
	\mbB_{11}\mbc_2 +  \mbB_{12}\mbc_3\\
	\mbB_{21}\mbc_2 +  \mbB_{22}\mbc_3
	\end{bmatrix}= \mbc_{\mbb}^\prime \in \realNum^{2\times1}.
	\end{align} 
	to give $\mbc^\prime = \begin{bmatrix}\mbc_{\mba1}^\prime & \left(\mbc_{\mba2}^\prime + \mbc_{\mbb1}^\prime\right) & \mbc_{\mbb2}^\prime \end{bmatrix}^T$ as the basis upon which we build our implementation  of the local system matrix-vector multiplications during PCG. 
	So if the local solution vectors $\mbc_{\mba}^\prime$ and $\mbc_{\mbb}^\prime$ are computed independently, then it is possible to later merge results into the `global' vector $\mbc^\prime$. 
	We use this property to solve the linear system $\mathbb{H} \mbd = \mbg$ without ever forming the global system $\mathbb{H} \mbc$ at each iteration of PCG.

	\paragraph{Parallel local matrix-vector multiplication}
	We assign one thread per scalar entry in a local system matrix (\ie~in \mbA~or in~\mbB), and apply a CUDA-warp level reduction scheme (\algref{matReduction}) to compute the components of the local solution vector (\ie~$\mbc_{\mba}^\prime$, or $\mbc_{\mbb}^\prime$) in parallel. 
	Briefly, threads in a warp are each mapped to one scalar entry in a local system matrix; the thread will then multiply the matrix entry to the corresponding component of the input vector $\mbc$. 
	The multiplication result is then stored in private register memory, which (using CUDA builtins \texttt{clz}, \texttt{shfldown} \etc) is then accumulated within each warp before being added to the global output vector $\mbc^\prime$. 
	
	However, the inconsistency between the CUDA thread block size (can only be $2^n$, where n > 4) and the local Hessian dimension ($3n\times 3n$, where $n = 4, 3, 2$), as well as the warp size (32), unavoidably results in a challenge for executing the process of multiplying the matrix-vector elements and summing them in parallel. 
	Ideally, the elements of the same row of Hessian are managed by the same thread warp, as then the summation of the matrix-vector elements can be done by warp reduction, which is significantly faster than summing the product individually by atomic operations. 
	As shown in \figref{warp-reduction}, the inconsistent sizes of the thread block/thread warp/matrix dimensions easily cause each row/matrix to be managed across  different thread blocks/warps.  
	
	To maximize parallelism, we define a data structure that represents the size and range of reduction in each row of the matrix, which consists of the start position and the length of the reduction.   
	For example, in  \figref{warp-reduction}, 
	in the first row of Hessian-0, the start position is 0, and the length is 3, while in the second row the start position is 4 and the length is 2. We can maximize the parallelism and minimize the branching, such that approximately one thread per row (of any local system matrix)  will actually commit data to the global memory (see \algref{matReduction} for the details).

	The local system matrices from our IPC formulation will at-most have dimensions $\realNum^{12\times12}$ but allocating such space for all of our contact pairs (potentially millions) is wasteful since some pairs require much less memory (\eg~\textit{point-point} with $\realNum^{6\times6}$) .
	Thus, we also arrange these matrices into four storage buffers based on the \textit{type} of contact pairs to minimize storage costs and improve parallelism by grouping workloads with same Hessian dimensions.
	In addition, we also design the warp reduction scheme (\algref{matReduction}) to assume that all local system matrices assigned to each block/group of CUDA threads have the same local system matrix dimensions to reduce booking-keeping and minimize divergent execution flow.
	
	\section{Supplementary unit tests}\label{sec:unit-tests}
\begin{figure}[t]
		\resizebox{\columnwidth{}}{!}{
			\begin{tabular}{ccccc}
				\includegraphics[width=\columnwidth, trim=100 0 100 0, clip]{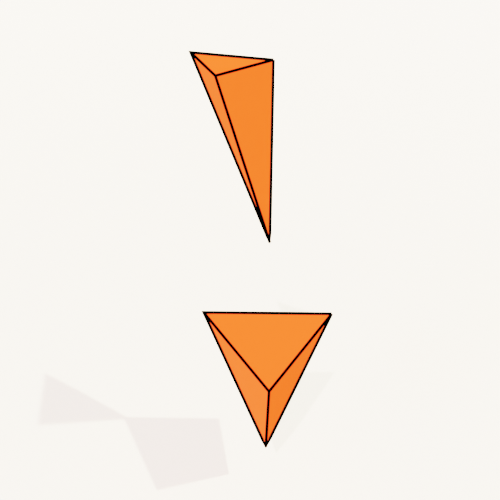} & \includegraphics[width=\columnwidth, trim=100 0 100 0, clip]{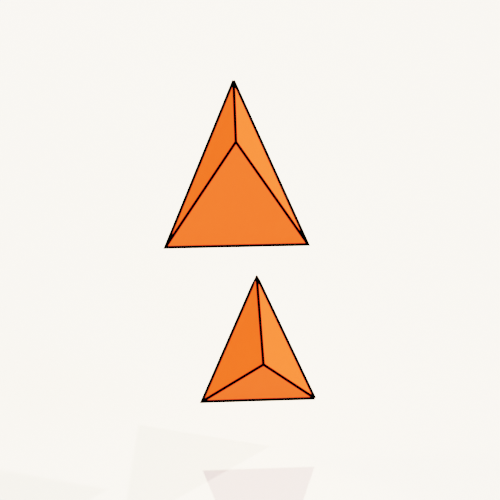} & 
				\includegraphics[width=\columnwidth, trim=100 0 100 0, clip]{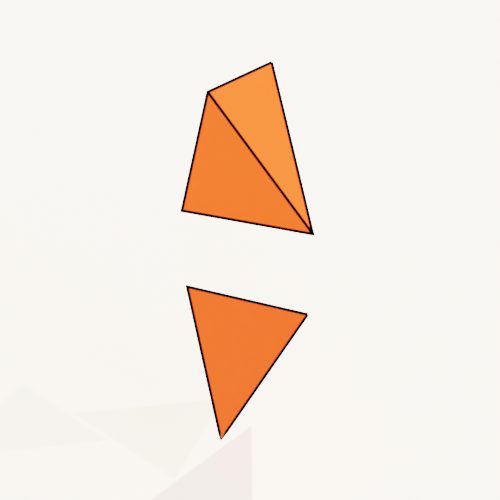} & \includegraphics[width=\columnwidth, trim=100 0 100 0, clip]{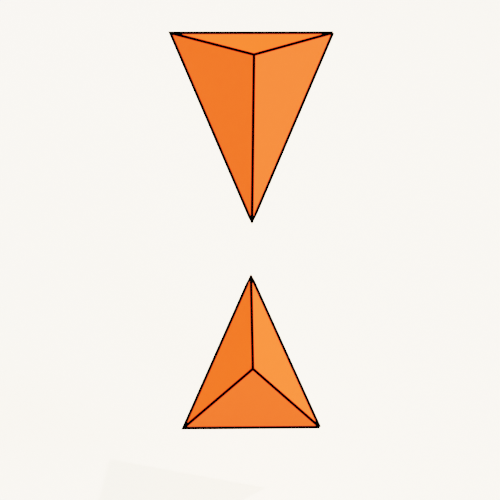} &
				\includegraphics[width=\columnwidth, trim=100 0 100 0, clip]{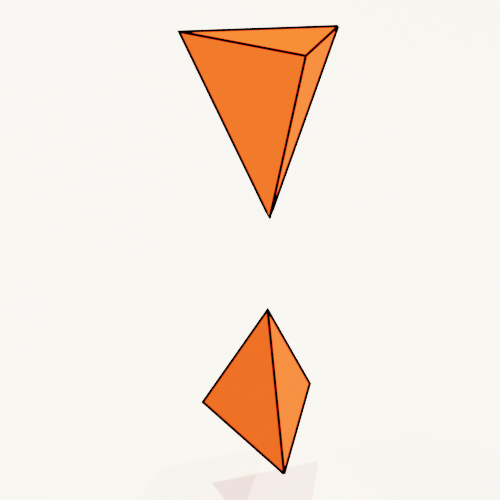}\\
				\includegraphics[width=\columnwidth, trim=100 0 100 150, clip]{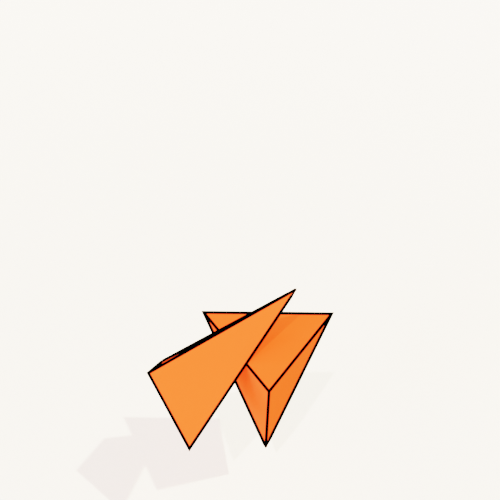} & \includegraphics[width=\columnwidth, trim=100 0 100 150, clip]{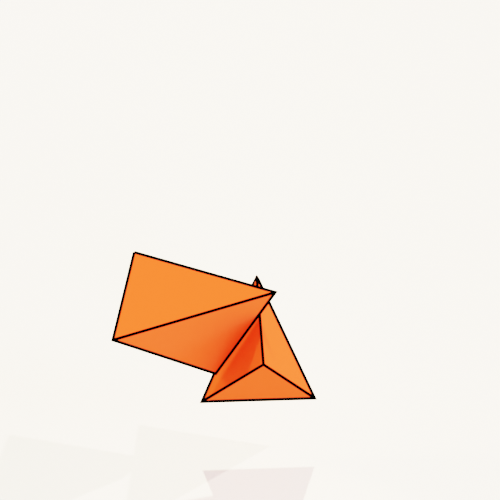} &  
				\includegraphics[width=\columnwidth, trim=100 0 100 150, clip]{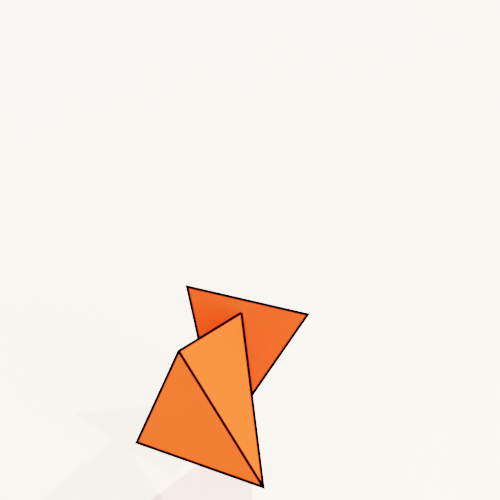}& 
				\includegraphics[width=\columnwidth, trim=100 0 100 150, clip]{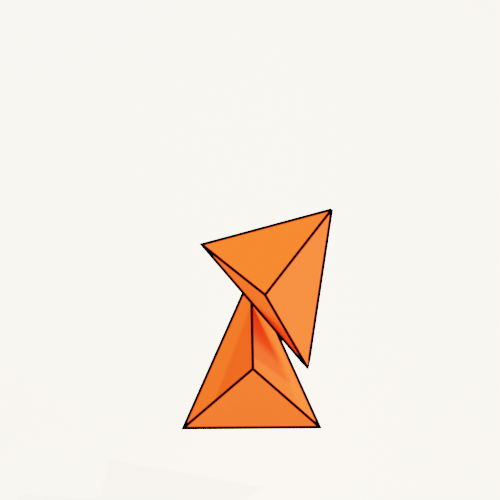} &
				\includegraphics[width=\columnwidth, trim=100 0 100 150, clip]{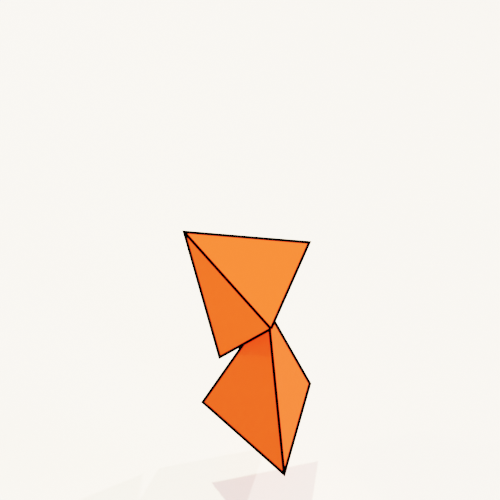} 
			\end{tabular}
		}
		\caption{\label{fig:unit-tests}Unit tests: From left to right we have the point-triangle, edge-edge, parallel edge-edge, point-edge and point-point case. We robustly pass all these tests.}
	\end{figure}

\begin{figure}[t]
		\centering
		
		\begin{subfigure}[b]{0.45\columnwidth}
			\centering
			\includegraphics[width=1.0\columnwidth, trim=0 130 0 80, clip]{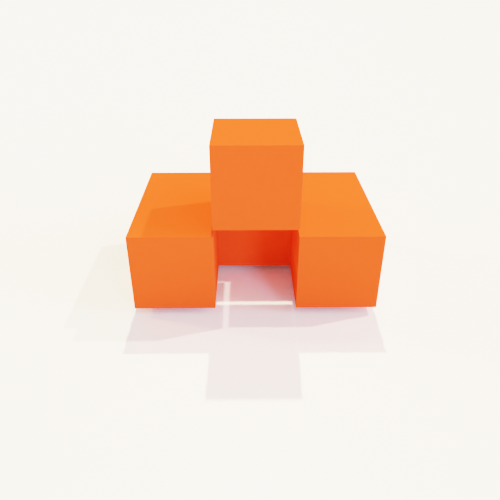}  
		\end{subfigure}
		\begin{subfigure}[b]{0.45\columnwidth}
			\centering
			\includegraphics[width=1.0\columnwidth, trim=0 130 0 80, clip]{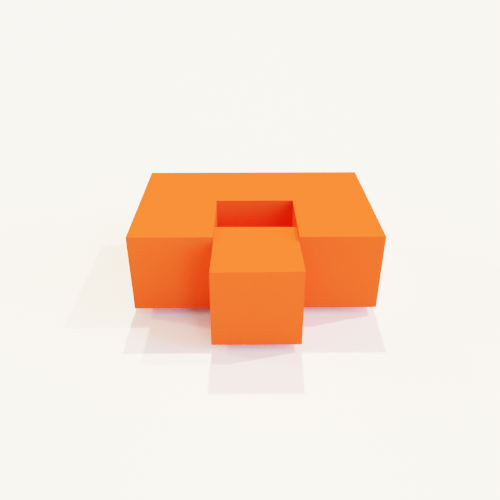}  
		\end{subfigure}
		\caption{\label{fig:cube-1}Aligned, close and nonsmooth contact test: conforming collisions are also accurately and stably resolved.}
	\end{figure}

	This section provides supplementary results, which show additional units tests that where used to evaluate our implementation. These results are shown in \figref{unit-tests} and \figref{cube-1}. Readers are also referred to our supplementary video.


	\section{Comparison of barrier methods on CPU}
	\label{sec:barrier-tests}
	Different aspects of implementation such as hardware choice (\eg~GPU or CPU) and the choice of linear solver will inevitably influence the performance enhancements observed in our barrier method. 
	Thus, to validate the accuracy of the improvements described in the paper, we conduct additional evaluations of our barrier method within the CPU-IPC framework. This involved implementing our barrier method in CPU-IPC while using the same direct solver, CHOLMOD, to measure performance improvement. The purpose of this evaluation was to demonstrate improvements across various time steps and material stiffness parameter settings. The detailed experimental settings and the corresponding evaluation results are provided in \tabref{statscpuVcpu}, which shows the improvements of our method as described in the paper (see column {\texttt{\#i}}). Our method leads to a consistently lower number of Newton iterations on average. The advantage of our approach is most evident when simulating with highly stiff materials, reducing the number of iterations by up-to 3$\times$.

	\begin{table}[htbp]
		\centering
		\caption{Solution tolerance versus the number of Newton iterations, and the gradient-norm at convergence.}
		\resizebox{\columnwidth}{!}{
			\begin{tabular}{lllllll}
				\toprule
				& 1e-2 & 1e-3 & 1e-4 & 1e-5 & 1e-6 & 1e-7 \\ 
				\midrule
				Iterations & 3159 & 5772 & 9300 & 11123 & 14474 & 16631 \\ 
				Gradient-norm &$3\mathrm{e}{-2}$ & $3\mathrm{e}{-3}$ & $3\mathrm{e}{-4}$ & $3\mathrm{e}{-5}$ & $3\mathrm{e}{-6}$ & $3\mathrm{e}{-7}$ \\ 
				\bottomrule
			\end{tabular}
		}
		\label{tab:sol-tol}
	\end{table}
	
	\begin{table}[htbp]
		\centering
		\caption{Newton Iterations w.r.t the gradient-norm tolerance in standard CPU-IPC \cite{10.1145/3386569.3392425}.}
		\resizebox{\columnwidth}{!}{
			\begin{tabular}{lllllll}
				\toprule
				& $3\mathrm{e}{-2}$ & $3\mathrm{e}{-3}$ & $3\mathrm{e}{-4}$ & $3\mathrm{e}{-5}$ & $3\mathrm{e}{-6}$ & $3\mathrm{e}{-7}$ \\ 
				\midrule
				Iterations & 3582 & 6292 & 9765 & 11910 & 15105 & 16962 \\ 
				\bottomrule
			\end{tabular}
		}
		\label{tab:grad-tol}
	\end{table}
	
	\begin{table}[htbp]
		\centering
		\caption{Newton iterations w.r.t  PCG tolerance and Newton tolerance (based on gradient-norm). The data is plotted in Fig. \ref{fig:pcg-newton-iteration}.
		}
		\resizebox{\columnwidth}{!}{
			\begin{tabular}{ccccccc}
				\toprule
				{PCG Tolerance} & \multicolumn{6}{c}{Newton solver tolerance (gradient-norm)}\\
				\cmidrule{2-7}
				& {$3\mathrm{e}{-2}$} & {$3\mathrm{e}{-3}$} & {$3\mathrm{e}{-3}$} & {$3\mathrm{e}{-5}$} & {$3\mathrm{e}{-6}$} & {$3\mathrm{e}{-7}$} \\ 
				\midrule
				{$1\mathrm{e}{-1}$} & 3981 & 7058 & 10418 & 13726 & 14970 & 19021 \\ 
				{$1\mathrm{e}{-2}$} & 3894 & 6585 & 10573 & 13257 & 15213 & 18534 \\ 
				{$1\mathrm{e}{-3}$} & 3698 & 6381 & 10340 & 12747 & 15915 & 18651 \\ 
				{$1\mathrm{e}{-4}$} & 3557 & 6116 & 10051 & 12163 & 14508 & 17367 \\ 
				{$1\mathrm{e}{-5}$} & 3846 & 6344 & 9896 & 12075 & 15320 & 17454 \\ 
				{$1\mathrm{e}{-6}$} & 3766 & 6288 & 9068 & 12695 & 14414 & 17491 \\ 
				{$1\mathrm{e}{-7}$} & 3666 & 6235 & 9921 & 12059 & 15120 & 17222 \\ 
				\bottomrule
		\end{tabular}}
		\label{tab:newt-solv-pcg}
	\end{table}
	
	\begin{table}[htbp]
		\centering
		\caption{Newton Iterations w.r.t the gradient-norm tolerance, using our barrier method (with mollification) implemented the CPU-IPC framework.}
		\resizebox{\columnwidth}{!}{
			\begin{tabular}{lllllll}
				\toprule
				& $3\mathrm{e}{-2}$ & $3\mathrm{e}{-3}$ & $3\mathrm{e}{-4}$ & $3\mathrm{e}{-5}$ & $3\mathrm{e}{-6}$ & $3\mathrm{e}{-7}$ \\ 
				\midrule
				Iterations & 2322 & 5425 & 8280 & 11231 & 14407 & 16148 \\ 
				\bottomrule
			\end{tabular}
		}
		\label{tab:barrier-haccuracy}
	\end{table}

	\begin{table*}[t]
		\centering
		\resizebox{\textwidth}{!}{
			\begin{tabular}{r|ccccc|cccccccccccccc|c}
				\toprule
				
				& \texttt{v},\texttt{t},\texttt{f}  & $\rho$,$E$,$\upsilon$ & $\hat{d}$, $\varepsilon_d$ & $\mu$, $\epsilon_{v}$ & $\Delta{t}$, \#$\Delta{t}$& \multicolumn{2}{c}{\texttt{buildCP}} &  \multicolumn{2}{c}{\texttt{buildGH}} & \multicolumn{2}{c}{\texttt{solve}} & \multicolumn{2}{c}{\texttt{CCD}}   & \multicolumn{2}{c}{\texttt{\#i}} &   \multicolumn{2}{c}{\texttt{misc}} &\multicolumn{2}{c}{\texttt{timeTot}} & \textbf{speedup} \\ 
				& & & & & & cpu-Li&cpu-Our & cpu-Li&cpu-Our & cpu-Li&cpu-Our & cpu-Li&cpu-Our & cpu-Li&cpu-Our & cpu-Li&cpu-Our & cpu-Li&cpu-Our & (\textit{cpu-Our vs. cpu-Li}) \\
				\toprule
				Fig. (10)(paper) & \makecell{32k, 135k, 38k}  &  \makecell{1e3,1e4,0.49\\1e3,1e5,0.49\\1e3,1e6,0.49\\1e3,1e7,0.49}&1e-3, 1e-2 & - & \makecell{0.01, 95\\0.01, 80\\0.01, 100\\0.01, 81} & \makecell{6.33e2\\4.09e2\\4.66e2\\4.62e2} & \makecell{4.39e2\\2.34e2\\2.27e2\\1.44e2} & \makecell{2.97e2\\2.05e2\\2.39e2\\2.33e2} & \makecell{1.95e2\\1.08e2\\1.01e2\\6.28e1} & \makecell{3.39e3\\2.32e3\\2.74e3\\2.62e3} & \makecell{2.68e3\\1.51e3\\1.42e3\\8.76e2} & \makecell{5.41e2\\3.58e2\\4.04e2\\3.94e2} & \makecell{3.75e2\\2.07e2\\2.27e2\\1.23e2} & \makecell{33.2\\26.5\\24.8\\29.8} & \makecell{24.4\\16.5\\12.7\\9.93} & \makecell{1.93e1\\1.81e1\\2.14e1\\3.11e1} & \makecell{1.61e1\\1.30e1\\1.25e1\\1.10e1} & \makecell{4.88e3\\3.31e3\\3.87e3\\3.74e3} & \makecell{3.71e3\\2.07e3\\1.99e3\\1.22e3} & \makecell{\textbf{1.32}$\times$\\\textbf{1.60}$\times$\\\textbf{1.94}$\times$\\\textbf{3.06}$\times$} \\
				
				\midrule
				Fig. (14)(paper) & \makecell{8k, 36k, 10k(dolphin)\\30k, -, 60k(funnel)}  & 1e3,1e4,0.40&1e-3, 1e-2 & - & \makecell{0.01, 300\\0.02, 150\\0.03, 100\\0.04, 75\\0.05, 60} & \makecell{2.64e3\\1.62e3\\1.24e3\\9.69e2\\8.86e2} & \makecell{2.35e3\\1.25e3\\9.50e2\\7.18e2\\6.55e2} & \makecell{5.47e2\\4.27e2\\3.62e2\\3.28e2\\3.13e2} & \makecell{2.16e2\\1.52e2\\1.44e2\\1.16e2\\1.21e2} & \makecell{3.46e3\\2.55e3\\2.13e3\\1.72e3\\1.59e3} & \makecell{3.10e3\\1.92e3\\1.69e3\\1.29e3\\1.24e3} & \makecell{2.06e3\\1.36e3\\1.09e3\\8.93e2\\8.28e2} & \makecell{1.75e3\\1.03e3\\8.47e2\\6.96e2\\6.88e2} & \makecell{22.7\\29.6\\36.8\\38.4\\43.6} & \makecell{20.1\\23.2\\28.4\\28.6\\32.7} & \makecell{1.84e2\\1.03e2\\6.82e1\\5.38e1\\4.47e1} & \makecell{1.67e2\\8.68e1\\5.68e1\\4.41e1\\3.63e1} & \makecell{8.89e3\\6.06e3\\4.89e3\\3.97e3\\3.66e3} & \makecell{7.58e3\\4.45e3\\3.69e3\\2.87e3\\2.74e3} & \makecell{\textbf{1.15}$\times$\\\textbf{1.36}$\times$\\\textbf{1.32}$\times$\\\textbf{1.38}$\times$\\\textbf{1.33}$\times$} \\
				
				\bottomrule
			\end{tabular}
		}
		\caption{\label{tab:statscpuVcpu}Performance summary using comparison \citet{10.1145/3386569.3392425}. The columns are as follows: number of vertices, including the interior for tetrahedral meshes (\texttt{v}); number of tetrahedra (\texttt{t}); number of surface triangles (\texttt{f}); time step size in seconds ($\Delta{t}$); material density($\rho$), Young's modulus ($E$) in units of pascals Pa, and Poisson's ratio ($\upsilon$); computational accuracy target in meters ($\hat{d}$) which is set w.r.t. to the scene bounding box diagonal length l; Newton Solver tolerance threshold ($\epsilon_d$); friction coefficient ($\mu$) and velocity magnitude bound ($\epsilon_{v}$);Total number of time steps (\#$\Delta{t}$); Total time to build/find contact pairs (\texttt{buildCP}); Total time to build energy gradients and Hessians for all types (\texttt{buildGH}); Total linear solver time (\texttt{solve}); Total CCD time (\texttt{CCD});  Average number of Newton iterations per time step (\texttt{\#i}); Total time for remaining miscellaneous tasks (\texttt{misc}); Total simulation compute time (\texttt{timeTot}); We replace the barrier method of CPU-IPC (\citet{10.1145/3386569.3392425}) with ours to estimate the speedup of our method. All time measurements are presented in seconds. 
		}
	\end{table*}
	
	\section{comparison between direct solver and PCG}
	\label{sec:linear-solver-tests}
	In this section, we present and discuss results based on an apple-to-apple comparison between the direct solver and PCG method in terms of Newton solver iterations, tested with the scene shown in Fig. (10 (a)) in the paper. 
	This experiment is based on the setup described in \secref{barrier-tests}, using our PCG method within the CPU-IPC framework \cite{10.1145/3386569.3392425}.
	The objective is to illustrate that the PCG tolerance employed in our experiments (\ie~to give $\delta_{new} < $ 1e-4$\delta_{0}$ as the termination condition of PCG), as mentioned in the paper, is sufficient for generating simulations of comparable accuracy to those achieved with direct solver in CPU-IPC.
	
	In \tabref{sol-tol} and \tabref{grad-tol}, we provide reference data showing the number of Newton iterations w.r.t tolerance, where this tolerance is defined based on the solution and the gradient, respectively.
	We obtain this data by running standard CPU-IPC \cite{10.1145/3386569.3392425} with a direct solver (CHOLMOD) in two phases. In the first phase, we run with six distinct settings of the solution tolerance and record the number of Newton iterations and gradient-norm at convergence.
	We then repeat the simulation in the second phase using the recorded gradient-norm as the solver tolerance to analyse the number of iterations required to converge. 
	The solution threshold with the highest accuracy in our experimental setup is approximately $1\mathrm{e}{-7}$ m/s. 
	The gathered data in \tabref{sol-tol} and \tabref{grad-tol} is used as reference to compare against the results shown in \tabref{newt-solv-pcg}. It is important to note that the convergence condition is determined by $\frac{\|\mbd\|{\infty}}{l\Delta{t}} \le \varepsilon_d$ for the solution tolerance and $\frac{\|\mbg\|{\infty}}{l\Delta{t}^2} \le \varepsilon_g$ for the gradient tolerance.
	
	\begin{figure}[htbp]
		\centering 
		\begin{subfigure}[b]{1\columnwidth}
			\centering
			\includegraphics[width=\columnwidth]{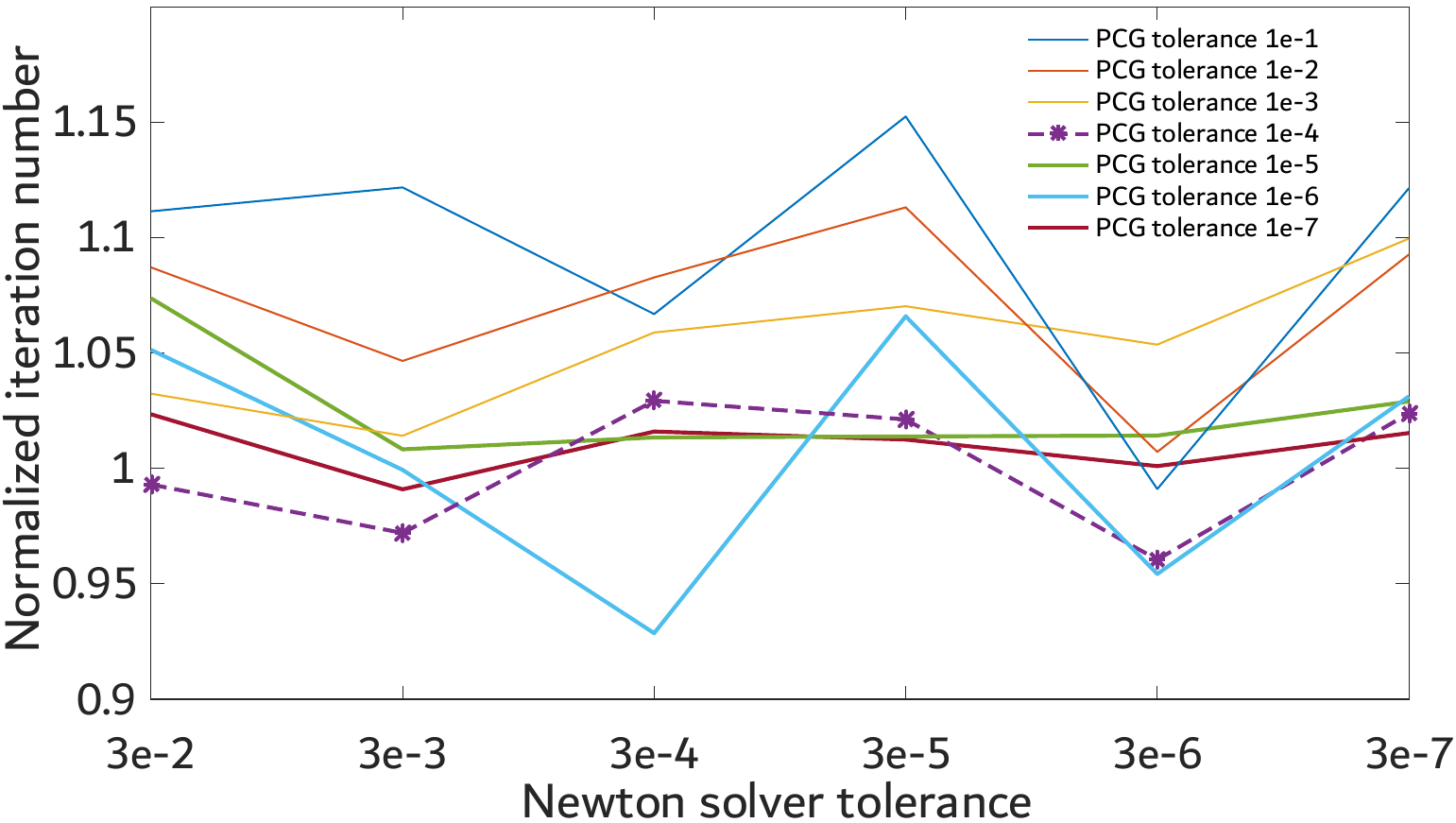}
		\end{subfigure}
		\caption{ 
			{Normalized plot of the rows in \tabref{newt-solv-pcg}: Newton-iteration number when using PCG w.r.t tolerance. We normalize by the corresponding iteration number for the direct solver (\cf~\tabref{grad-tol}). The figure demonstrates that the difference in Newton-iteration number between using PCG and the direct solver is typically within a margin of approximately $5\%$ when PCG tolerance is smaller than $1\mathrm{e}{-4}$. Using a higher accuracy tolerance for PCG does not necessarily lower the Newton-iteration count, as locally accurate solutions (\ie per Newton iteration) do not necessarily translate into a globally optimal convergence rate. Our data further indicates that even with a relatively low-accuracy PCG tolerance, the Newton solver can still converge. This follows the fact that PCG provides an optimal search direction and step size within the current Krylov subspace, which corresponds to the direction of energy descent in that subspace, even with a relatively low-accuracy tolerance}. 
			\label{fig:pcg-newton-iteration}}
	\end{figure}
	
	In \tabref{newt-solv-pcg}, we provide the data we obtain by running standard CPU-IPC with PCG instead of a direct solver, which we use to demonstrate that CPU-IPC converges to the same accuracy with either of these two linear solvers.
	The gradient-norm is used for determining whether the Newton solver has converged in this experiment, which we do in order to ensure a fair comparison because using the solution for validation may be less reliable given that it varies with different linear solvers (potentially impacting convergence/Newton iterations). 
	Crucially, the data of \tabref{newt-solv-pcg} and \figref{pcg-newton-iteration} provide further evidence that our choice of solution tolerance ($1\mathrm{e}{-4}$) for PCG in the paper yields an equivalent number of Newton iterations when compared to higher-accuracy thresholds.
	
	In order to demonstrate that our barrier method, using the PCG solver with $1\mathrm{e}{-4}$ tolerance, maintains the accuracy of the IPC method, we conducted additional tests. We replaced the barrier method with our approach and continued to use the gradient-norm threshold to assess the convergence capability of our method across different simulation accuracies. The results, presented in \tabref{barrier-haccuracy}, indicate that our barrier method is capable of performing high-accuracy simulations as well.
	
	\section{Impact of friction Hessian projection}
	\label{sec:friction-impact}
	
	\begin{figure}[htbp]
		\centering 
		\begin{subfigure}[b]{0.8\columnwidth}
			\centering
			\includegraphics[width=\columnwidth]{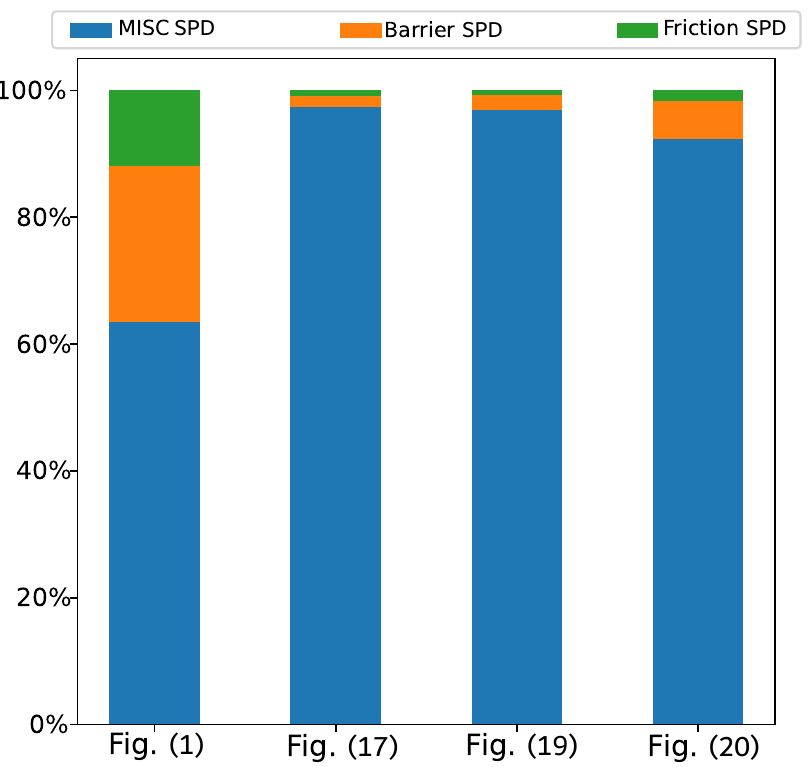}
		\end{subfigure}
		\caption{Percentage of time on Hessian projection:  Time breakdown considering the local Hessian projections in our simulator, which include those of the elastic energy. Here we particularly focus on projections for friction- and barrier Hessians. The x-axis refers to the figures/demos provided in the paper, which are the examples involving friction (full simulation).  }
		\label{fig:friction-breakdown}
	\end{figure}
	
	\begin{figure}[htbp]
		\centering 
		\begin{subfigure}[b]{1\columnwidth}
			\centering
			\includegraphics[width=\columnwidth]{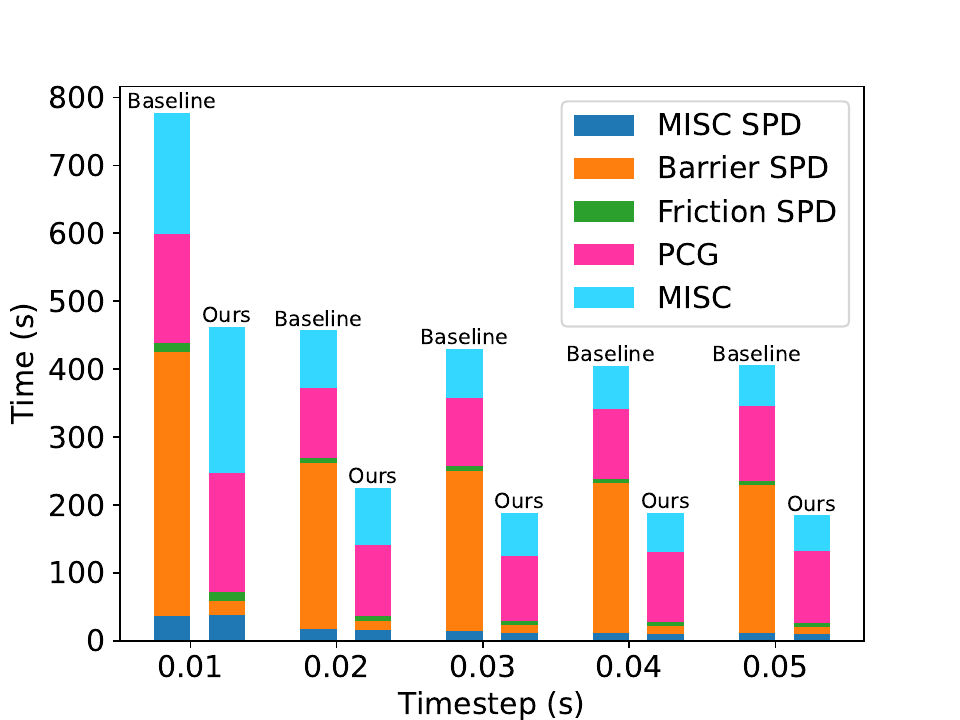}
		\end{subfigure}
		\caption{By employing the Funnel test showcased in Fig. (14) and conducting the comparison illustrated in Fig. (9) of the main paper \textit{with friction added}, we include a time breakdown of the friction- and barrier Hessian projections, along with the remaining miscellaneous components of our simulator.}
		\label{fig:friction-breakdown2}
	\end{figure}

	\figref{friction-breakdown} provides a comprehensive breakdown of compute time between local friction- and barrier Hessian projection alongside the remaining miscellaneous local projections that take place in our simulator (e.g. elastic energy Hessian). The results show that projection of the friction hessian has negligible overhead when compared to our approximated analytic barrier Hessian projection. We have found friction Hessian projection to merely require one-third to one-half the time needed to compute barrier Hessian projection. 
	
	To provide a deeper understanding of the role friction plays within large-scale contact simulations, we also run a duplicate experiment of the demo shown in Fig. (14) of the paper, setting the friction coefficient to $1e{-2}$  (\ie~$\mu$ parameter in \tabref{statscpuVcpu} or Tab. (5) in the paper). A breakdown w.r.t total simulation time is presented in \figref{friction-breakdown2}, where it can be observed that the time expenditure associated with friction Hessian projection has overall minimal footprint.

	\bibliographystyle{ACM-Reference-Format}
	\bibliography{bibliography}